\documentclass[aps,pra,amsmath,amssymb,
superscriptaddress, 
reprint]{revtex4-1}
\usepackage{CJK}

\usepackage{graphicx}
\usepackage{dcolumn}
\usepackage{bm}

\usepackage[separate-uncertainty = true,multi-part-units=single]{siunitx} 
\usepackage{xcolor} 
\usepackage{ulem}

\begin{document}
\begin{CJK*}{GB}{}

\title{{Kerr enhanced backaction cooling in magnetomechanics}}
\author{D.~Zoepfl}
\email{david.zoepfl@uibk.ac.at}
\affiliation{Institute for Quantum Optics and Quantum Information, Austrian Academy of Sciences, 6020 Innsbruck, Austria
}
\affiliation{Institute for Experimental Physics, University of Innsbruck, 6020 Innsbruck, Austria}

\author{M.~L.~Juan}
\affiliation{Institut Quantique and D\'epartement de Physique,
Universit\'e de Sherbrooke, Sherbrooke, Qu\'ebec, J1K 2R1, Canada}

\author{N.~Diaz-Naufal}
\affiliation{Dahlem Center for Complex Quantum Systems and Fachbereich Physik, Freie Universit\"{a}t Berlin, 14195 Berlin, Germany}

\author{C.~M.~F.~Schneider}
\affiliation{Institute for Quantum Optics and Quantum Information, Austrian Academy of Sciences, 6020 Innsbruck, Austria
}
\affiliation{Institute for Experimental Physics, University of Innsbruck, 6020 Innsbruck, Austria} 

\author{L.~F.~Deeg}
\affiliation{Institute for Quantum Optics and Quantum Information, Austrian Academy of Sciences, 6020 Innsbruck, Austria
}
\affiliation{Institute for Experimental Physics, University of Innsbruck, 6020 Innsbruck, Austria} 

\author{A.~Sharafiev}
\affiliation{Institute for Quantum Optics and Quantum Information, Austrian Academy of Sciences, 6020 Innsbruck, Austria
}
\affiliation{Institute for Experimental Physics, University of Innsbruck, 6020 Innsbruck, Austria} 

\author{A.~Metelmann}
\affiliation{Dahlem Center for Complex Quantum Systems and Fachbereich Physik, Freie Universit\"{a}t Berlin, 14195 Berlin, Germany}
\affiliation{Institute for Theory of Condensed Matter, Karlsruhe Institute of Technology, 76131 Karlsruhe, Germany}
\affiliation{Institute for Quantum Materials and Technology, Karlsruhe Institute of Technology, 76344 Eggenstein-Leopoldshafen, Germany}

\author{G.~Kirchmair}
\email{gerhard.kirchmair@uibk.ac.at}

\affiliation{Institute for Quantum Optics and Quantum Information, Austrian Academy of Sciences, 6020 Innsbruck, Austria
}
\affiliation{Institute for Experimental Physics, University of Innsbruck, 6020 Innsbruck, Austria}

\date{\today}

\begin{abstract}
{
Optomechanics is a prime example of light matter interaction, where photons directly couple to phonons, allowing to precisely control and measure the state of a mechanical object. This makes it a very appealing platform for testing fundamental physics or for sensing applications.
Usually, such mechanical oscillators are in highly excited thermal states and require cooling to the mechanical ground state for quantum applications, which is often accomplished by utilising optomechanical backaction. However, while massive mechanical oscillators are desirable for many tasks, their frequency usually decreases below the cavity linewidth, significantly limiting the methods that can be used to efficiently cool. Here, we demonstrate a novel approach relying on an intrinsically nonlinear cavity to backaction-cool a low frequency mechanical oscillator. We experimentally demonstrate outperforming an identical, but linear, system by more than one order of magnitude. Furthermore, our theory predicts that with this approach we can also surpass the standard cooling limit of a linear system. By exploiting a nonlinear cavity, our approach enables efficient cooling of a wider range of optomechanical systems, opening new opportunities for fundamental tests and sensing.

}

\end{abstract}

\maketitle
\begin{figure}[!t]
    \centering
    \includegraphics[width = 8.8cm]{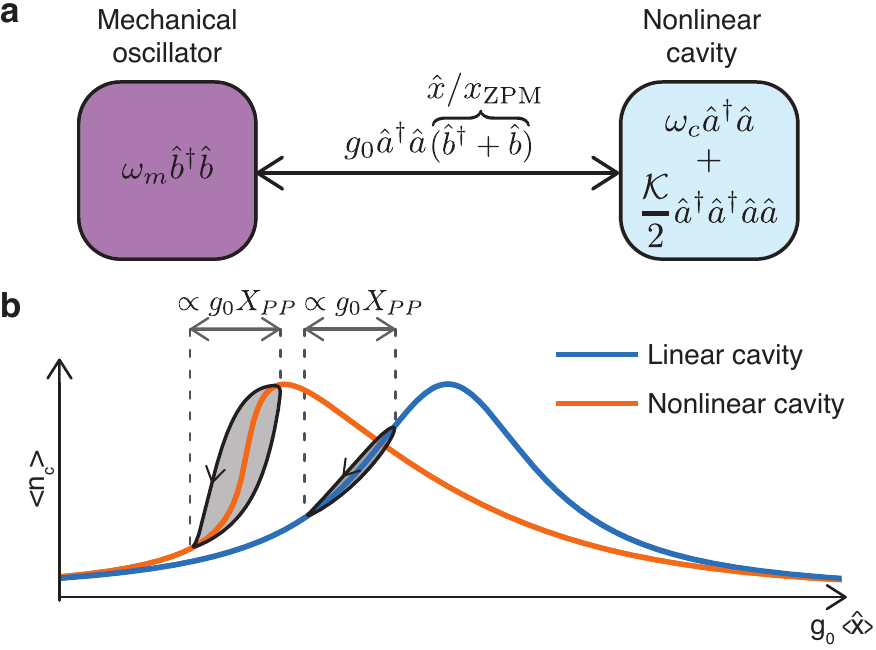}
    \caption{Nonlinear cooling illustration. \textbf{a,} Optomechanical interaction between a mechanical oscillator and a nonlinear cavity. \textbf{b,} To illustrate the enhanced cooling, we compare the response of the cavity photon number, $n_c$, of a linear to an identical nonlinear cavity. The cavity frequency changes due to the optomechanical interaction ($g_0 \langle \hat{x} \rangle$), changing the probe-cavity detuning. For the linear case we recover a symmetric response, while for the nonlinear one the typical nonlinear response. This is identical to sweeping the probe tone itself through a fixed frequency cavity. The shaded area indicates the cooling work done on a mechanical system within a cycle by the cavity for parameters from the experiment. $X_{PP}$ denotes the peak-to-peak amplitude of the mechanical resonator. The cooling enhancement provided by the nonlinearity is clearly visible.}
    \label{fig:nonLinCoolingSketch}
\end{figure}

{
Cooling mechanical modes in, or close to, their motional ground state is central for quantum applications. Even at cryogenic temperatures most systems are highly populated and further cooling is necessary. Such cooling can be achieved with feedback cooling~\cite{manciniOptomechanicalCoolingMacroscopic1998b,cohadonCoolingMirrorRadiation1999b,tebbenjohannsMotionalSidebandAsymmetry2020,rossiMeasurementbasedQuantumControl2018b}, or utilising a cavity to perform sideband cooling~\cite{marquardtQuantumTheoryCavityAssisted2007a,wilson-raeTheoryGroundState2007a} which works best in the so called good cavity regime, where the mechanical frequency exceeds the cavity decay rate ($\omega_m \gg \kappa$). There, a linear cavity is desirable to allow for high photon numbers and cooling to the ground state has been shown several years ago~\cite{teufelSidebandCoolingMicromechanical2011b,chanLaserCoolingNanomechanical2011b}. As mechanical systems increase in size, their frequency naturally decreases, which inevitably brings them into the bad cavity regime ($\omega_m \ll \kappa$). There, the same cooling mechanism still applies, it is however limited to a finite phonon occupation due to unwanted backaction~\cite{aspelmeyerCavityOptomechanics2014}. Different schemes to overcome this limitation have already been proposed, which include using two mechanical modes~\cite{ojanenGroundstateCoolingMechanical2014}, two cavity modes~\cite{yangGroundstateCoolingMechanical2019, liuCoupledCavitiesMotional2015}, frequency modulated light~\cite{wangOptomechanicalCoolingQuantum2018} or entirely different coupling mechanisms, such as either coupling to the cavity decay rate~\cite{elsteQuantumNoiseInterference2009b} instead of the usual dispersive coupling, or coupling the mechanical system additionally to two level systems~\cite{genesMicromechanicalOscillatorGroundstate2009}. Another approach that gained much attention is to use squeezed light created outside or even inside the cavity to improve the cooling performance~\cite{xiongStrongSqueezingDuffing2020,asjadOptomechanicalCoolingIntracavity2019,asjadSuppressionStokesScattering2016,huangEnhancementCavityCooling2009,ganIntracavitySqueezedOptomechanicalCooling2019,clarkSidebandCoolingQuantum2017b}.


Here, we present a fundamentally different, yet very simple, approach by using an intrinsically nonlinear cavity dispersively coupled with a mechanical system, Fig.~\ref{fig:nonLinCoolingSketch}a, proposed in~\cite{laflammeQuantumlimitedAmplificationNonlinear2011,nationQuantumAnalysisNonlinear2008b}. We show that the optomechanical cooling is much more efficient than an - otherwise identical - linear system. Interestingly, the benefits of this nonlinear cooling scheme arise in the bad cavity regime, in contrast with another recent experiment utilising a nonlinear system in the good cavity regime to demonstrate cooling using four-wave mixing~\cite{bothnerFourwavecoolingSinglePhonon2022}.
}

{
A nonlinear cavity dispersively coupled to a mechanical resonator, Fig.~\ref{fig:nonLinCoolingSketch}a, can be described by~\cite{nationQuantumAnalysisNonlinear2008b,diaz-naufal}:
\begin{equation}
    \hat{H}/\hbar = \omega_c \hat{a}^\dagger \hat{a} + \omega_m \hat{b}^\dagger \hat{b} + \frac{\mathcal{K}}{2} \hat{a}^\dagger \hat{a}^\dagger \hat{a} \hat{a} + g_0 \hat{a}^\dagger \hat{a} (\hat{b}^\dagger + \hat{b}) + \hat{H}_{\text{d}}
    \label{equ:H}
\end{equation}
Here, $\hat{a}^\dagger (\hat{a})$ and $\hat{b}^\dagger (\hat{b})$ are the creation (annihilation) operators of the cavity and the mechanical resonator and the respective frequencies are given by $\omega_c$ and $\omega_m$. The nonlinearity of the cavity is introduced by the Kerr constant $\mathcal{K}$, leading to a frequency shift per photon. The coupling strength between the two systems is given by the single-photon coupling strength $g_0$ and $\hat{H}_{\text{d}}$ is an external drive. The position operator of the mechanical mode, translates as $\hat{x} = x_{\text{zpm}} (\hat{b}^\dagger + \hat{b})$, where $x_{\text{zpm}}$ is the mechanical zero point motion. Additional information is given in the supplementary material~\cite{SeeSupplementaryMaterial}.

For an intuitive picture of the nonlinear cooling we consider the cavity response to a fixed frequency probe tone, while the optomechanical interaction changes the probe-cavity detuning, Fig.~\ref{fig:nonLinCoolingSketch}b. In case of a completely linear system, a symmetric response is recovered, identical to the response when sweeping a probe tone over a fixed frequency cavity. The origin of the cooling can be understood as a time lag of the cavity photons due to a finite cavity lifetime~\cite{marquardtQuantumTheoryOptomechanical2008}. The shaded area depicts the cooling work done on the mechanical system within one cycle for an ideal red detuned probe tone, which we simulate with a simple model~\cite{SeeSupplementaryMaterial} using parameters closely related to the experiment. Now, let us consider a cavity with a negative Kerr, such that the cavity is close to bistability~\cite{muppallaBistabilityMesoscopicJosephson2018} (i.e. where the response of the cavity gets infinitely steep at a certain detuning) using otherwise the same parameters as for the linear case discussed previously. Probing this system with a fixed frequency tone, we obtain the typical nonlinear response. For lower drive strengths the cavity would be effectively linear, while for higher drives bistability is reached and two metastable states appear in a certain range of detunings. As the enclosed area within one cycle increases, it is evident that the cooling is enhanced compared to the linear case. Due to the nonlinear line shape, small changes of the cavity frequency related to the mechanical motion, induce a large variation of the cavity photon number, which not only increases the cooling itself, but also suppresses the unwanted backaction heating. Working on the blue side (i.e. frequency of the probe tone above the cavity frequency), the cavity slope is effectively more shallow compared to the linear case, leading to a decrease of the heating backaction. Alternatively, the nonlinear enhanced cooling can also be explained via the usual scattering picture. Due to the nonlinearity, the density of states of the cavity is asymmetric, which impacts the Stokes and anti-Stokes rates. Additionally, due to the mechanical oscillation, the photon number changes, which is given by the very asymmetric shape of the intracavity photon number. The combination of both effects leads to the enhancement of the cooling performance, using a nonlinear cavity~\cite{diaz-naufal}. For a positive Kerr, this effect would be entirely reversed.
}

\begin{figure}[!t]
    \centering
    \includegraphics[width = 8.8cm]{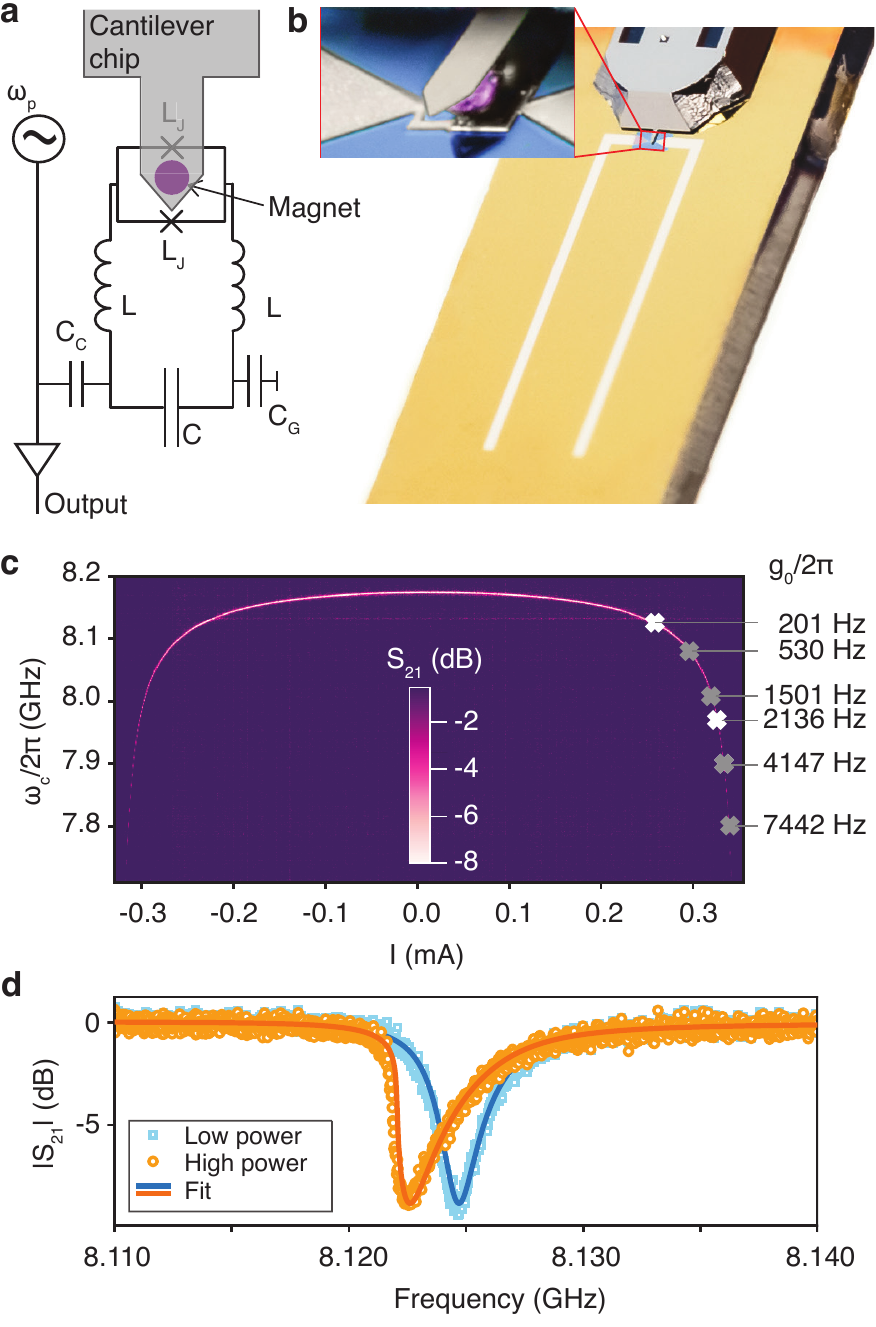}
    \caption{Setup and characterisation. \textbf{a,} Schematic depiction of our setup. $L$ and $L_J$ are the cavity and the junction inductance, $C$, $C_c$ and $C_G$ denote the self capacitance of the cavity, the coupling and the ground capacitance. \textbf{b,} Picture of the microstrip cavity (white) on top of a Silicon substrate (golden). The cantilever chip (gray) is glued to the Silicon substrate. An optical microscope picture (zoom in) shows the cantilever with false coloured magnet above the SQUID loop. \textbf{c,} Change of cavity frequency when applying external magnetic flux. Indicated are the measured coupling values $g_0$ as the flux sensitivity changes. White crosses symbolise measurements presented in the main text, grey crosses in~\cite{SeeSupplementaryMaterial}. \textbf{d,} Characterisation of the cavity nonlinearity. Close to bistability, the response is in good agreement with the model for a Kerr nonlinear cavity~\cite{SeeSupplementaryMaterial}. For low drive powers we effectively recover a linear response.}
    \label{fig:setup}
\end{figure}
{
The setup consists of a superconducting microstrip cavity coupled to a single clamped beam - a cantilever - with a magnet on its tip, Fig.~\ref{fig:setup}(a,b), similar to the setup discussed in~\cite{zoepflSinglePhotonCoolingMicrowave2020a,zoepflCharacterizationLowLoss2017}. A superconducting quantum interference device (SQUID) embedded in the cavity makes it sensitive to magnetic fields, mediating the inductive coupling to the cantilever. Recently also other experimental realisations using inductively coupled optomechanical systems have been demonstrated~\cite{bothnerFourwavecoolingSinglePhonon2022, beraLargeFluxmediatedCoupling2021, luschmannMechanicalFrequencyControl2022}. Our setup is mounted to the base plate of a dilution refrigerator, kept at \SI{100}{\milli \kelvin} for most of the experiments, where the system is very stable and the mechanical mode is well thermalised. The cavity has a frequency of $\omega_c / 2 \pi  = \SI{8.176}{\giga \hertz}$ with a linewidth of $\kappa / 2 \pi = \SI{3.5}{\mega \hertz}$, while the cantilever has a frequency of $\omega_m / 2 \pi = \SI{274.41}{\kilo \hertz}$ with a linewidth of approximately $\Gamma_m /2 \pi = \SI{0.4}{\hertz}$, which means that the setup resides deep in the bad cavity regime. In Fig.~\ref{fig:setup}c we show the dependence of the cavity frequency on the magnetic field, which also tunes its sensitivity and determines - up to a prefactor - the coupling rate. So far we directly measured single-photon coupling strengths of up to \SI{7.4}{\kilo \hertz}, while the sensitivity of the cavity allows couplings exceeding \SI{90}{\kilo \hertz}~\cite{SeeSupplementaryMaterial}. Excessive flux noise does not allow for a stable operation at those flux sensitivities, where we estimate a flux noise of \SI{302(19)} {\micro \phi_0}, which is mainly induced by mechanical vibrations~\cite{SeeSupplementaryMaterial}.

Another aspect of a SQUID is its nonlinearity with regard to input power, since its inductance also depends on the number of photons circulating in the cavity. As the number of photons in the cavity increases, the frequency shifts to lower values, leading to the nonlinear response, when scanning a sufficiently strong probe tone across, Fig.~\ref{fig:setup}d. This effect leads to an enhanced cooling, as discussed previously, Fig.~\ref{fig:nonLinCoolingSketch}.
}

\begin{figure}[!t]
    \centering
    \includegraphics[width = 8.8cm]{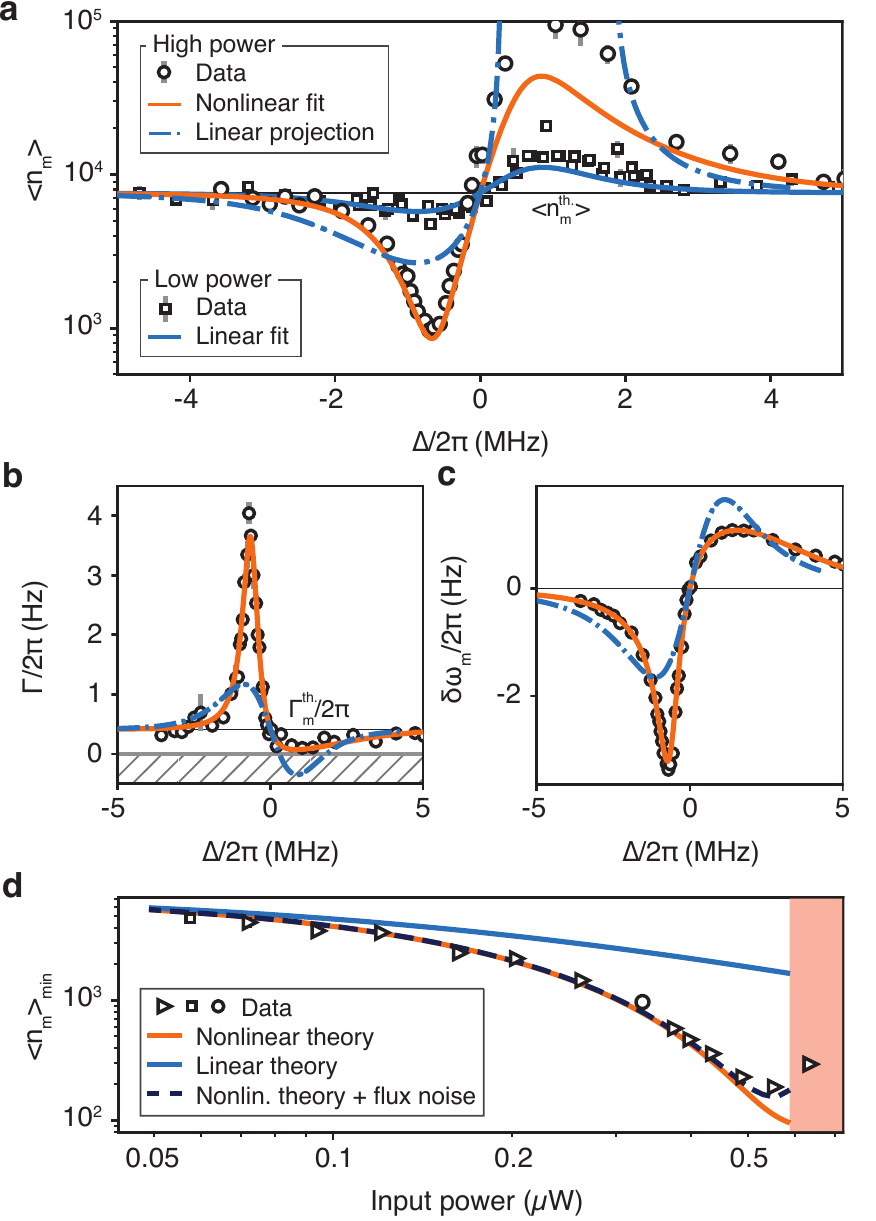}
    \caption{Nonlinear cooling measurement at $g_0 /2 \pi = \SI[separate-uncertainty = true]{201(3)}{\hertz}$. \textbf{a,} Phonon number $\langle n_m \rangle$ against probe-cavity detuning, $\Delta$, for two powers. \textbf{b-c,} Mechanical linewidth and frequency against probe-cavity detuning for the high power case. The suppressed heating prevents us from entering the instable regime ($\Gamma / 2 \pi < 0$). \textbf{d,} Lowest phonon number, $\langle n_m \rangle_{\text{min}}$, for different input power. Right before bistability (shaded region), the cooling is limited by flux noise, which is captured in a simple model~\cite{SeeSupplementaryMaterial}. The errors shown are the standard errors~\cite{SeeSupplementaryMaterial}.}
    \label{fig:nonLinCoolingData}
\end{figure}

We measure the cantilever motion by taking a homodyne noise spectrum of the probe tone, where the mechanical signature appears as an amplitude (phase) modulation sideband~\cite{zoepflSinglePhotonCoolingMicrowave2020a} and fit this sideband with the model of a damped harmonic oscillator~\cite{gorodetksyDeterminationVacuumOptomechanical2010a}. To investigate the backaction, we measure the mechanical cantilever for different detunings between the probe tone and the cavity. Taking such cooling traces for several powers up to bistability reveals the cooling enhancement due to the nonlinearity. For this measurement, we work at a moderate coupling of \SI[separate-uncertainty = true]{201(3)}{\hertz} to avoid limiting effects from flux noise, where the cavity shows a Kerr nonlinearity of $\mathcal{K}/2 \pi = \SI[separate-uncertainty = true]{-12.2(1)}{\kilo \hertz}\text{/Photon}$. Fig.~\ref{fig:nonLinCoolingData}a shows such cooling traces for two different powers. For the low power measurement, the cavity is effectively in the linear regime and thus the cooling curve agrees well with linear theory~\cite{safavi-naeiniLaserNoiseCavityoptomechanical2013a,marquardtQuantumTheoryCavityAssisted2007a}. Conversely, in the high power regime only the nonlinear theory describes the measurement data accurately. Assuming a linear cavity, but otherwise identical parameters would predict much weaker cooling and much stronger heating than what we observe. We note that in this regime the cooling happens over a much narrower range of detunings such that a fit with the linear theory is not in good agreement with the data~\cite{SeeSupplementaryMaterial}. In Fig.~\ref{fig:nonLinCoolingData}b,c we plot the change of mechanical linewidth and frequency for the high power measurement. Again, we only observe good agreement with the nonlinear theory. We further clearly demonstrate the asymmetry in backaction strength when either working red detuned (cooling) or blue detuned (heating) with respect to the cavity as expected for a negative Kerr, already illustrated in Fig.~\ref{fig:nonLinCoolingSketch}.

In Fig.~\ref{fig:nonLinCoolingData}d we show the lowest phonon number measured for each cooling trace~\cite{SeeSupplementaryMaterial} against the input power together with the predictions from the linear and nonlinear theory. For increasing power, we clearly see the cooling enhancement due to the nonlinear cavity, reaching strongest cooling just before bistability. There we outperform a conventional linear system by more than an order of magnitude, making the nonlinear cooling a very efficient cooling scheme. To properly model the impact of flux noise at high input power, we include a Gaussian distribution for the detuning instead of a fixed value, which reproduces the measurement data well~\cite{SeeSupplementaryMaterial}.


\begin{figure}[!t]
    \centering
    \includegraphics[width = 8.8cm]{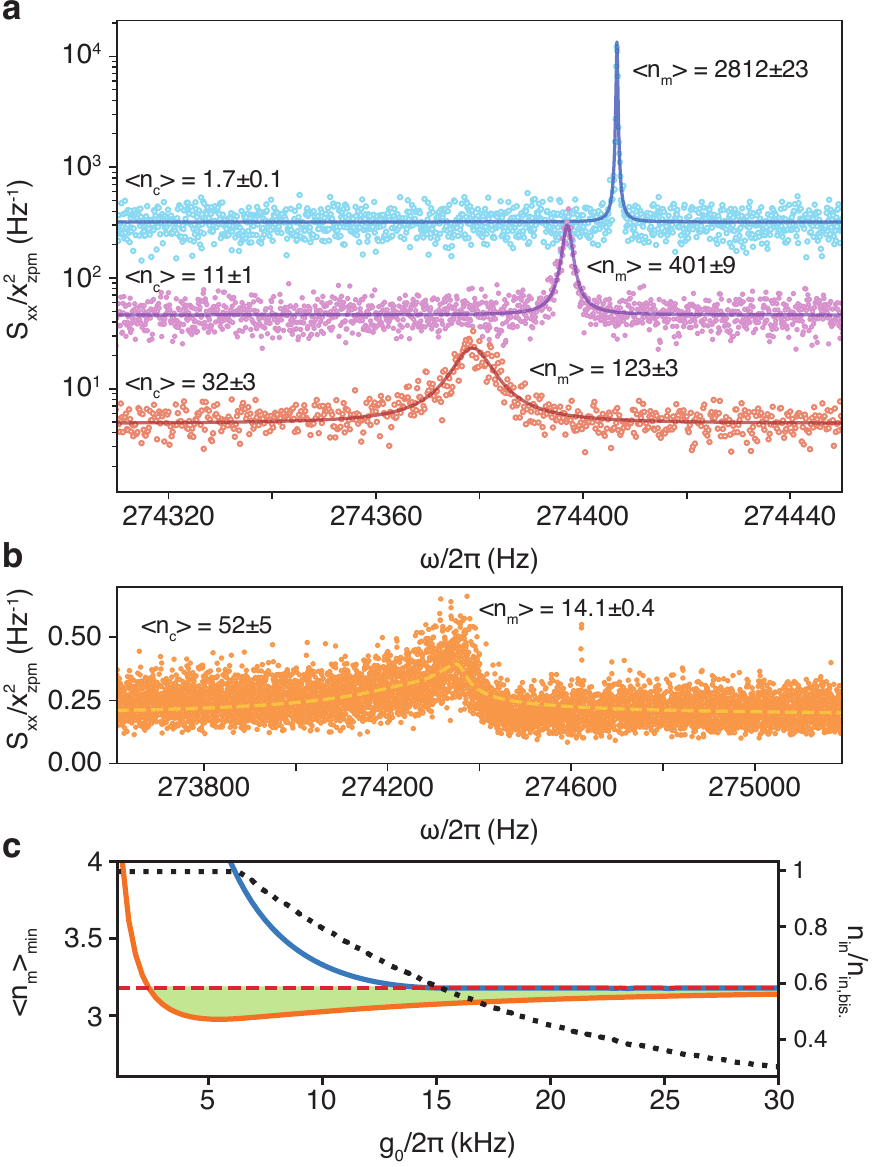} %
    \caption{Best cooling using higher $g_0$. \textbf{a,} Mechanical noise spectral densities and corresponding fits at \SI{40}{\milli \kelvin} and $g_0 /2 \pi = \SI[separate-uncertainty = true]{2136(26)}{\hertz}$ for increasing backaction by changing the input power and the probe-cavity detuning. The noise floor decreases with increasing backaction, which is also a result of our cavity acting as a parametric amplifier~\cite{hatridgeDispersiveMagnetometryQuantum2011a}. $\left< n_{\text{c}} \right>$ are the average number of photons circulating in the cavity for the measurement, the uncertainties arise from uncertainties in the input photon number, linewidth and detuning. The uncertainties for the phonon numbers are fit errors. \textbf{b,} Spectral density for one of the lowest mechanical occupation numbers we measured. We fit this trace with a model including flux noise (dashed,~\cite{SeeSupplementaryMaterial}). The uncertainty on the phonon number comes from the uncertainty of the offset determination for the numeric integration~\cite{SeeSupplementaryMaterial}). \textbf{c,} Theory prediction for the lowest phonon occupation reachable when increasing $g_0$ always using the optimal input power for the nonlinear case ($\mathcal{K}/2 \pi = \SI{-12}{\kilo \hertz}/n_c$) up to bistability (orange, the dotted line is the input power). For the comparison to linear theory we either use the same power (blue line) or the ideal power for the linear case (red dashed). All other parameters remain constant. Nonlinear cooling allows to cool to $\langle n_m \rangle = 2.97$, while in the linear case cooling is limited to 3.18 phonons.}
    \label{fig:BestCooling}
\end{figure}
To explore the limits of the system, we move to higher couplings, where we expect an increased backaction and thus increased cooling, but at the cost of higher sensitivity to flux noise. To reach the lowest phonon occupation currently possible in our setup, we use a coupling of $g_0 /2 \pi = \SI[separate-uncertainty = true]{2136(26)}{\hertz}$ and decrease the temperature of our cryostat to \SI{40}{\milli \kelvin}. This is doubly beneficial since the thermal phonon occupation and the linewidth of the cantilever become smaller, reducing the impact of the environment~\cite{SeeSupplementaryMaterial}. What further helps us, is a slight anomaly in the Kerr. In this region we still have the discussed benefits of a nonlinear system but with a smaller Kerr, which allows us to drive the system even harder, together with a high coupling strength. The disadvantage here is that the anomalous Kerr prevents us from modelling the full system as in Fig.~\ref{fig:nonLinCoolingData}~\cite{SeeSupplementaryMaterial}.

In Fig.~\ref{fig:BestCooling}a we show three mechanical noise spectral densities with increasing cooling backaction by increasing the input power and changing the probe-cavity detuning, starting with a thermal spectral density. In Fig.~\ref{fig:BestCooling}b we plot a spectral density for one of the lowest phonon occupation numbers we can currently reach, which is heavily influenced by flux noise. Thus, instead of the usual fit, we extract the phonon number in two different ways: by directly integrating the area below the peak and by fitting the data using a model including flux noise~\cite{SeeSupplementaryMaterial}. Both methods show good agreement and give an occupation of around 14 phonons, around 200 times below the thermal occupation.

Finally, we theoretically investigate the full capabilities of nonlinear cooling without the limiting factors of flux noise and hence also the possibilities of increasing $g_0$ further. In Fig.~\ref{fig:BestCooling}c we show the lowest phonon number achievable for increasing coupling. For the nonlinear theory, shown in orange, we vary the power (dotted line) to minimise the phonon number while operating to at most 99\% of the bistable power. For comparison, the linear cooling using the same power is shown in blue, providing at low $g_0$ a less efficient cooling, something which was already observed in the experimental data, Fig.~\ref{fig:nonLinCoolingData}. Remarkably, when not limiting the power used for the linear cavity to minimise the phonon occupation (red dashed line), the nonlinear cavity still provides better a cooling. Thus, nonlinear cooling is not only more efficient, but also allows to reach a lower phonon number in the bad cavity limit.

To conclude, we demonstrate a novel way of cooling an optomechanical system by using an intrinsically nonlinear cavity. We show that the nonlinearity has to be crucially taken into account when describing optomechanical backaction on the mechanical cantilever and demonstrate a ten fold cooling enhancement compared to a - otherwise identical - linear system. By increasing the coupling, we show cooling of the mechanical occupation from 2800 thermal phonons to 14 phonons. However, not only does flux noise prevent us from cooling to lower occupation, it also restricts us from operating at even higher couplings, while the sensitivity of our cavity allows for a single-photon coupling strength exceeding \SI{90}{\kilo \hertz}. There we would reach single-photon quantum cooperativity exceeding unity when working at \SI{40}{\milli \kelvin} and assuming that we reduce our flux noise by a factor of 2.

Interestingly, for some region of coupling strengths, the nonlinear cooling even beats the standard cooling limit of a linear system by a small amount, where the region can be tuned by the nonlinearity. This combines an optomechanical system in the bad cavity regime with a nonlinear cavity, which are separately often considered as unfavourable. Furthermore, previous work~\cite{clarkSidebandCoolingQuantum2017b} has experimentally demonstrated that externally generated squeezed light injected into the cavity can eliminate unwanted backaction. In this context, it will be interesting to study whether or not internal squeezing that can be generated in Kerr cavities can be applied as a resource for enhanced cooling. This approach is especially relevant when working with massive mechanical systems, which are essential for many tests of fundamental physics and sensing applications~\cite{gelySuperconductingElectromechanicsTest2021,romero-isartQuantumSuperpositionMassive2011a,pikovskiProbingPlanckscalePhysics2012,arndtTestingLimitsQuantum2014,whittleApproachingMotionalGround2021,krauseHighresolutionMicrochipOptomechanical2012}.

\begin{acknowledgments}
\textbf{Acknowledgement:}  We want to thank Hans Huebl and John D. Teufel for fruitful discussion. Further we thank our in-house mechanical workshop. In addition, we want to thank the referees for their useful comments, which helped to improved our article. DZ is funded by the European Union$'$s Horizon 2020 research and innovation program under grant agreement No. 736943 and grant agreement No. 101080143. MLJ acknowledges funding by the Canada First Research Excellence Fund. CMFS and LD are supported by the Austrian Science Fund FWF within the DK-ALM (W1259-N27). AM acknowledges funding by the Deutsche Forschungsgemeinschaft through the Emmy Noether program (Grant No. ME 4863/1-1) and NDN is supported by the Deutsche Forschungsgemeinschaft project CRC 183.
\end{acknowledgments}

\nocite{probstEfficientRobustAnalysis2015,gardinerDrivingQuantumSystem1993,HttpsScipyOrg,gardinerInputOutputDamped1985a,HttpsScikitlearnOrg,burnhamModelSelectionMultimodel2004,kornevNumericalSimulationJosephsonjunction1997,polonskyPSCANPersonalSuperconductor1991}


%

\clearpage
\onecolumngrid

\begin{center}
\textbf{\large Supplemental information: Kerr enhanced backaction cooling in magnetomechanics}
\end{center}

\makeatletter
\renewcommand{\theequation}{S\arabic{equation}}
\renewcommand{\thefigure}{S\arabic{figure}}
\renewcommand{\thesection}{S\arabic{section}}

\section{Setup}
\label{secS:setup}
\begin{figure*}[h]
    \centering
    \includegraphics[width = 16cm]{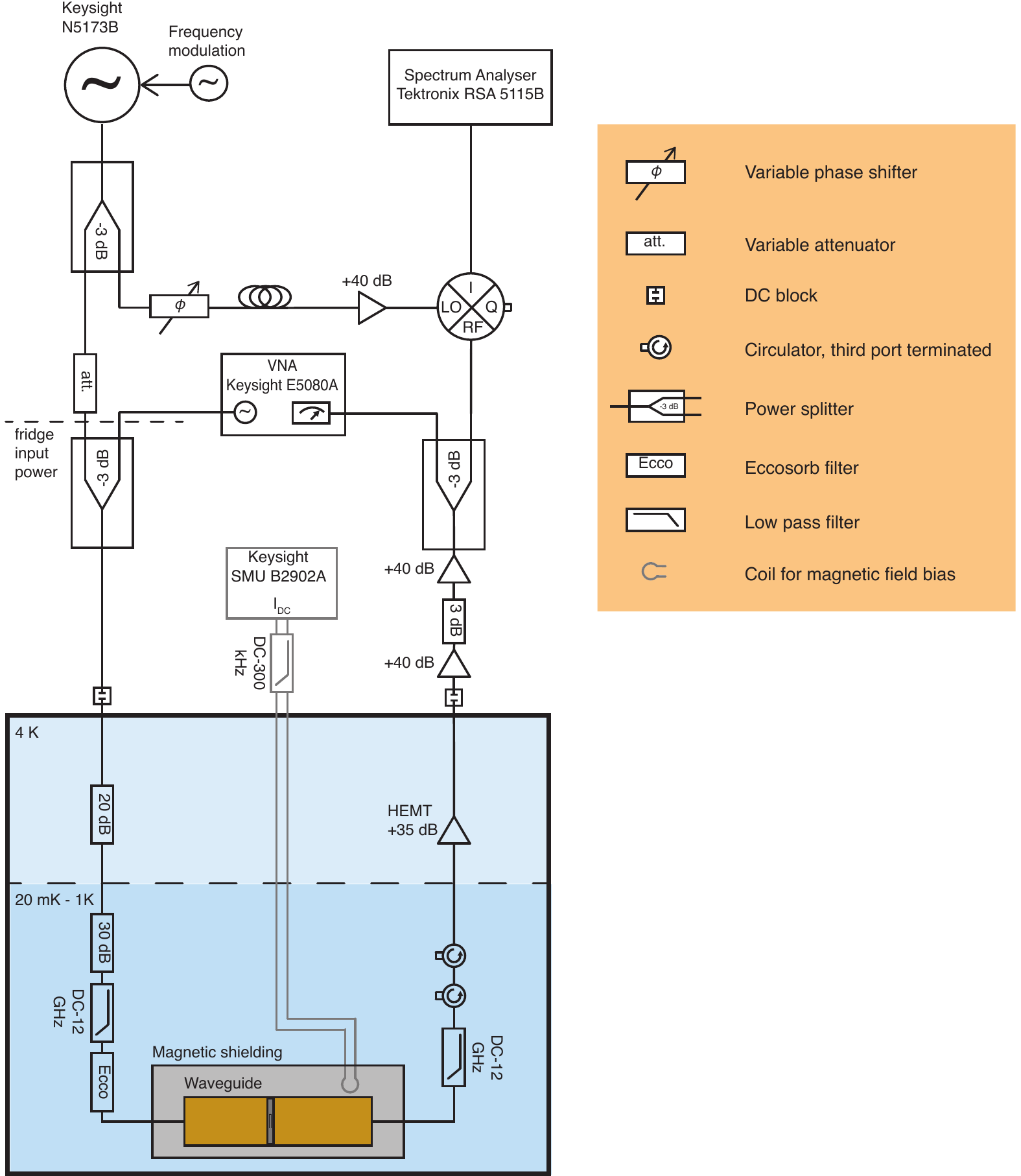}
    \caption{Full measurement setup.}
    \label{figS:setup}
\end{figure*}

The sample we use as well as the measurement setup (Fig.~\ref{figS:setup}) is very similar to the one described in~\cite{SzoepflSinglePhotonCoolingMicrowave2020a}. We also use the same calibration method as described in~\cite{SgorodetksyDeterminationVacuumOptomechanical2010a} to determine the optomechanical coupling rate. In contrast to~\cite{SzoepflSinglePhotonCoolingMicrowave2020a} we use an additional room temperature amplifier before downmixing the signal, to improve the robustness of our calibration method. Also we use a different variable phase shifter having a higher return loss, which we compensate for by operating the signal generator at -3~dBm compared to -7~dBm in~\cite{SzoepflSinglePhotonCoolingMicrowave2020a}. When referring to an input power in the following figures, we consider the input into the setup (see Fig.~\ref{figS:setup}).

\section{Characterisation of the microwave cavity}
We characterise the cavity by a low power transmission measurements using the VNA, followed by an analysis using the circle fit routine~\cite{SprobstEfficientRobustAnalysis2015}:
\begin{align}
     S_{21}^{\text{full}} (\omega) = (a e^{i \alpha} e^{- i \omega \tau}) S_{21} (\omega) \\
     S_{21} (\omega) =  1 - \frac{Q_l/|Q_c| e^{i \phi_0}}{1 + 2 i Q_l \frac{\omega - \omega_c}{\omega_c}}.
     \label{equS:S21_full}
\end{align}
Here, $a$ and $\alpha$ are the effects of the environment. Another effect of the environment is the electrical delay $\tau$. Further, $Q_l$ is the loaded (total) quality factor, $|Q_c|$ the absolute value of the coupling quality factor and $\phi_0$ quantifies the impedance mismatch. The cavity frequency is given by $\omega_c$ and the probe frequency by $\omega$.
\begin{figure*}[h]
    \centering
    \includegraphics[width = 10cm]{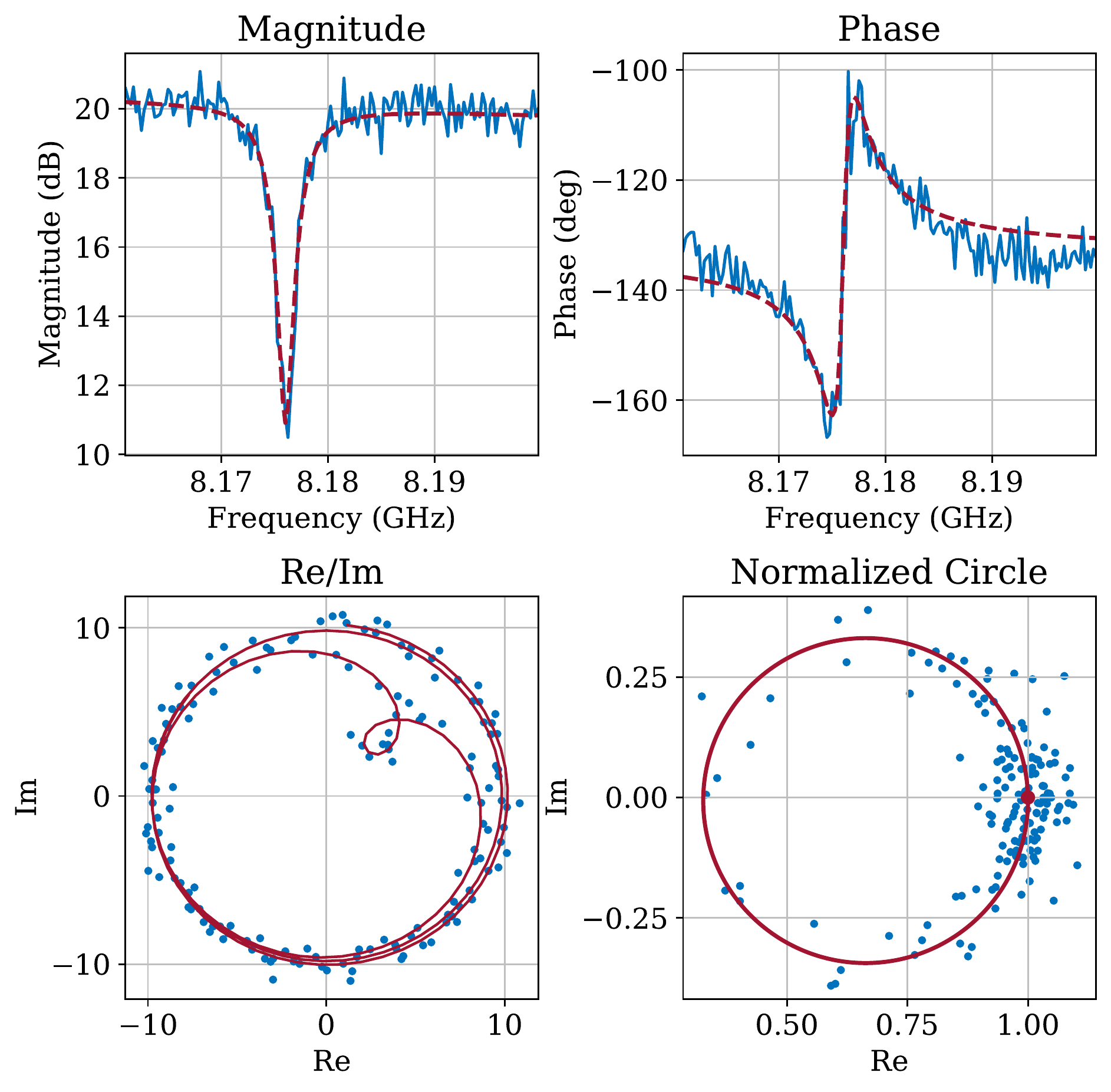}
    \caption{Circle fit to the cavity measured in transmission. Blue is the measurement data and red the fit. Top left: Direct magnitude $|S_{21}|$ of the measurement data and fit. Top right: Measured phase of $S_{21}$ with the electrical delay already subtracted. Bottom left: Direct measurement data in the complex plane and fit of the full model. Bottom right: Data and fit in the complex plane after subtraction of the environment.}
    \label{figS:full_circle_fit}
\end{figure*}
In Fig.~\ref{figS:full_circle_fit} we show the data along with the full circle fit, the results are given in table~\ref{tabS:fit_res_cav}.

\begin{table}[ht]
    \caption{Parameters obtained by a circle fit to the cavity. The internal quality factor $Q_{int}$ is calculated from the fit parameters, as $Q_l^{-1} = Q_c^{-1} + Q_{\text{int}}^{-1}$.}
    \centering
    \begin{tabular}{c|c|c|c|c|c|c|c|c}
    Fit parameter & $a$ & $\alpha$ & $\tau$ & $\omega_c/ 2 \pi$ & $\phi_0$ &  $Q_l$ & $Q_c$ & $Q_{\text{int}}$ \\ \hline
    Value & 10.1  & -2.33 rad & \SI{73.7}{ns} & \SI{8.1760}{GHz} & \SI{0.02}{rad} & 2349 & 3485  & 7209 
    \end{tabular}
    \label{tabS:fit_res_cav}
\end{table}

\begin{figure*}[h]
    \centering
    \includegraphics[width = 12cm]{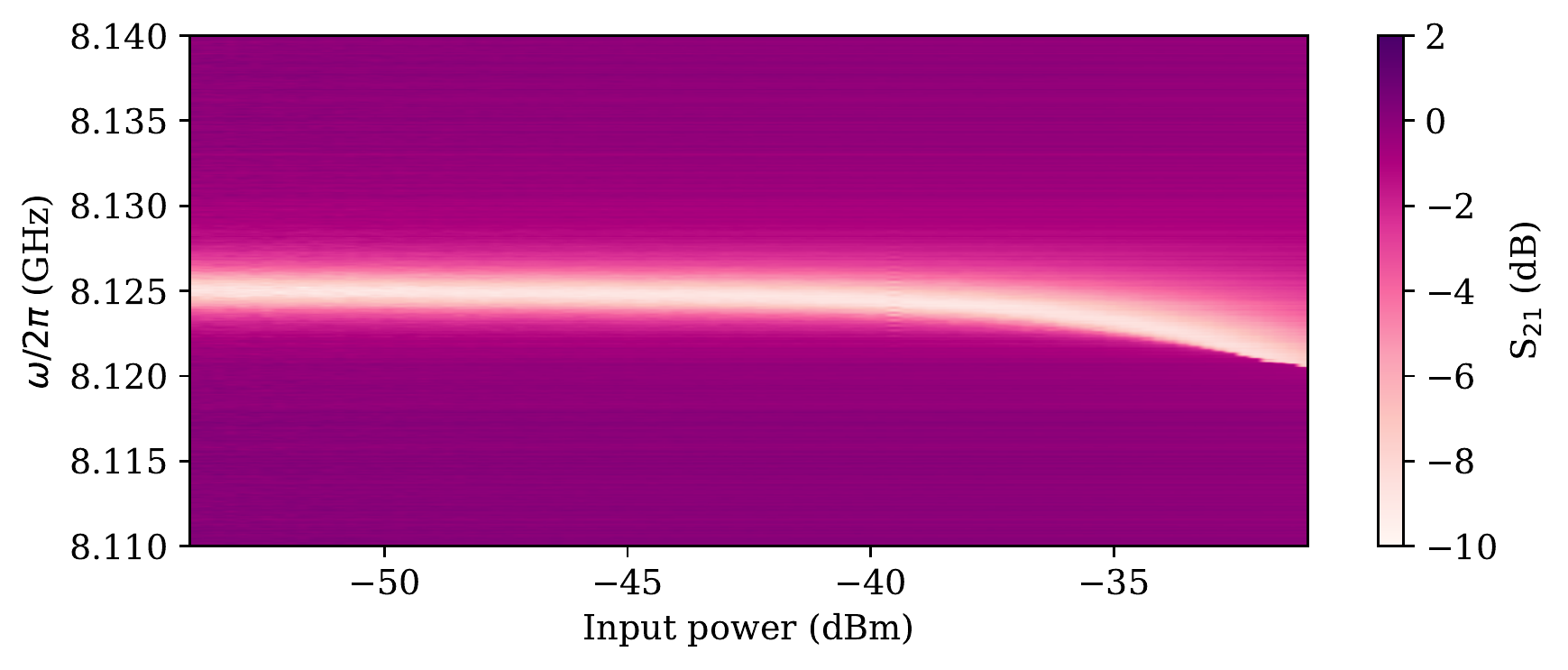}
    \caption{Response of the microwave cavity for increasing drive strengths when sweeping a probe tone across.}
    \label{figS:kerrShift_8p125}
\end{figure*}

Performing a transmission measurement of the cavity with increasing power, we can investigate the nonlinearity of the cavity, as discussed in the main article. In Fig.~\ref{figS:kerrShift_8p125} we show the response of the cavity for sweeping a probe tone with increasing power across the cavity.
Clearly the frequency of the cavity shifts to lower values with increasing drive strength. In Fig.~2d of the main article, we show the measurement at an input power of -42~dBm, which is in the linear regime, and the measurement at -33.5~dBm, deep in the nonlinear regime. We fit those measurements by modelling the nonlinear response of the cavity~\cite{SmuppallaBistabilityMesoscopicJosephson2018}. The nonlinearity arises from the SQUID embedded in the cavity, as its inductance changes with the number of circulating photons. We can compare the strength of the nonlinearity obtained in this measurement (Fig.~\ref{figS:kerrShift_8p125}) to the nonlinearity obtained when fitting a nonlinear cooling trace (Fig.~\ref{figS:nonLinCooling_201Hz}), which we convert and determine from the frequency shift per circulating cavity photon (Kerr shift). We calibrate the input power to photon number estimating the attenuation from the input of the fridge to the cavity and convert this to circulating photon number. With this we get an estimated Kerr of $$\mathcal{K}/2 \pi = [-10.26 \pm 0.02 (\text{fit}) \pm 2.58 (\text{sys.})]\SI{}{\kilo \hertz} / \text{Photon}.$$ This is in good agreement to the value obtained from fitting the cooling traces (Fig.~\ref{figS:nonLinCooling_201Hz}). The systematic error comes from the uncertainty in the input power attenuation, where we estimate an uncertainty of \SI{3}{\decibel}. It should be noted that we used a similar cavity frequency for this power sweep as well as for the cooling traces, as we typically observed an increasing Kerr with lower cavity frequency (and thus higher coupling, Sec.~\ref{secS:fluxmaps}).

\section{\label{sec:workcycles}Cavity time dynamics (work cycle)}

The time evolution of the cavity field can be calculated using input-output theory~\cite{SgardinerDrivingQuantumSystem1993}
\begin{equation}
\frac{\mathrm{d}}{\mathrm{d} t} \alpha=\left[i (\Delta - g_0\frac{x(t)}{x_\mathrm{ZPM}} - \mathcal{K}|\alpha|^{2}) - \frac{\kappa}{2}\right] \alpha -\sqrt{\kappa} \alpha_{\mathrm{in}},
\end{equation}
with the detuning $\Delta=\omega - \omega_c$, the coupling strength $g_0$, the linewidth $\kappa$, and the Kerr factor $\mathcal{K}$. This is a driven cavity, whose frequency is modulated by the mechanical oscillator. We assume the mechanical oscillator to be in a coherent state $\mathcal{B}(t) = \sqrt{\bar{n}_\mathrm{m}}e^{-i\omega_m t}$, leading to a displacement 
\begin{equation*}
x(t)=x_\mathrm{ZPM}(\mathcal{B} + \mathcal{B}^*)=2\sqrt{\bar{n}_\mathrm{m}}x_\mathrm{ZPM}\cos(\omega_m t).
\end{equation*}
We solve this complex ordinary differential equation with the scipy package from Python to obtain the cavity population as a function of time~\cite{SHttpsScipyOrg}. We can plot this over the mechanical displacement to obtain the work cycles shown in the main text.

\section{Nonlinear cooling theory}
\label{secS:nonLinTheo}
Let us consider a mechanical mode which is coupled to a Kerr-resonator, where the latter corresponds to a nonlinear oscillator which therefore can be driven into a bistable regime. The Hamiltonian of the system \cite{SnationQuantumAnalysisNonlinear2008b,Sdiaz-naufal} reads 
\begin{align}
    \hat{\mathcal{H}}_\text{tot} = \omega_c \hat{a}^\dagger \hat{a} + \omega_m \hat{b}^\dagger \hat{b} + \frac{\mathcal{K}}{12} \left( \hat{a} + \hat{a}^\dagger \right)^4 + \frac{g_0}{2} \left( \hat{a} + \hat{a}^\dagger \right)^2 \left( \hat{b} + \hat{b}^\dagger \right) + \hat{\mathcal{H}}_\text{d},
\end{align}
where $\hat{a}(\hat{b})$ and $\hat{a}^\dagger(\hat{b}^\dagger)$ are the annihilation and creation operator of the optical(mechanical) mode, respectively; $\omega_c(\omega_m)$ is the resonance frequency of the cavity(mechanical oscillator); $\mathcal{K}$ is the Kerr constant, typically assumed to be $\mathcal{K} < 0$, and $g_0$ denotes the bare optomechanical coupling strength. Moreover, the external drive driving the system is encoded in the last term as $\hat{\mathcal{H}}_\text{d} = \alpha_p e^{- i \omega_p t} \hat{a}^\dagger + \text{h.c}$, where $\omega_p$ and $\alpha_p$ are their resonant frequency and amplitude, respectively. Assuming a weak enough coupling $g_0$ and nonlinearity $\mathcal{K}$, the system can be simplified by neglecting small and constant shifts, and counter-rotating terms yielding
\begin{align}
     \hat{\mathcal{H}} = \omega_c \hat{a}^\dagger \hat{a} + \omega_m \hat{b}^\dagger \hat{b} + \frac{\mathcal{K}}{2} \hat{a}^\dagger \hat{a}^\dagger \hat{a}\hat{a} + g_0 \hat{a}^\dagger \hat{a} \left( \hat{b} + \hat{b}^\dagger \right) + \hat{\mathcal{H}}_\text{d}.
    \label{Ham_Approx1}
\end{align}

Under the assumption of strong driving at frequency $\omega_p$, we can perform a displacement transformation of the cavity mode operator: $\hat{a} = \alpha e^{- i \omega_p t} + \hat{d}$ with $\alpha = \langle \hat{a} \rangle$ the classical drive amplitude and $\hat{d}$ the fluctuations around this amplitude. Hence, moving into a rotating frame with respect to the drive frequency, we obtain the \textit{linearized} Hamiltonian
\begin{align}
    \hat{\mathcal{H}}_\text{eff} = -\tilde{\Delta} \hat{d}^\dagger \hat{d} + \omega_m \hat{b}^\dagger \hat{b} + \frac{1}{2} \left[ \Lambda \hat{d}^\dagger \hat{d}^\dagger + \Lambda^* \hat{d} \hat{d} \right] + \left( G \hat{d}^\dagger + G^*\hat{d} \right) \left( \hat{b} + \hat{b}^\dagger \right)
    \label{Ham_lin}
\end{align}
where we introduced the modified detuning $\tilde{\Delta} \equiv \Delta - 2 |\alpha|^2 \mathcal{K}$ with $\Delta \equiv \omega_p - \omega_c$ the optical detuning, the single-mode squeezing strength $\Lambda  = |\alpha|^2 \mathcal{K} e^{i \phi_\Lambda}$ and the photon enhanced optomechanical coupling strength $G =  |\alpha| g_0 e^{i \phi_G}$. 

\subsection{Classical Dynamics}
To study the dynamics of the classical solution we move the Hamiltonian in \eqref{Ham_Approx1} into a rotated frame with respect to the drive frequency $\omega_p$. We furthermore assume that the cavity is coupled to external waveguides with rate $\kappa$ (and that there are no additional losses) and use the standard \textit{input-output} theory \cite{SgardinerInputOutputDamped1985a} to obtain the equation of motion for the cavity average amplitude, namely
\begin{align}
    \frac{d}{dt} \alpha =  \left( i \Delta - \frac{\kappa}{2} \right) \alpha - i \mathcal{K} |\alpha|^2 \alpha - i \sqrt{2} g_0 \langle \hat{q} \rangle  \alpha - \sqrt{\kappa} \alpha_\text{in},
    \label{class_cav}
\end{align}
where we defined $\hat{q} = ( \hat{b} + \hat{b}^\dagger)/ \sqrt{2}$ as the position quadrature of the mechanical oscillator, and $\alpha_\text{in}$  is the coherent drive amplitude. Analogously, the classical dynamics of the mechanical mode are given by
\begin{align}
\begin{split}
\frac{d}{dt} \langle \hat{q} \rangle &= \omega_m \langle \hat{p} \rangle - \frac{\gamma_m}{2} \langle \hat{q} \rangle,
    \\
    \frac{d}{dt} \langle \hat{p} \rangle &= -\omega_m \langle \hat{q} \rangle - \frac{\gamma_m}{2} \langle \hat{p} \rangle - \sqrt{2} g_0 |\alpha|^2
\end{split}
\label{mech_eq}
\end{align}
with $\hat{p} = i( \hat{b}^\dagger - \hat{b})/ \sqrt{2}$ and $\gamma_m$ the momentum quadrature and decay rate of the mechanical oscillator, respectively.

Since the oscillation of the mechanical position is small enough to weakly modulate the optical field, we solve \eqref{mech_eq} in the long time limit and find the steady state of the mechanical position operator
\begin{align}
    \langle \hat{q} \rangle_s = - \frac{\sqrt{2} g_0 \omega_m |\alpha|^2}{\omega^2_m + \frac{\gamma_m^2}{4}},
\end{align}
which we insert into the equation for the classical cavity amplitude \eqref{class_cav} and obtain
\begin{align}
    \frac{d}{dt} \alpha = \left( i \Delta - \frac{\kappa}{2} \right) \alpha - i \mathcal{K}_\text{eff} \alpha |\alpha|^2 - \sqrt{\kappa} \alpha_\text{in},
    \label{steady_sol_cav}
\end{align}
with the effective Kerr constant
\begin{align}
    \mathcal{K}_\text{eff} \equiv  \mathcal{K} - \frac{2 g_0^2 \omega_m }{\omega^2_m + \frac{\gamma_m^2}{4}},
\end{align}
which includes a mechanical induced nonlinearity. 
The multiplication of the steady state solution of \eqref{steady_sol_cav} with its complex conjugate yields the average photon occupation
\begin{align}
    \bar{n}_c \left[ \left( -\Delta + \mathcal{K}_\text{eff} \bar{n}_c  \right)^2 + \left( \frac{\kappa}{2}\right)^2 \right] = \kappa \bar{n}_\text{in}
\end{align}
with $\bar{n}_c \equiv|\alpha_s|^2$ and $\bar{n}_\text{in} \equiv |\alpha_\text{in}|^2$. The cubic equation has one real root for weak driving, but for large enough input power a bistable regime exists. It can be shown~\cite{SlaflammeQuantumlimitedAmplificationNonlinear2011,Sdiaz-naufal} that the bifurcation occurs at a critical drive amplitude $\bar{n}_\text{in,bi} = \kappa^2/(3 \sqrt{3} \mathcal{K}_\text{eff})$.

\subsection{Dynamics and occupation of the mechanical mode}
Using the linearized Hamiltonian \eqref{Ham_lin} we can derive the equations of motion of the fluctuations within the input-output theory. For the cavity mode we have
\begin{align}
    \begin{split}
        \frac{d}{dt} \hat{d} =  \left( i \tilde{\Delta} - \frac{\kappa}{2} \right) \hat{d} - i \Lambda \hat{d}^\dagger - i G \left( \hat{b} + \hat{b}^\dagger \right) - \sqrt{\kappa} \hat{d}_\text{in}
    \end{split}
    \label{EOM_cav}
\end{align}
and for the mechanical mode
\begin{align}
    \begin{split}
        \frac{d}{dt} \hat{b} = - \left( i \omega_m + \frac{\gamma_m}{2} \right) \hat{b} - i \left( G^* \hat{d} + G \hat{d}^\dagger \right) - \sqrt{\gamma_m} \hat{b}_\text{in},
    \end{split}
    \label{EOM_mech}
\end{align}
with the input noise operator $\hat{f}_\text{in}$ with associated non-zero correlators $\langle \hat{f}_\text{in}(t) \hat{f}^\dagger_\text{in}(\tau) \rangle = (\bar{n}^T + 1 ) \delta( t - \tau)$ and $\langle \hat{f}^\dagger_\text{in}(\tau) \hat{f}_\text{in}(t) \rangle = \bar{n}^T \, \delta( t - \tau)$ for $\hat{f} \in \{ \hat{d}, \hat{b} \}$  and where $\bar{n}^T$ is the thermal occupation. 

Our aim is to decouple the previous set of equations and find an effective description of the mechanical mode. For this, we express \eqref{EOM_cav} and \eqref{EOM_mech} in frequency space using the Fourier transform, such that $\hat{o}[\omega] = \int_{-\infty}^\infty dt\, e^{i \omega t} \hat{o}(t)$ is the Fourier transform of $\hat{o}(t)$. Hence, from \eqref{EOM_cav} we have that the cavity dynamics in frequency space reads
\begin{align}
\begin{bmatrix}
\mathcal{X}^{-1}_c[\omega] &  i \Lambda  \\
-i \Lambda^*  & \mathcal{X}^{*-1}_c[-\omega]
\end{bmatrix}
\begin{bmatrix}
\hat{d}[\omega] \\
\hat{d}^\dagger[\omega]
\end{bmatrix}
=
- i 
\begin{bmatrix}
G & G \\
-G^* & -G^*
\end{bmatrix}
\begin{bmatrix}
\hat{b}[\omega]\\
\hat{b}^\dagger[\omega]
\end{bmatrix}
- 
\sqrt{\kappa}
\begin{bmatrix}
\hat{d}_\text{in}[\omega]\\
\hat{d}_\text{in}^\dagger[\omega]
\end{bmatrix},
\label{d_langevin}
\end{align}
where $\mathcal{X}^{-1}_c[\omega] = -i(\omega + \tilde{\Delta}) + \kappa/2$ is the cavity susceptibility. Analogously, we find that the dynamics of the mechanics is described by
\begin{align}
\begin{bmatrix}
- i (\omega - \omega_m) + \frac{\gamma_m}{2} & 0 \\
0 & - i (\omega + \omega_m) + \frac{\gamma_m}{2}
\end{bmatrix}
\begin{bmatrix}
\hat{b}[\omega] \\
\hat{b}^\dagger[\omega]
\end{bmatrix}
=
-i 
\begin{bmatrix}
G^* & G \\
-G^* & -G
\end{bmatrix}
\begin{bmatrix}
\hat{d}[\omega]\\
\hat{d}^\dagger[\omega]
\end{bmatrix}
- 
\sqrt{\gamma_m}
\begin{bmatrix}
\hat{b}_\text{in}[\omega]\\
\hat{b}_\text{in}^\dagger[\omega]
\end{bmatrix}.
\label{b_langev}
\end{align}
In order to find the effective dynamics of the mechanical mode we will algebraically eliminate the cavity operators in \eqref{b_langev}. For this, we first solve \eqref{d_langevin} for $\hat{d}$ and $\hat{d}^\dagger$, and substitute its solution in the second term on the RHS of equation \eqref{b_langev}. Finally, due to the modification of the nonlinear cavity, the oscillator's dynamics becomes
\begin{align}
    \begin{bmatrix}
    - i (\omega - \omega_m) + \frac{\gamma_m}{2} - i \Sigma_c[\omega] & - i \Sigma_c[\omega] \\
    i \Sigma_c[\omega] & - i (\omega + \omega_m) + \frac{\gamma_m}{2} + i \Sigma_c[\omega]
    \end{bmatrix}
    \begin{bmatrix}
    \hat{b}[\omega] \\
    \hat{b}^\dagger[\omega]
    \end{bmatrix}
    = - \sqrt{\gamma_m}
    \begin{bmatrix}
    \hat{B}_\text{in}[\omega] \\
    \hat{B}^\dagger_\text{in}[\omega]
    \end{bmatrix},
    \label{effective_mech}
\end{align}
where the modification induced by the cavity is compactly described by $\Sigma_c[\omega] = -2|G|^2 \left\{ \tilde{\Delta} + |\Lambda| \cos(\varphi) \right\} \tilde{\mathcal{X}}_c[\omega]$ with $\tilde{\mathcal{X}}_c^{-1}[\omega] = \mathcal{X}_c^{-1}[\omega] \mathcal{X}_c^{*-1}[-\omega] - |\Lambda|^2$ and the phase $\varphi = 2 \phi_G - \phi_\Lambda$. Taking the dependencies on the classical cavity amplitude it follows here that $\phi_G = \phi_\Lambda/2$, hence $\varphi = 0$, such that the phases of the coupling strengths are not independent. Here, the optical cavity induces both a frequency shift and an asymmetric optical damping given by $\delta \omega_m = \Re \{ \Sigma_c[\omega] \}$ and $\Gamma = \pm 2\Im \{ \Sigma_c[\omega] \}$, respectively. Additionally, it can be seen that the interaction with the cavity introduces single-mode squeezing in the oscillator's dynamics. On the other hand, the modified mechanical noise is given by 
\begin{align}
    \hat{B}_\text{in}[\omega] = \hat{b}_\text{in}[\omega] - i|G|\sqrt{\frac{\kappa}{\gamma_m}}\tilde{\mathcal{X}}_c[\omega] 
    \left\{ e^{-i \phi_G}\left(\mathcal{X}^{*-1}_c[-\omega] + i |\Lambda| \right) \hat{d}_\text{in}[\omega] + e^{+i \phi_G}\left(\mathcal{X}^{-1}_c[\omega] - i |\Lambda| \right) \hat{d}^\dagger_\text{in}[\omega] \right\}, 
\end{align}
which includes the optical noise contribution.

So far, we have studied the modification of the mechanical oscillator's dynamics due to its coupling to a nonlinear cavity yielding \eqref{effective_mech}. In fact, our main interest is to analyze the cooling benefits that arise when both classical and quantum dynamics are formally dealt with. Finally, the occupation of the mechanical mode can be obtained via 
\begin{align}
    \bar{n}_m = \int_{-\infty}^\infty \frac{d\omega}{2\pi} \int_{-\infty}^\infty  \frac{d\Omega}{2\pi} \langle  \hat{b}^\dagger[\Omega] \hat{b}[\omega]  \rangle,
\end{align}
where the mode operators are given by the solution of \eqref{effective_mech} and the nonzero noise autocorrelators in frequency space read $\langle \hat{f}_\text{in}[\omega] \hat{f}^\dagger_\text{in}[\Omega] \rangle = 2 \pi (\bar{n}^T + 1 )\, \delta[\omega + \Omega]$ and $\langle \hat{f}^\dagger_\text{in}[\Omega] \hat{f}_\text{in}[\omega] \rangle = 2 \pi \bar{n}^T \, \delta[\omega + \Omega]$ for $\hat{f} \in \{ \hat{d}, \hat{b} \}$.

\section{Data taking routine}
\label{secS:dataTaking}
Here, we describe how we measure the mechanical cantilever. As the setup is highly sensitive to mechanical vibrations, we have to switch off the pulse tube cooler of the cryostat during our measurements, giving us an approximate five to ten minutes time window. The data taking routine takes around five minutes for a data point. To determine the coupling rate we use a calibration tone~\cite{SgorodetksyDeterminationVacuumOptomechanical2010a} via frequency modulation of our probe tone. While this scheme directly gives the transmission coefficient of the setup, it requires very precise electrical length matching of the signal through the experiment (RF port of the mixer) to the signal through the delay line (LO port). Also it is highly sensitive to internal leakage of the mixer and therefore only allows us to work at specific frequencies. Thus, we flux tune the cavity to different values to change the probe-cavity detuning.

\begin{figure*}[h]
    \centering
    \includegraphics[width = 15.8cm]{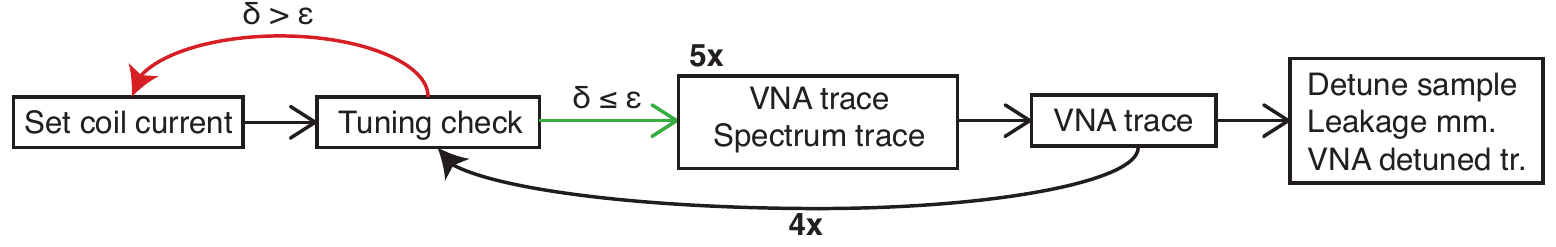}
    \caption{Diagram of the measurement process. Here, $\delta$ is the measured detuning, while $\epsilon$ is the detuning threshold we require to start the measurement. See text for details.}
    \label{figS:mmDia}
\end{figure*}

Before doing the measurement itself, we take a local flux map over the range which we are measuring. To measure a data point (see diagram Fig.~\ref{figS:mmDia}), we start by checking the frequency of the cavity and retune it according to the flux map until it is within a certain threshold (i.e. \SI{20}{\kilo \hertz} for the measurement at $g_0/2 \pi = \SI{201}{Hz}$). We always take a pair of traces, where we first measure the frequency of the cavity with a low power VNA trace. Then we switch on the microwave generator and measure the cantilever using the spectrum analyser. We usually take five of those pairs before taking another VNA trace and re-checking the detuning as explained above. In total we do four of those sets, giving us 20 measurement traces of the mechanical mode per data point. Afterwards we switch the pulse tube on, detune the sample to measure the leakage of the calibration tone and a background VNA trace. Ideally we would measure no calibration tone as the cavity is detuned.

For the spectrum analyser we typically use a bandwidth of \SI{0.1}{\hertz}, which gives us the lowest noise floor possible, but is also required for most measurements due to the narrow linewidth of the cantilever mode. We always take 8001 points, leading to a span of \SI{800}{\hertz}.

We also note that the data presented in the main article is taken from two different cooldown runs. The data shown in Figures 2d and 3 of the main article was taken during the first run. For the second run, the setup was unchanged and data presented in figures 2c and 4(a,b) was taken during that run. There was no noticeable difference whether data was taken in the first or second run.

We also uploaded all of our raw data for the measurements discussed in the main manuscript to Zenodo~\cite{SDOI105281}.

\section{Data analysis}
\label{secS:dataAnalysis}
\subsection{Data treatment}
In this section, we explain how we treat the data for our cooling traces. We use the same routines across all dataset, however change some goodness of fit criteria, which is explained in Sec.~\ref{secS:goodnessOfFit}.

\begin{figure*}[h]
    \centering
    \includegraphics[width = 9cm]{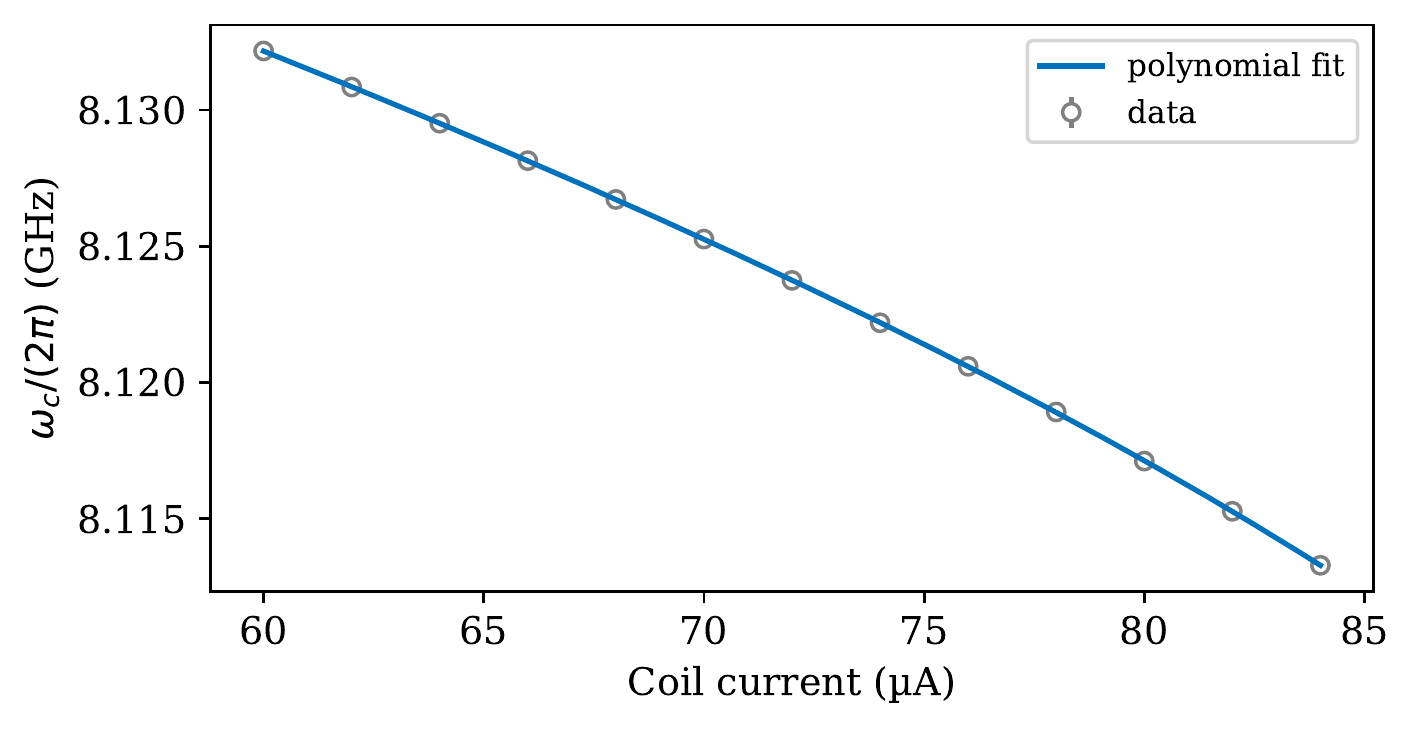}
    \caption{Change of the cavity frequency with tuning of the coil. The fit is a 4-th order polynomial.}
    \label{figS:fluxMap1}
\end{figure*}

\begin{figure*}[h]
    \centering
    \includegraphics[width = 9cm]{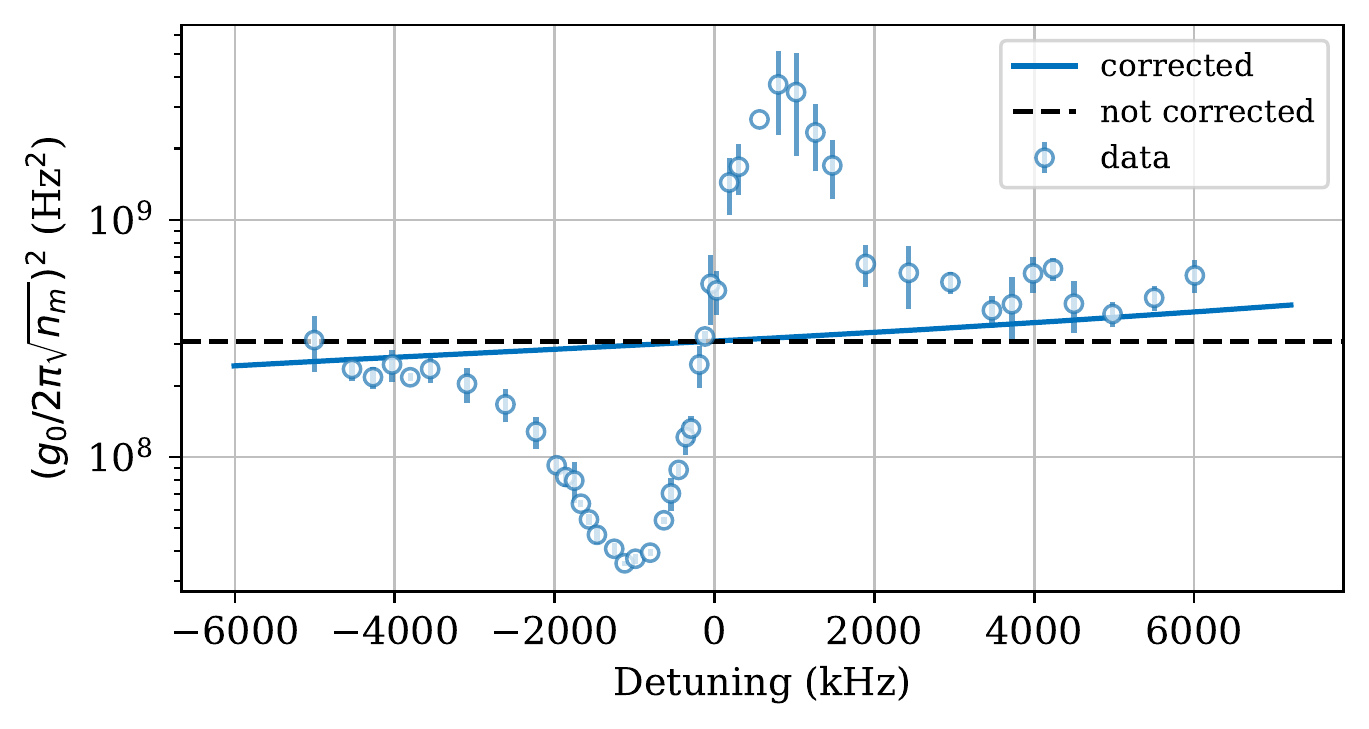}
    \caption{We plot $g_0/2 \pi \times \sqrt{n_{\text{m}}}$ to demonstrate the necessity for correcting for the change of $g_0$ over the measurement range. In the not corrected version we assume a constant $g_0$, while in the corrected version, we correct for the slightly changing slope of the flux map for different detunings (Fig.~\ref{figS:fluxMap1}). This data is taken from the high power set at $g_0/2 \pi = \SI{201}{\hertz}$ presented in the main article. The error shown is the standard error of multiple data points.}
    \label{figS:g0n_corrVsNocorr}
\end{figure*}

As we shift the frequency of the cavity, instead of changing the frequency of the pump, $g_0$ is slightly different depending on the exact detuning. To compensate for this, we take a flux map of the cavity and evaluate $g_0$ via the slope of the flux map, Section~\ref{secS:fluxmaps}. In Fig.~\ref{figS:fluxMap1} we show such a flux map with the corresponding polynomial fit. To find out the best order for the polynomial, we use the Akaike criterion~\cite{SburnhamModelSelectionMultimodel2004}.
In Fig.~\ref{figS:g0n_corrVsNocorr} we compare using a fixed $g_0/2 \pi$ at \SI{201}{\hertz} (black) to a changing $g_0$ (blue) as estimated from Fig.~\ref{figS:fluxMap1} and thus show the importance for correcting this.

\begin{figure*}[h]
    \centering
    \includegraphics[width = \textwidth]{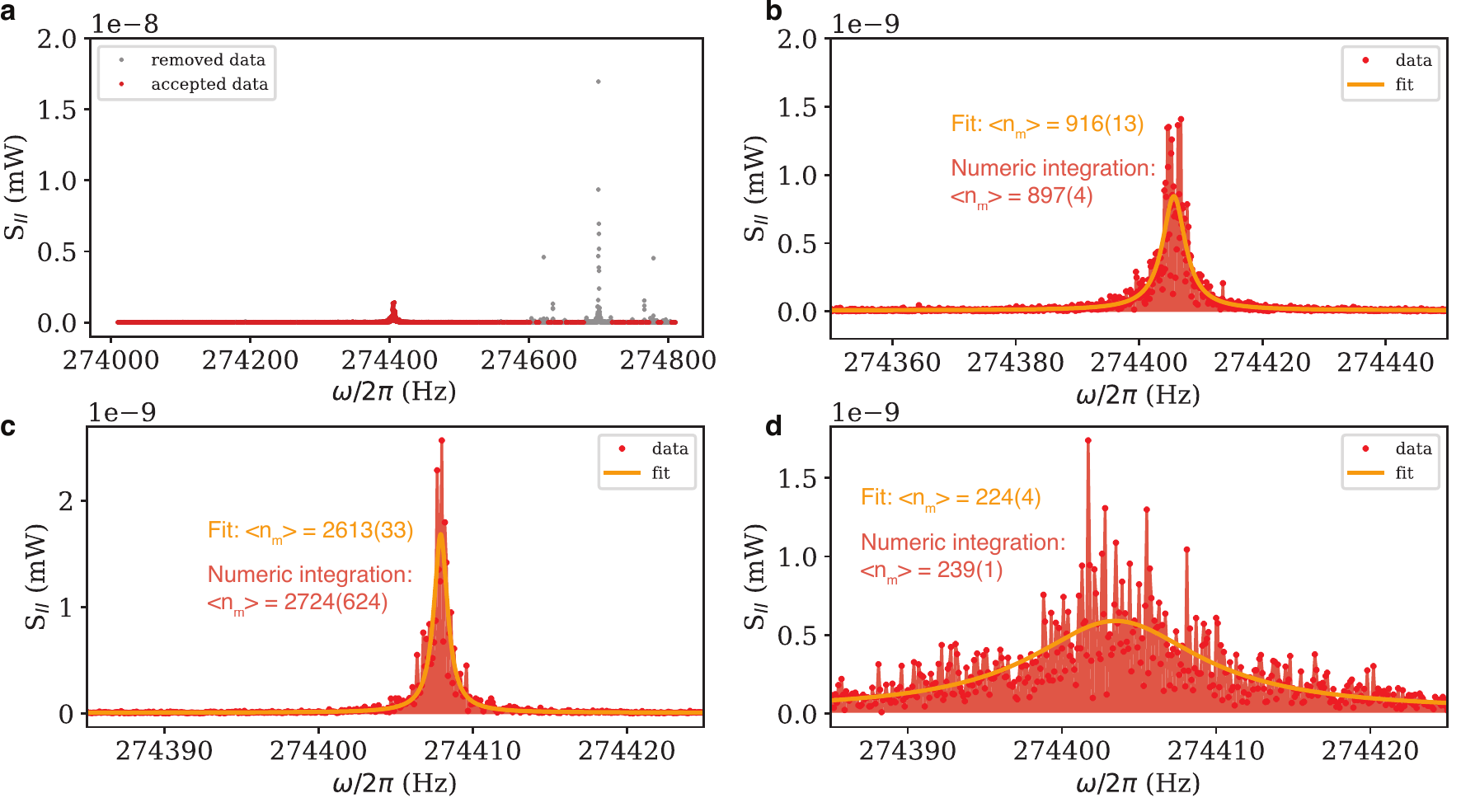}
    \caption{Here, we show some aspects of the analysis on data from one of the cooling traces taken at $g_0/2 \pi = \SI{201}{\hertz}$ close to bistability. \textbf{a,} Outlier removal. As described in the text, we remove outlying peaks far from the mechanical ones, which is shown here (grey). Those peaks often appear around the calibration peak, where we believe that those peaks are upmixed flux noise. \textbf{b-d,} Comparison of fit and numeric integration for traces with medium cooling (b), weak cooling (c) and strong cooling (d). Coloured is the area below the mechanical peak, which we integrate over. The two methods show good agreement. For the fit, we give the propagated fitting error. For the numerical integration, we propagate the standard error of the noise floor. All traces shown here, are averages of four traces measured with the spectrum analyser.}
    \label{figS:manInt}
\end{figure*}

As discussed in Section~\ref{secS:dataTaking}, due to the small mechanical linewidth, we operate our spectrum analyser at the lowest bandwidth it allows of \SI{0.1}{\hertz}, which leads to a measurement time of \SI{10}{\second}. The cavity frequency can change during a measurement due to flux noise, which changes the detuning between the probe tone and the cavity frequency and as a consequence the backaction. Especially operating close to bistability and in the region with highest cooling we are very sensitive to such changes. While we cannot compensate for flux noise happening on shorter time scales than our typically \SI{10}{\second} spectrum analyser traces, we can partly compensate for slower flux drifts, by using the measurements of the cavity directly before each measurement. To do this, we bin our data together by applying a k-means binning~\cite{SHttpsScikitlearnOrg}, which groups the traces having most similar cavity frequencies. Doing this, we regroup the data, but do not change the total number of bins for the cooling trace. This means that the number of traces per bin is determined by the binning routine, but as we do not change the number of total bins, on average there are 20 traces within a bin. In the next step we fit the mechanical traces using the damped harmonic oscillator model~\cite{SgorodetksyDeterminationVacuumOptomechanical2010a}. First, we average all data within a bin and fit it on a limited span, where we expect the mechanical frequency. This helps to get good initial parameters for the subsequent fits, but is also required, as we remove residual peaks away from the mechanical resonance. We then treat the individual traces in groups of four traces averaged on top of each other, which increases our signal to noise while still provides us some statistics. Before fitting those spectra, we remove residual peaks, where we identify a peak if it is 6 standard deviations above the noise, and we exclude an area of \SI{100}{\hertz} to \SI{200}{\hertz} around the mechanical frequency depending on the mechanical linewidth (Fig.~\ref{figS:manInt}a). This is required, as the fit of the mechanical trace is performed over the whole measurement range. We also numerically integrate the area under the curve as an additional check (Fig.~\ref{figS:manInt}(b-d)). We usually estimate the error as the standard error between the groups within a bin. We also compared this to the propagated fit error and obtained very similar results.

\begin{figure*}[h]
    \centering
    \includegraphics[width = 9cm]{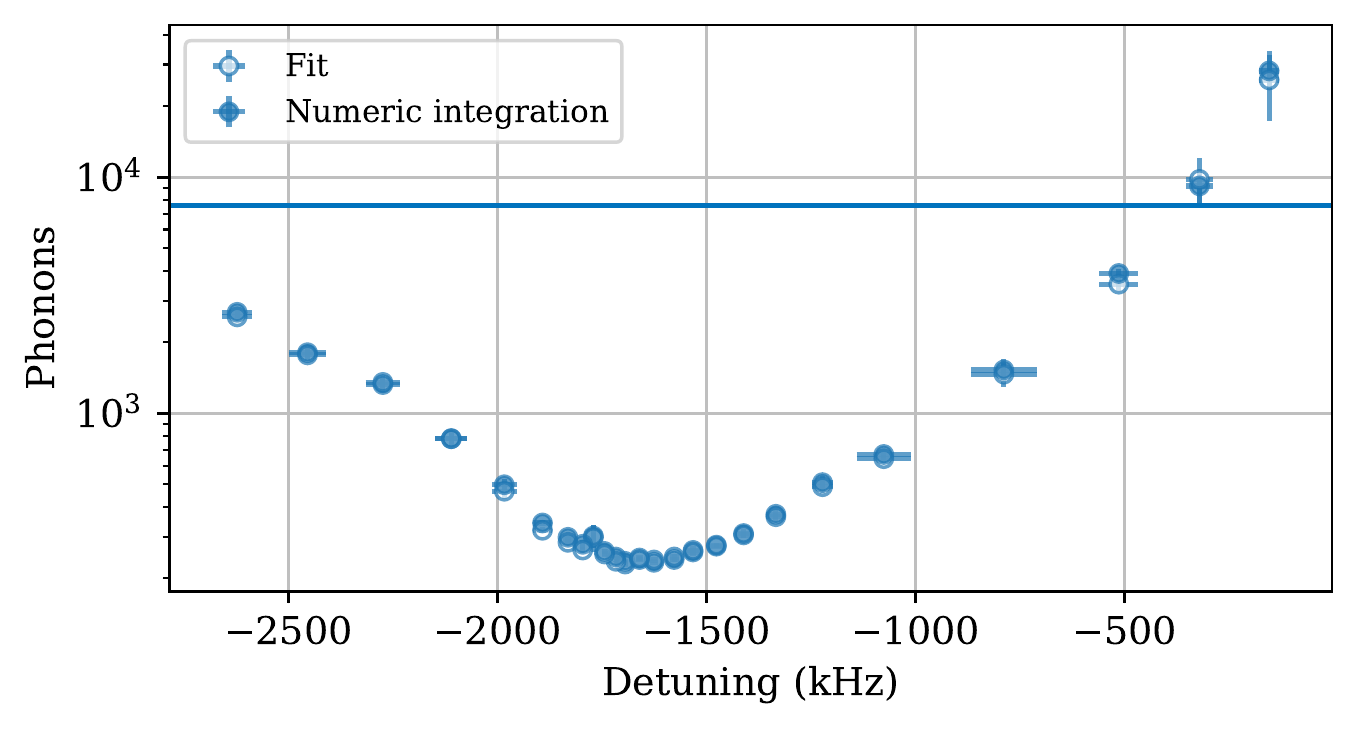}
    \caption{Cooling trace, comparing phonon numbers extracted via numerical integration and fitting. In Fig.~\ref{figS:manInt} we already show some spectra corresponding to data points here. We see very good agreement between the fit and numeric integration. The plotted error is the standard error of multiple data points for each detuning (see main text).}
    \label{figS:16dB_coolingTrace}
\end{figure*}

In Fig.~\ref{figS:16dB_coolingTrace} we show the complete cooling trace, of which we already show a few spectra in Fig.~\ref{figS:manInt}. We clearly see that the numerical integration and the fit agree very well. Typically we show the phonon number obtained from the fit to the data.

\subsection{Goodness of fit criteria}
\label{secS:goodnessOfFit}
We apply several independent criteria to check if we can trust the data and treatment, which we explain here. These are the automated tests we perform on our data:
\begin{enumerate}
    \item Check the leakage of the calibration tone. This checks the credibility of the calibration tone, if we measure a significant calibration tone, even though the cavity is detuned we cannot trust the data point, and therefore ignore the data. As we usually check the calibration before starting the measurement, this is a rare occurrence.
    
    \item We get the detuning for each data trace by fitting the cavity response directly measured before the measurement with the spectrum analyser. In case this fit fails, we have to neglect this trace. This only happens for the highest couplings we measured and usually happens due to a distorted cavity response due to excessive flux noise.
    
    \item We check if the cavity frequency moved too much between subsequent data traces. In case it did, we assume that there was excessive flux noise during the measurement of the mechanical mode and neglect the data. On top of that, we also check if the standard deviation of the cavity frequencies within a subgroup of four averaged mechanical traces is below a threshold.
    
    \item When the data passed the above criteria, we fit the mechanical spectrum. Afterwards we compare the phonon number from the fit to the one obtained from numerical integration, and neglect the data if the phonon number does not agree within a factor 1.5. Also we neglect the data if the fit is not more than \SI{3}{\decibel} above the noise.
    
    \item In a final step, we neglect data, if the obtained linewidth is above \SI{300}{\hertz} or below \SI{5}{\micro \hertz}, which points to an issue within the fitting routine. 
\end{enumerate}

The first criterion aims at checking the credibility of our calibration tone, which is crucial for trusting the phonon number. Criteria two and three aim at assuring that we can trust the frequency of the cavity we extract and that no excessive flux noise occurred during the measurement. With criteria four and five we want to make sure that the fit gives an accurate description of the data, also requiring that we have enough signal to get faithful information doing the fit.

We keep most parameters the same for all the data sets treated (i.e. sets at different $g_0$). The only parameter we systematically change according to the change of coupling, is how far the cavity can move in frequency between subsequent data traces (criterion no. 3), as flux noise increases with increasing $g_0$. Also, for the data of highest cooling, we relax criterion no. 4 comparing the phonon number obtained by numerical integration and the fit, as the spectra are heavily influenced by flux noise and we cannot trust the fit anymore (see Section~\ref{secS:bestCooling}). While there are several checks done on the data, especially for low $g_0$ only a small fraction of the data is removed, as the system is very well behaved. For instance, for the high power set at $g_0/2 \pi = \SI{201}{\hertz}$ presented in the main article more than 85\% of the data is accepted. For other data this acceptance is also far above 90\%. For data at higher $g_0$ this naturally changes, while we still have typical acceptance rates of 60\%.

\section{Temperature ramp}
\label{secS:tempRamp}
\begin{figure*}[h!]
    \centering
    \includegraphics[width = \textwidth]{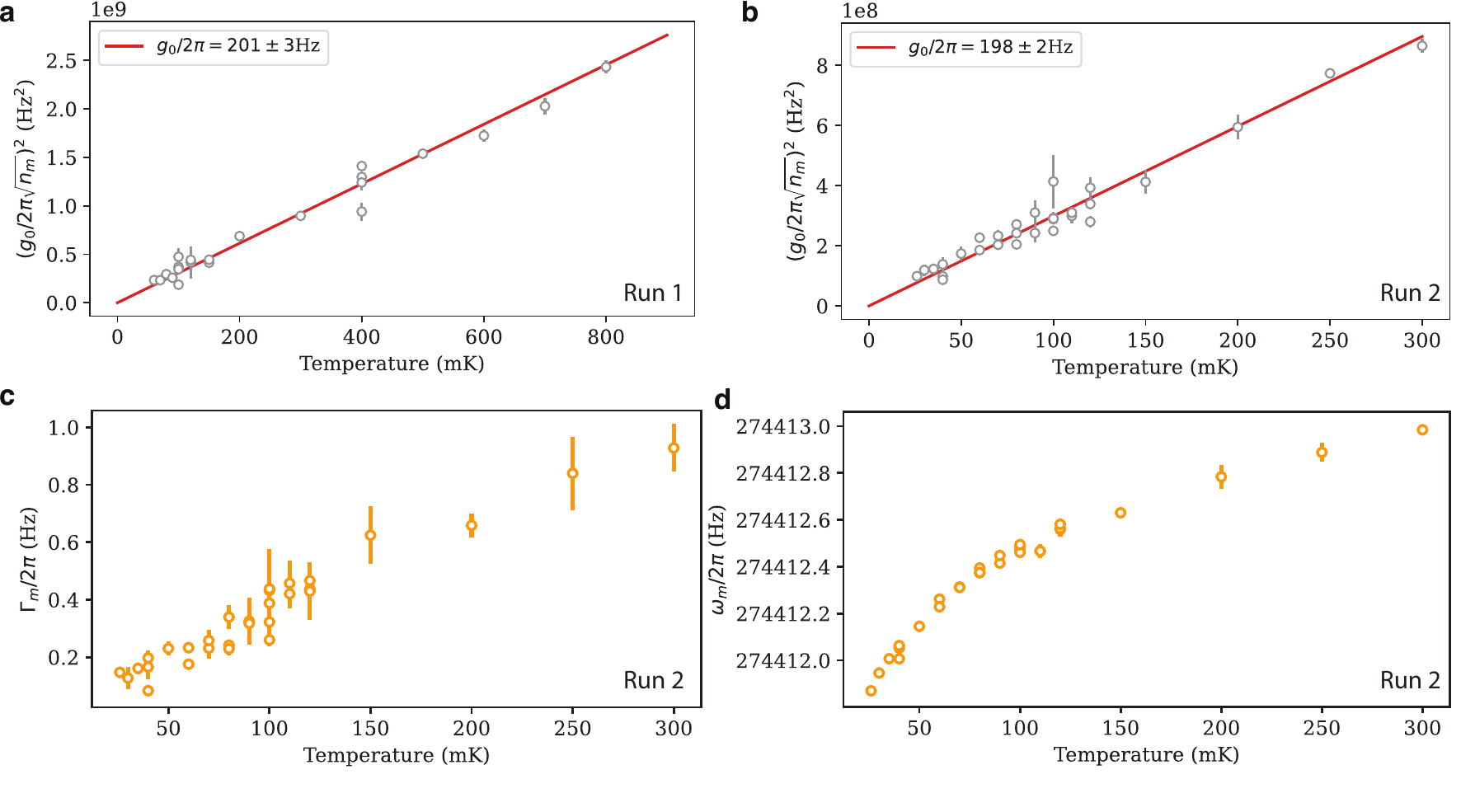}
    \caption{Measurements of the mechanical cantilever at different temperatures to characterise the system and to check its thermalisation. \textbf{a-b,} Measuring $g_0/2 \pi \times \sqrt{n_{\text{m}}}$ for run 1 (a) and run 2 (b). We see that the mechanical system is well thermalised with the cryostat until its base temperature of around \SI{25}{\milli \kelvin} and get a calibration for $g_0$. The value for $g_0$ is slightly different in both runs, as a slightly different flux bias point was used. \textbf{c,} Mechanical linewidth versus temperature.  \textbf{d,} Mechanical frequency versus temperature. The error shown in all plots is the standard error of multiple data points.}
    \label{figS:tempRamp}
\end{figure*}
In this section we discuss measurements of the mechanical cantilever at different temperatures of the cryostat. The temperatures quoted here are the ones measured with the temperature sensor at the baseplate of the cryostat. Those measurements are important to ensure that the cantilever is thermalised to its environment, but also to determine the optomechanical coupling rate $g_0$, as in the measurement we only have access to $g_0/ 2 \pi \times \sqrt{n_m}$. Thus when the cantilever is out of its thermal equilibrium during measurements with backaction, we crucially need to know $g_0$ in order to extract the phonon number.

In Fig.~\ref{figS:tempRamp}(a,b) we show temperature ramps for both runs considered within this manuscript. In both runs the mechanical system is thermalised until lowest temperatures. The coupling we extract is slightly different in both cases due to a slightly different flux bias point. In Fig.~\ref{figS:tempRamp}c we see that the linewidth increases with temperature, and measure a linewidth of below \SI{0.2}{\hertz} for lowest temperatures. From Fig.~\ref{figS:tempRamp}d we obtain a slight change of the mechanical frequency with temperature.

\section{Flux maps and change of coupling with flux bias point}
\label{secS:fluxmaps}
In this part we give additional information on the flux sensitivity of our cavity, which crucially also influences the coupling rate. We compare the couplings obtained from the cooling traces to the slope of the complete flux map and also estimate at which couplings we could operate according to the sensitivity of the cavity, in case we had less flux noise.

\begin{figure*}[h]
    \centering
    \includegraphics[width = 9cm]{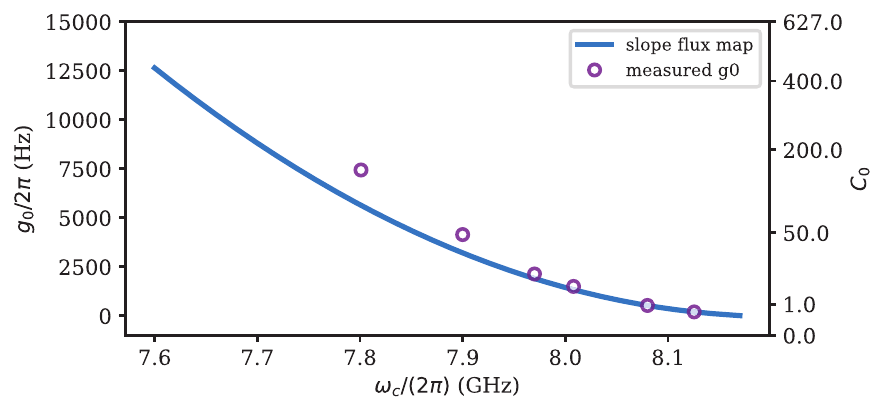}
    \caption{Slope of the flux map, which - upon a calibration factor - gives $g_0$. Plotted on top the values of $g_0$ where we took cooling traces. There we determine $g_0$ using a fixed calibration factor together with a local flux map (e.g. Fig.\ref{figS:fluxMap1}). We also show the single-photon cooperativities on a secondary axis.}
    \label{figS:g0_fluxMap}
\end{figure*}

The coupling strength in our system is given by~\cite{SzoepflSinglePhotonCoolingMicrowave2020a}:
\begin{equation}
    g_0 = \frac{\partial \omega}{\partial \phi} \frac{\partial \phi}{\partial x} x_{\text{ZPM}}
    \label{equS:g0}
\end{equation}
The second part, $\partial \phi/\partial x$, cannot be changed during the experiment, and depends mainly on the size/magnetisation of the magnet and the distance between SQUID and cantilever The first part, $\partial \omega/\partial \phi$, the sensitivity to flux of the cavity, can be changed by changing the flux bias point. In Fig.~\ref{figS:g0_fluxMap} we plot the slope of the flux map, which is the sensitivity of the cavity and gives - upon a calibration factor - directly $g_0$. To determine the calibration factor we use the temperature ramp, which gives us $g_0$ for a given cavity frequency, Section~\ref{secS:tempRamp}. For couplings larger than \SI{600}{\hertz} we reach a regime where single-photon cooperativities exceeds unity.

\begin{figure*}[h]
    \centering
    \includegraphics[width = 12cm]{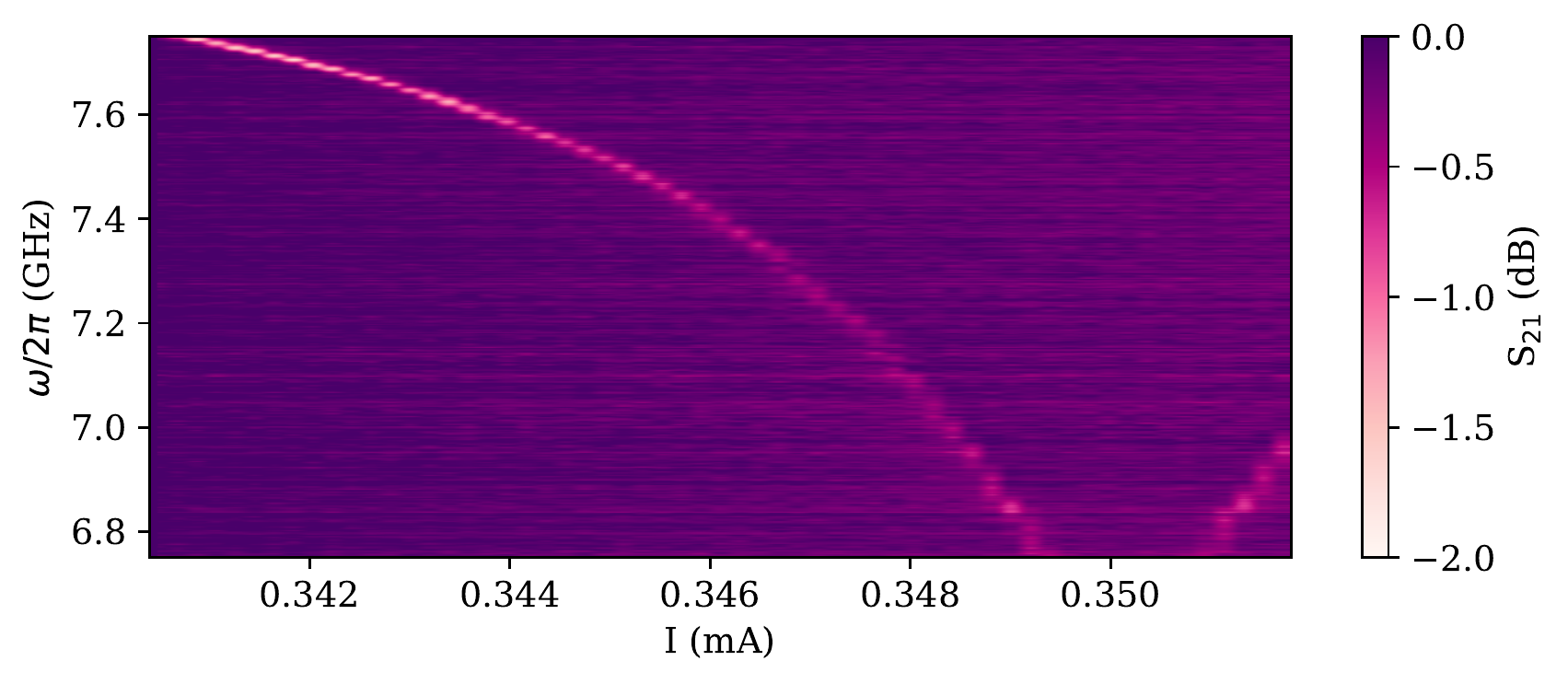}
    \caption{Flux map for highest sensitivities. Measuring to even lower frequencies is not possible as we reach the cutoff of our waveguide.}
    \label{figS:fm_detail_lowFreq}
\end{figure*}

In Fig.~\ref{figS:fm_detail_lowFreq} we plot the flux map for highest sensitivities (lowest cavity frequencies). Here, the cavity is extremely sensitive to flux noise, which prevents us from directly measuring the mechanical system. However using the slope of the flux map together with the calibration factor, we can estimate $g_0$ in this regime. That results in a $g_0/2 \pi  \simeq \SI{90}{\kilo \hertz}$, which might be only a lower boundary, as the cavity still tunes to even lower frequency, but the cutoff of the waveguide reduces the signal too much and we cannot access this region. Remarkable, this is only a factor of three from the mechanical frequency. Such a high coupling would amount to a single-photon cooperativity of $C_0 = 22.5 \times 10^3$ and would thus put us in the regime of single-photon strong quantum cooperativity with $$C_0^{\text{qu}} = C_0 / \langle n_{\text{m}}^{\text{th}} \rangle \simeq 3$$ already at \SI{100}{\milli \kelvin}. Going to even lower temperatures, we would expect even higher values, due to the lower thermal occupation as well as the reduced mechanical linewidth and expect $C_0^{\text{qu}} \approx 50$ at \SI{25}{\milli \kelvin}.

\section{Flux noise origin}
\label{secS:FN_ori}
In this part the origin of our flux noise as well as its strength estimated via the change of the cavity linewidth will be discussed.

\begin{figure*}[h]
    \centering
    \includegraphics[width = 9cm]{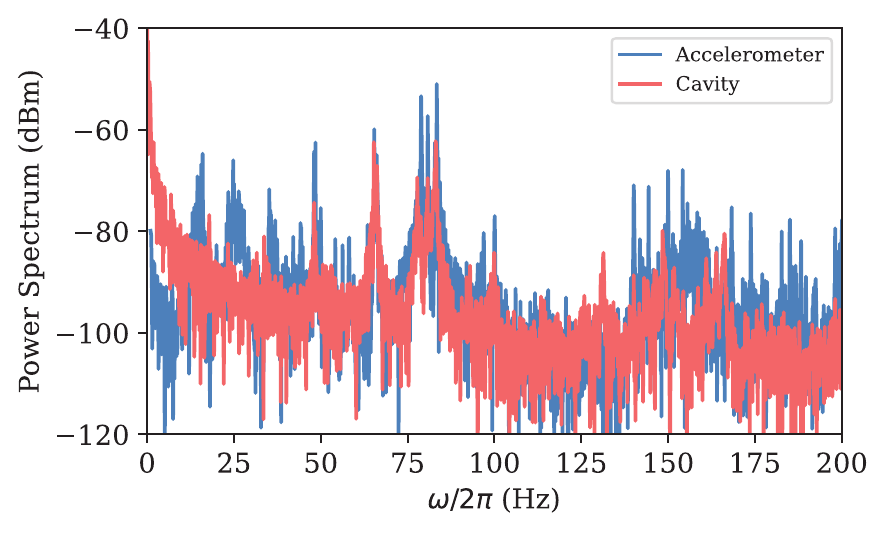}
    \caption{Comparison of the spectrum taken using a microwave probe tone at the cavity, which corresponds to flux variations and a spectrum taken using an acceleormeter at the top of the cryostat (this spectrum is offsetted, such that the baselines approximately agree). All major peaks, which are visible as flux noise in the cavity are also present in the mechanical spectrum. This is a strong indication, that the flux noise is of mechanical origin and gets converted to magnetic noise via the cantilever. Also the majority of the noise peaks occur below \SI{100}{\hertz}.}
    \label{figS:accVsCav_spec}
\end{figure*}

Fig.~\ref{figS:accVsCav_spec} shows the comparison of two spectra. One is measured using a microwave signal probing the cavity, which tells us about the flux noise present in our cavity. A second spectrum shows mechanical noise, measured using an accelerometer. All major peaks present in the cavity as flux noise, are also measured using the accelerometer. This means that the origin of the flux noise is mechanical noise, and we believe that it gets converted to flux noise via the cantilever, as it is not visible on samples sitting besides the one without a cantilever. The most dominant peaks also appear well below \SI{100}{\hertz}, where the main components are around \SI{65}{\hertz} and \SI{80}{\hertz}. While this is very slow compared to the cavity decay rate, it is still fast compared to timescales of our measurements (Section~\ref{secS:dataTaking}).

\begin{figure*}[h]
    \centering
    \includegraphics[width = 9cm]{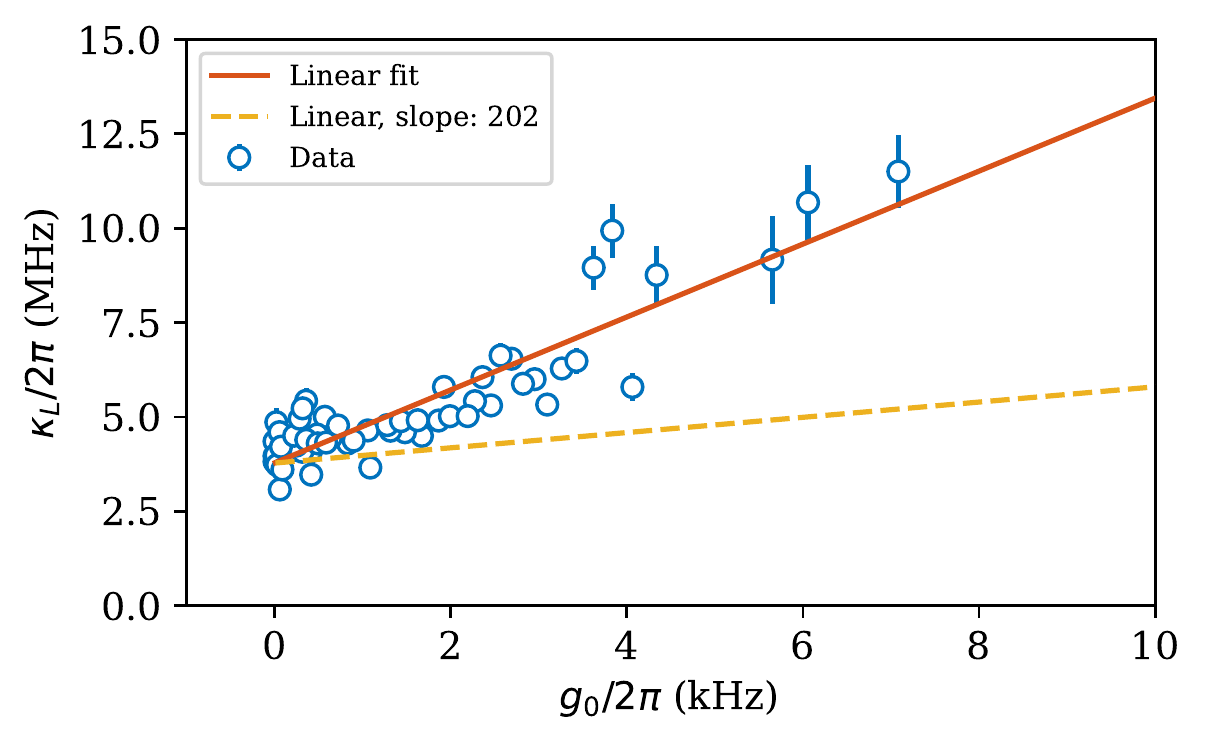}
    \caption{Change of cavity linewidth with flux sensitivity, parameterised using $g_0$ (i.e. Fig.~\ref{figS:g0_fluxMap}). We expect that the flux noise increases linearly with flux sensitivity, where we fit a slope of 794(50). This translates to a flux noise amplitude of \SI{302(19)}{\micro \phi0}. Additionally we also plot a line with a slope of 202, which is the extracted flux noise with the flux noise model discussed in Section~\ref{secS:FN_model}.}
    \label{figS:kappaL_vs_g0}
\end{figure*}

A useful measure for the strength of flux noise in our system is also the change of the cavity linewidth with increasing flux sensitivity. This is plotted in Fig.~\ref{figS:kappaL_vs_g0}, where the cavity linewidth is already plotted in terms of coupling strength (which is directly proportional to the flux sensitivity). We fit the linear dependence of the cavity linewidth on $g_0$ and obtain a slope of 794(50), corresponding to a flux noise of \SI[separate-uncertainty = true]{302(19)}{\micro \phi_0} (more on this conversion can be found in Section~\ref{secS:FN_model}, Eq.~\ref{equS:sigma}). Additionally we also plot a line with a slope of 202, corresponding to the flux noise extracted with the flux noise model, discussed in Section~\ref{secS:FN_model}.
From this data we can estimate, that at a coupling of \SI{90}{\kilo \hertz}, we have a cavity linewidth of \SI{70}{\mega \hertz} broadened by flux noise. To allow reaching the regime of single-photon strong quantum cooperativity at \SI{100}{\milli \kelvin} would required a more than ten fold reduction of flux noise, while when operating at \SI{40}{\milli \kelvin} we only have to half the flux noise due to the lower mechanical occupation and linewidth. Also we plan reducing the sensitivity to mechanical (flux) noise of our setup in the future.

\section{Flux noise model}
\label{secS:FN_model}
In this part we discuss how we model the influence of flux noise on our setup. As seen from data presented in the main article in Figures~3d and 4b, flux noise limits our cooling performance, and also influences the mechanical spectra we measure going towards the limits of our setup. Modelling the flux noise very accurately is highly challenging, as the flux noise seems to be an interplay between fast processes (compared to the measurement time and the mechanical linewidth) and processes on a timescale of the order of our measurement time. Furthermore the influence of flux noise also depends on the environment and is thus not constant over time. There are also observations, that the cavity stabilises using a strong drive~\cite{SbothnerFourwavecoolingSinglePhonon2022}, which makes it even more challenging to model.

\begin{figure*}[h]
    \centering
    \includegraphics[width = 18cm]{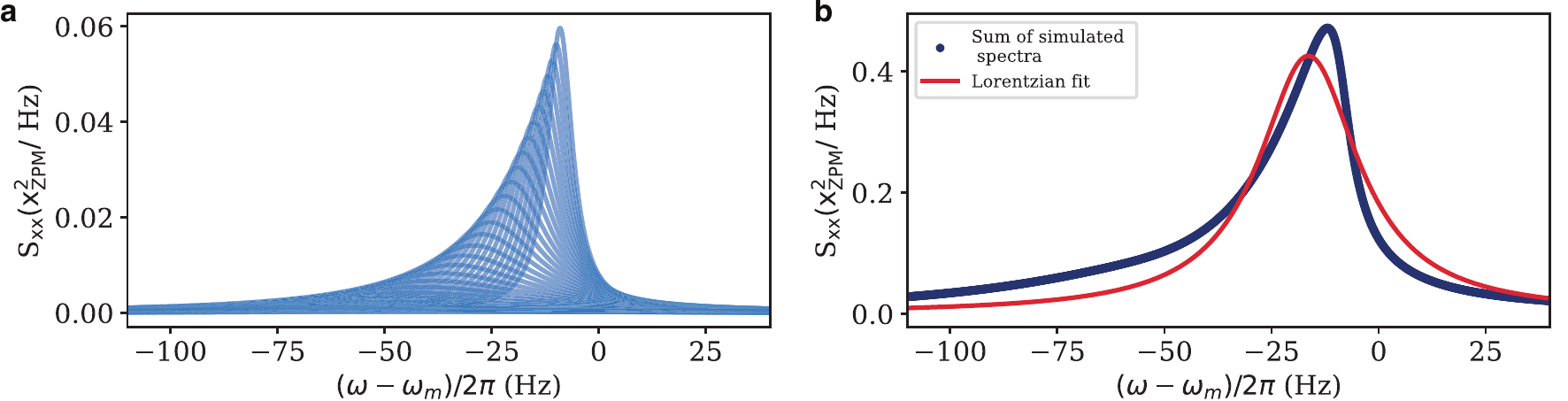}
    \caption{Calculated spectra to simulate flux noise for a case of strong cooling backaction under flux noise. \textbf{a,} Weighted spectra plotted on top. \textbf{b,} Sum of weighted spectra to get a single spectrum. We fit this spectrum with a Lorentzian, which is not a representative prediction of the line shape, but is most similar to what we do in the analysis of the experiment. From this fit we obtain the mechanical linewidth and frequency. We get the phonon number from each individual spectrum as seen in (a), together with the weighting.}
    \label{figS:FN_spectra}
\end{figure*}

Therefore, we decided to do a simple model by assuming Gaussian distributed noise, and get decent qualitative agreement for measurements where we have a low to medium influence from flux noise. As explained in section~\ref{secS:nonLinTheo}, we calculate the mechanical spectrum for given system parameters, including the probe-cavity detuning. In order to model the flux noise, we do not assume a single probe-cavity detuning anymore, but average multiple spectra with different detunings on top. For our model we assume Gaussian distributed noise, which is implemented by assigning a Gaussian distributed weight to each trace:
\begin{equation}
\begin{aligned}
    W_i (\Delta \omega_i) &= \frac{1}{\sqrt{2 \pi \sigma^2}} e^{-\frac{(\Delta \omega_i - \Delta \omega_0)^2}{2 \sigma^2}} \\
    w_i &= \frac{W_i}{\sum_i W_i}
\end{aligned}
    \label{equS:Gauss}
\end{equation}
Here, $w_i$ is the specific weight assigned to a spectrum at a given probe-cavity detuning $\Delta \omega_i$, while $\Delta \omega_0$ would be the intended probe-cavity detuning of the measurement. The width of the Gaussian, $\sigma$, gives an effective strength of flux noise. We calculate 50 spectra for equally distributed detunings up to $2 \sigma$ to each side of $\Delta \omega_0$. We obtain the phonon number as the weighted area of each spectrum, while we fit the final, averaged spectrum using a Lorentzian for estimating the mechanical linewidth and frequency. In Fig.~\ref{figS:FN_spectra} we show an example of how we obtain such a simulated spectrum influenced by flux noise. In (a) we show the weighted, normalised (Eq.~\ref{equS:Gauss}) spectra on top of each other. The asymmetry with a much flatter tail towards lower frequencies can be clearly seen and comes from the fact that for large cooling backaction, there is also the highest mechanical frequency shift and highest linewidth. This can be especially seen in sub panel (b) where we show the sum of the weighted spectra shown in (a) together with a Lorentzian fit. The reason to use a Lorentzian fit, even though it is not very accurate, is that it comes closest to what we also do in the analysis of the experiment. We also see the asymmetry of the mechanical spectrum in the data, which can be seen in Fig.~4b of the main manuscript, where we show a trace of high cooling, influenced by flux noise. In contrast however, the numerical integration, we also do within the analysis, does not suffer from flux noise in determining the phonon occupation number, as the total area below the curve matters.

\begin{figure*}[h]
    \centering
    \includegraphics[width = 18cm]{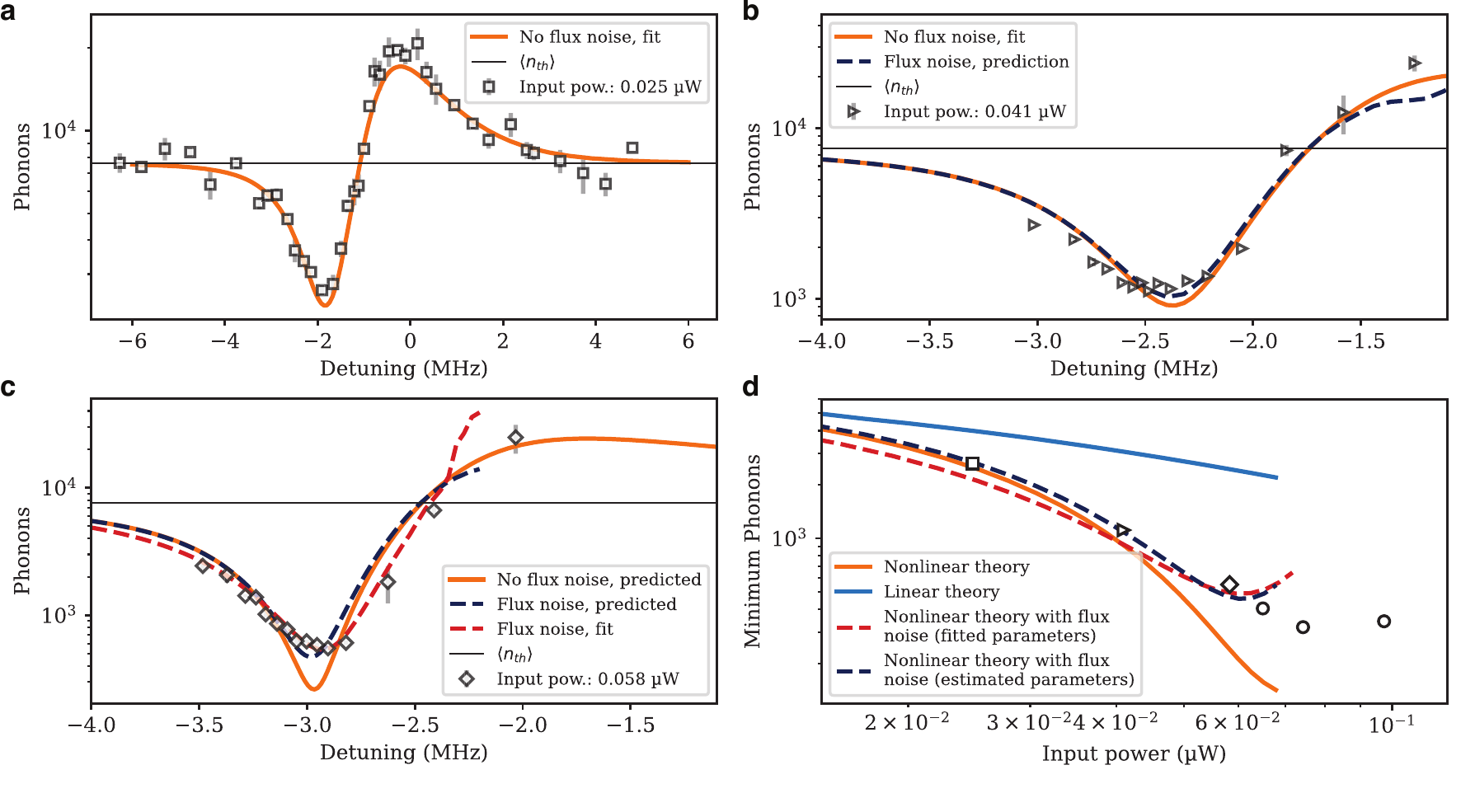}
    \caption{Here we show the influence of flux noise and the capabilities of our model to capture its influence. This is shown for a set of cooling traces at an intermediate $g_0/2 \pi$ of around $\SI{530}{\hertz}$. \textbf{a,} A trace measured at medium power, which is well in the nonlinear regime, but not limited by flux noise. \textbf{b,} A trace at higher power, which is already slightly limited by flux noise. Still we are able to perform the nonlinear fit and also plot what the flux noise model with a given strength of flux noise, but otherwise identical parameters would predict. \textbf{c,} Cooling trace at even higher power, which is clearly limited by flux noise. The theory plotted is not a fit, but a prediction based on the parameters obtained from lower power data. In red we show a fit using the flux noise model, where we fit the flux noise strength, the photon number as well as the Kerr. \textbf{d,} Best cooling for each power alongside the theory. The different symbols correspond to the symbols seen in the cooling traces (a-c). Besides the lowest powers, the cooling is clearly limited by flux noise. We show the prediction of flux noise, one time useing the predicted parameters form the low power sets (a-b) in blue. In red we show the results from fitting the flux noise model to (c), which only gives good agreement for this set. The predictions is not as accurate for that point, but captures the overall behaviour much better. Plotted errors in all panels are the standard errors from several data points.}
    \label{figS:FN_traces_500Hz}
\end{figure*}

We use a dataset measured at an intermediate $g_0/2 \pi$ of around \SI{530}{\hertz} to show the capabilities, but also the limits of our flux noise model. Such an intermediate coupling is favourable for doing this, as the influence of flux noise is low enough to get a trustworthy estimate of the system parameters, while it has a clear influence at higher powers. The sensitivity to flux noise increases approaching bistability, as the cooling backaction happens on an increasingly narrow range due to the nonlinearity. In Fig.~\ref{figS:FN_traces_500Hz} we show cooling traces for different powers, as well as the best cooling achieved for each power. In sub panel (a) we show a cooling trace with a weak enough power such that flux noise barely has an influence and fit the data with the nonlinear theory. Increasing the power (sub panel (b)), the cooling is already slightly limited due to flux noise, while we still manage to fit the data with the usual nonlinear theory. On top of that, we plot a trace including flux noise, taking the same parameters and assume a flux noise strength of $\sigma = 202.2 \cdot g_0$ used in the Gaussian distribution (Eq.~\ref{equS:Gauss}). This corresponds to an absolute flux noise strength of:
\begin{equation}
    \Delta \phi = \SI{76.8}{\micro \phi_0}.
    \label{equS:SigmaFN}
\end{equation}
This was found by re-writing $\sigma$ in terms of $g_0$ together with Eq.~\ref{equS:g0}. The second part of Eq.~\ref{equS:g0} is the flux change from a single mechanical excitation, which is related to $g_0$ via the flux sensitivity of the cavity. We calculated the change per zero point motion to: $\SI{0.38}{\micro \phi_0}$. After regrouping, we can re-express the flux noise strength $\sigma$ as:
\begin{equation}
    \sigma = \Delta \phi \left( \frac{\partial \phi}{\partial x} x_{\text{ZPM}}\right)^{-1} g_0 = \SI{76.8}{\micro \phi_0} \cdot \frac{1}{\SI{0.38}{\micro \phi_0}} g_0 = 202.2 \cdot g_0.
    \label{equS:sigma}
\end{equation}
We use this flux noise strength throughout all measured cooling traces where we show flux noise (e.g. Fig.~3d of the main article or Fig.~\ref{figS:15p5_FN} in here). We found this flux noise strength, by investigating different cooling traces and such a flux noise strength lead to a reasonable agreement. While this flux noise strength is about four times lower than what we obtain when fitting the change of cavity linewidth with increasing coupling (Fig.~\ref{figS:kappaL_vs_g0}) it is in the same order of magnitude. We also plot this flux noise strength in that figure. Especially in the region with a coupling of only a few $\SI{}{\kilo \hertz}$, also the estimated cavity linewidth with this lower flux noise strength is similar to the measured linewidth.

In sub panel Fig.~\ref{figS:FN_traces_500Hz}c we show a cooling trace for higher power, and thus increased sensitivity to flux noise. We plot the nonlinear theory based on predictions from lower power, as well as the flux noise model with the flux noise strength stated above. Moreover, we fit the full model including flux noise and also use the photon number and the Kerr as free fit parameters. There we get a very good agreement to the data, however extrapolating the parameters to other sets fails. In sub panel (d) we show the extraction of the lowest phonon number for different powers, together with the predictions from theory up to bistability. For the linear theory and the nonlinear theory (with and without flux noise), we use the parameters known from low power measurements (i.e. sub panel (b)). In addition, we show what the parameters obtained from the fit to the data shown in sub panel (c) would predict. Here, we already see the limits of our model. It manages good agreement with the data for low to medium influence from flux noise, but seems to fail when going to higher power, where flux noise has an increased impact. While we find parameters which agree well at a given power, agreement with traces from other powers is then not as good. As this is a just simple model, it is expected that the agreement is not perfect, especially in the regions of high sensitivity to flux noise. Among the shortcomings of this model is, that the flux noise is highly likely not Gaussian distributed and not constant over time. We further believe, that going into the nonlinear regime stabilises our cavity, effectively reducing the sensitivity to flux noise, something which was also observed in~\cite{SbothnerFourwavecoolingSinglePhonon2022}. So the effective flux noise would reduce with increasing power, which seems to agree with our data.

Overall we still get reasonable agreement between the data and the nonlinear theory including flux noise. Even though the model is very simple it captures the essential parts of what is happening. From this we can highly likely conclude that we are indeed limited by flux noise.

\section{Supplementary data on low coupling cooling traces}
Here we give additional information on the cooling traces measured at a low $g_0$, presented in Fig.~3 of the main article. We fit the change of mechanical linewidth and frequency to the model discussed in Section~\ref{secS:nonLinTheo}, having the circulating photon number on resonance and Kerr as free fit parameters. Using those results we then calculate the change of phonon number against detuning. We typically only fit the cooling region, as we reach into the region of mechanical instability for high powers in the heating region.
\begin{figure*}[h]
    \centering
    \includegraphics[width = 9cm]{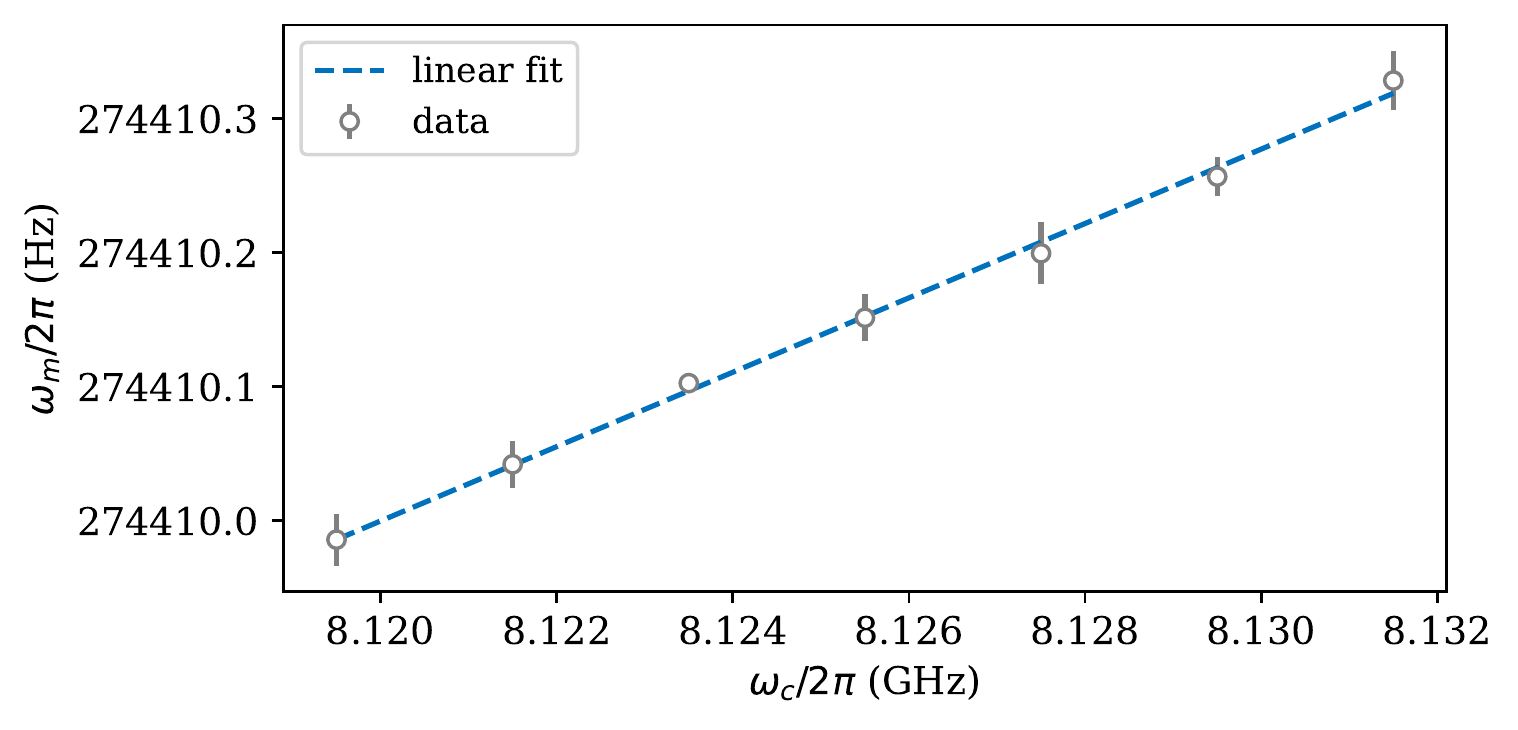}
    \caption{Change of the mechanical frequency with cavity frequency, fitted with a linear slope. This is used to calibrate the mechanical frequency shift for the cooling traces. The error shown is the standard error of multiple measurements.}
    \label{figS:mechFreqSlope}
\end{figure*}

As discussed in section~\ref{secS:dataTaking} we always use our probe tone at the same frequency, while changing the cavity frequency to measure at different probe-cavity detunings. This means that for each data point we have slightly different parameters (e.g. cavity linewidth, $g_0$, etc), which is only a higher order effect as those changes are small and do not limit our data treatment. Especially as the cooling backaction happens over a narrow range of frequencies, where the change of parameters is negligible. However, also the bare mechanical frequency changes with detuning, which proofed to be a limiting effect, especially when working at low powers and hence at low backaction. To avoid this, we measure the bare mechanical frequency for different detunings using a low enough power to avoid any backaction. The result is plotted in Fig.~\ref{figS:mechFreqSlope}, where we clearly see a linear dependence of the mechanical frequency on the cavity frequency and use this to calibrate the mechanical frequency shift. This effect can be modelled with the instantaneous (conservative) backaction from current circulating in the SQUID loop on the cantilever. As we bias the SQUID, a circulating current forms around the SQUID, which also influences the equilibrium position of the cantilever. When the cantilever oscillates around its resting position, the bias through the SQUID changes, which - instantaneously - leads to a force on the cantilever. This effect influences the frequency of the cantilever, while it is still conservative. Modelling this effect, we obtained an approximate linear dependence of the mechanical frequency on the cavity frequency with a comparable strength as seen in the measurement. Hence we believe that this effect leads to the observed change of the mechanical frequency.

\begin{figure*}[h!]
    \centering
    \includegraphics[width = 0.8\textwidth]{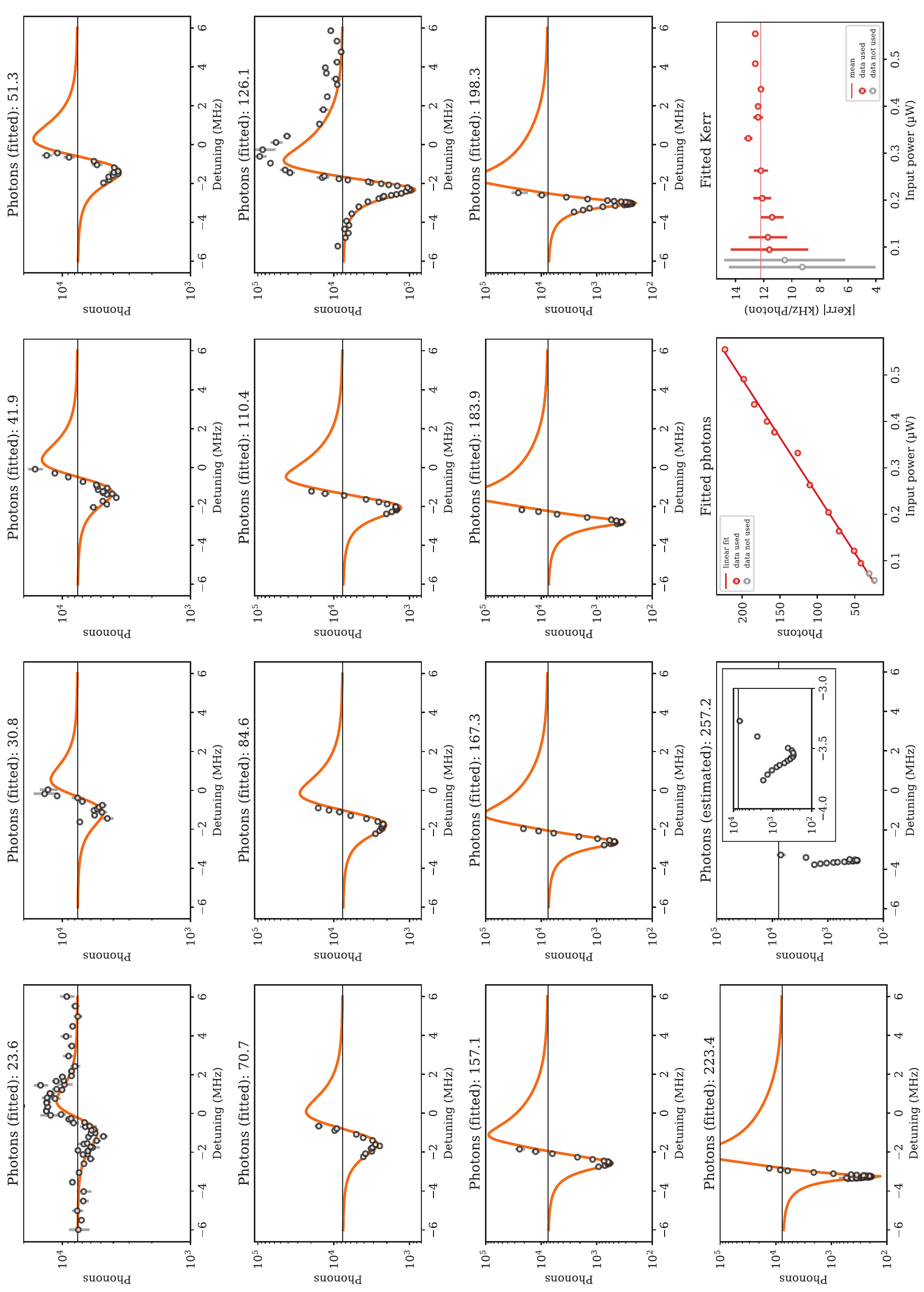}
    \caption{Cooling traces at \SI{201}{\hertz} $g_0$ for different powers. We show the traces together with an independent nonlinear fit for each trace. In the bottom right we show the extracted fit parameter against fridge input power, where we give the absolute value of the negative Kerr. More information is given in the text.}
    \label{figS:nonLinCooling_201Hz}
\end{figure*}
In Fig.~\ref{figS:nonLinCooling_201Hz} we plot all the cooling traces taken at a $g_0/2 \pi$ of \SI{201}{\hertz}. In the main paper we show the complete ones (where both tails, cooling and heating region were measured), which are plotted in the top left sub panel and in the second row most right sub panel. For all we show the nonlinear fit and give the extracted photon number. For the one with highest power, we do not show a fit, as this is measured above bistability and as shown in the inset, the cooling backaction happens on a very narrow range. It is evident, that the cooling backaction shifts to lower frequencies with increasing power, as expected from the nonlinearity (where we use the probe-cavity detuning with respect to the low power (linear) cavity on the x-axis). On top of that, the cooling region gets increasingly narrow approaching bistability. The nonlinear fit shows excellent agreement with the data. Only for the highest power before bistability, we seem to be limited by flux noise, which we will discuss below. In the two bottom right panels, we show the extracted Kerr and photon number, which are also the only two fit parameters. As expected, the photon number linearly increases with input power, while the Kerr remains the same over all powers. 

\begin{figure*}[h]
    \centering
    \includegraphics[width = \textwidth]{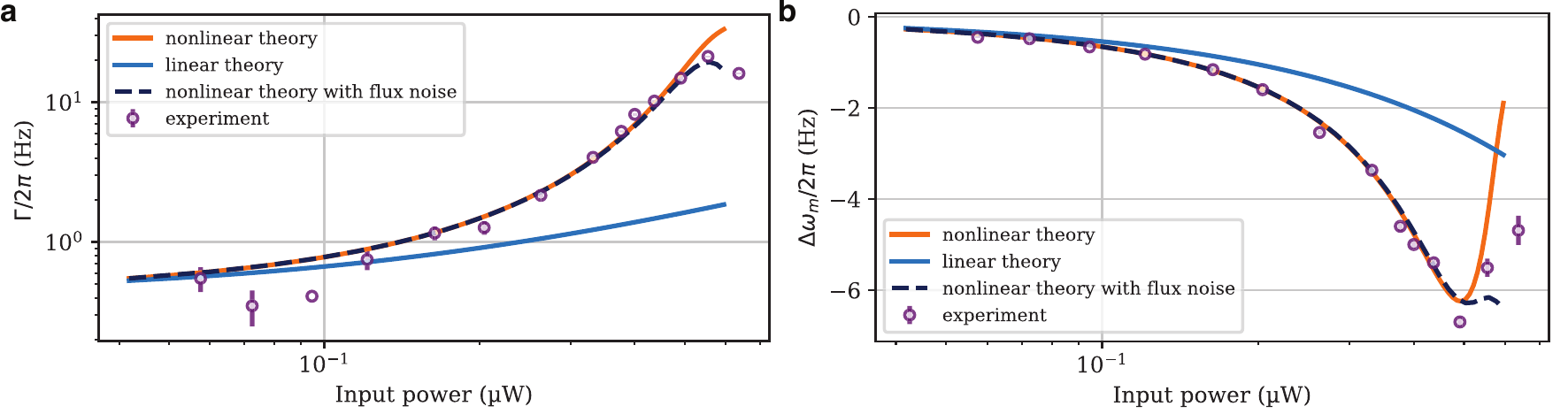}
    \caption{\textbf{a,} Mechanical linewidth and \textbf{b,} frequency shift with increasing power at the detuning we measure the lowest phonon number (Fig.~3d in the main article). We plot the predictions from linear and nonlinear theory as well as the prediction from the nonlinear theory including flux noise. The error shown is the standard error of multiple measurements.}
    \label{figS:201Hz_lw_fr}
\end{figure*}

In Fig.~\ref{figS:201Hz_lw_fr} we plot the change of mechanical linewidth and frequency with increasing input power at most cooling. In Fig.~3d of the main article we already plot the corresponding lowest phonon number we achieve for each power. To calculate the theory predictions, we use the average value of the Kerr and a linear fit for the conversion from input power to circulating photons (bottom right of Fig.~\ref{figS:nonLinCooling_201Hz}). However we ignore the lowest two powers due to their large uncertainties, as for those powers the cavity is still near the linear regime. We clearly see that only the nonlinear theory can accurately predict the experimental results. For the powers where we have the most cooling, as discussed in the main article, we are already limited by flux noise. While the model (Section~\ref{secS:FN_model}) seems to predict the influence of flux noise for the linewidth rather well, it works not as good for the prediction of the frequency shift. We estimate the strength of flux noise according to relation~\ref{equS:SigmaFN}.

\begin{figure*}[h]
    \centering
    \includegraphics[width = 9cm]{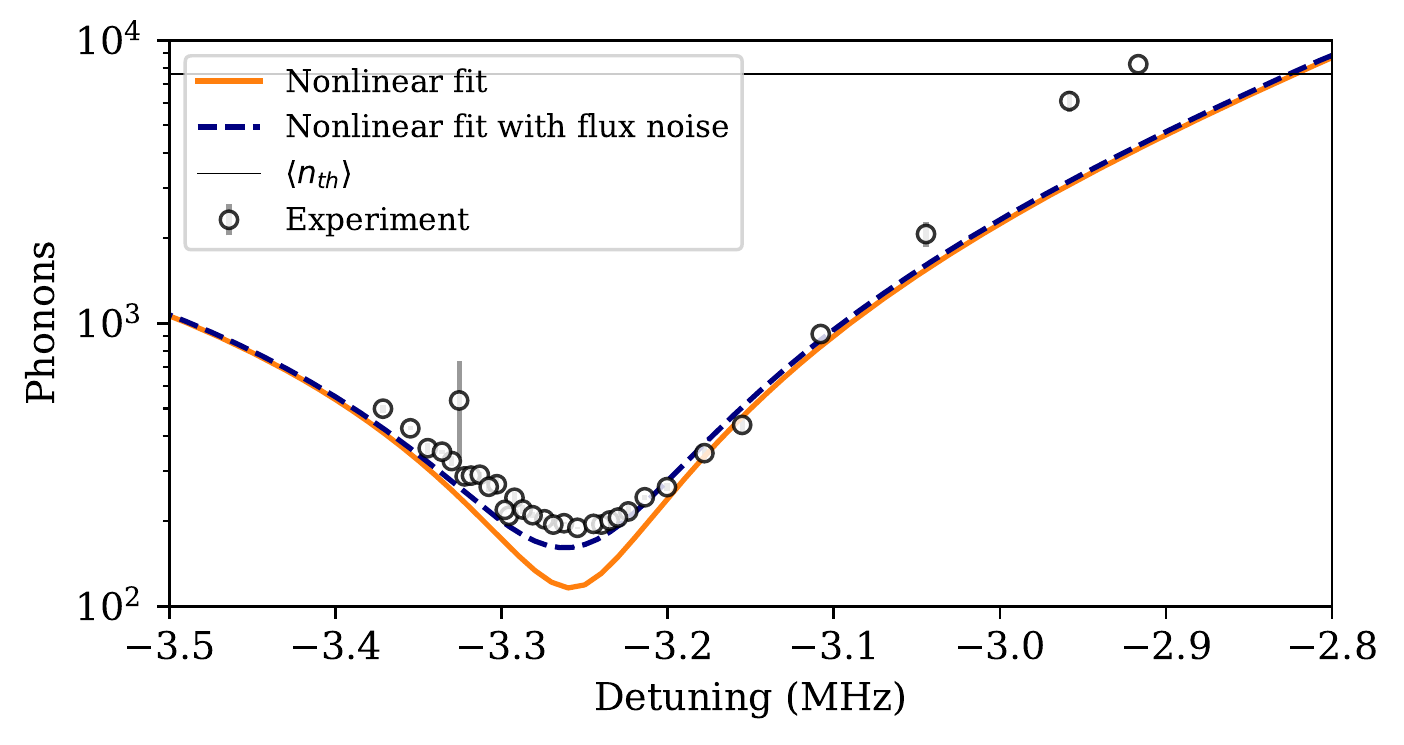}
    \caption{Prediction for the cooling trace with 224 photons in case of flux noise. The parameters are the same as obtained from the fit without flux noise to the trace, the strength of the flux noise is estimated from relation~\ref{equS:SigmaFN}.}
    \label{figS:15p5_FN}
\end{figure*}

Investigating the last trace before bistability closer (223.4 photons, Fig.~\ref{figS:nonLinCooling_201Hz}), we also see that the cooling is limited by flux noise. In Fig.~\ref{figS:15p5_FN} we plot this measurement together with a prediction for the cooling trace including flux noise. While there is not perfect agreement, the flux noise model is much closer to the data and also the shape is very similar. This is purely a prediction, using the parameters obtained from the fit of the full nonlinear model without flux noise together with the flux noise strength estimated from relation~\ref{equS:SigmaFN}.

\begin{figure*}[h]
    \centering
    \includegraphics[width = \textwidth]{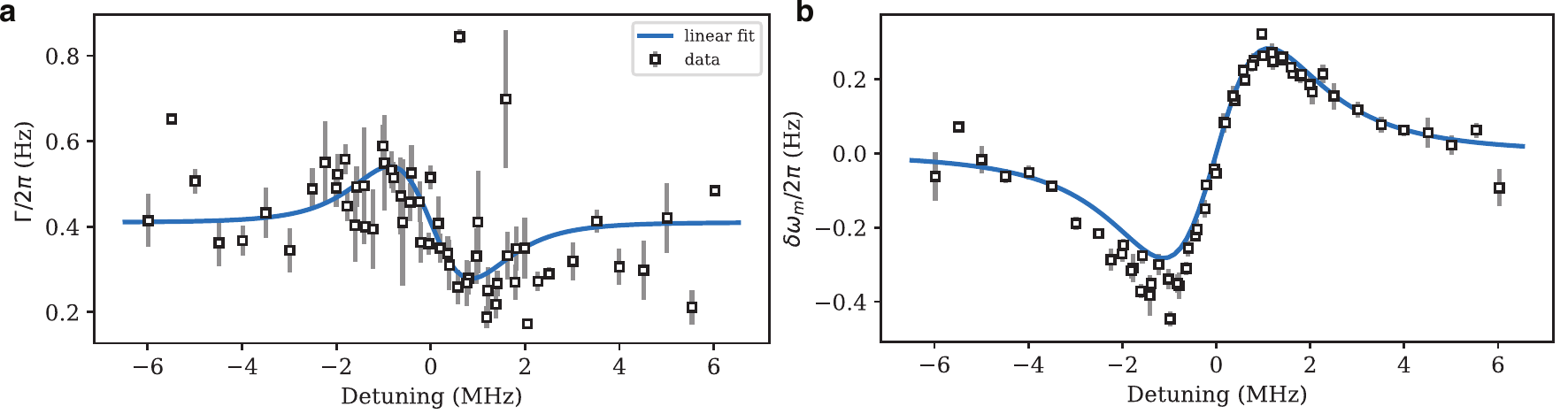}
    \caption{\textbf{a,} Change of mechanical linewidth and \textbf{b,} frequency against detuning for the low power dataset, where the change of phonon number is shown in the main article. We only have low backaction in this measurement, thus there is a large uncertainty to the change of linewidth, using the standard error of multiple data points. However the change of frequency (b) works more reliable, which also enables to get a faithful fit (linear theory) to the data.}
    \label{figS:24p5_linFits}
\end{figure*}

In Fig.~3a of the main article, we show the change of phonon number vs detunings for a low power measurement. Here, we additionally show the change of mechanical linewidth and frequency for this measurement, together with the linear fit, Fig.~\ref{figS:24p5_linFits}. While on the extraction of the linewidth, we have a large uncertainty, the change of the mechanical frequency is extracted with more accuracy, which also allows us to perform the fit. We see, that in the low power regime, the backaction is well described using the linear theory~\cite{Ssafavi-naeiniLaserNoiseCavityoptomechanical2013a,SmarquardtQuantumTheoryCavityAssisted2007a}, however the frequency shift is already systematically underestimated where we have the most cooling.

\begin{figure*}[h]
    \centering
    \includegraphics[width = 9cm]{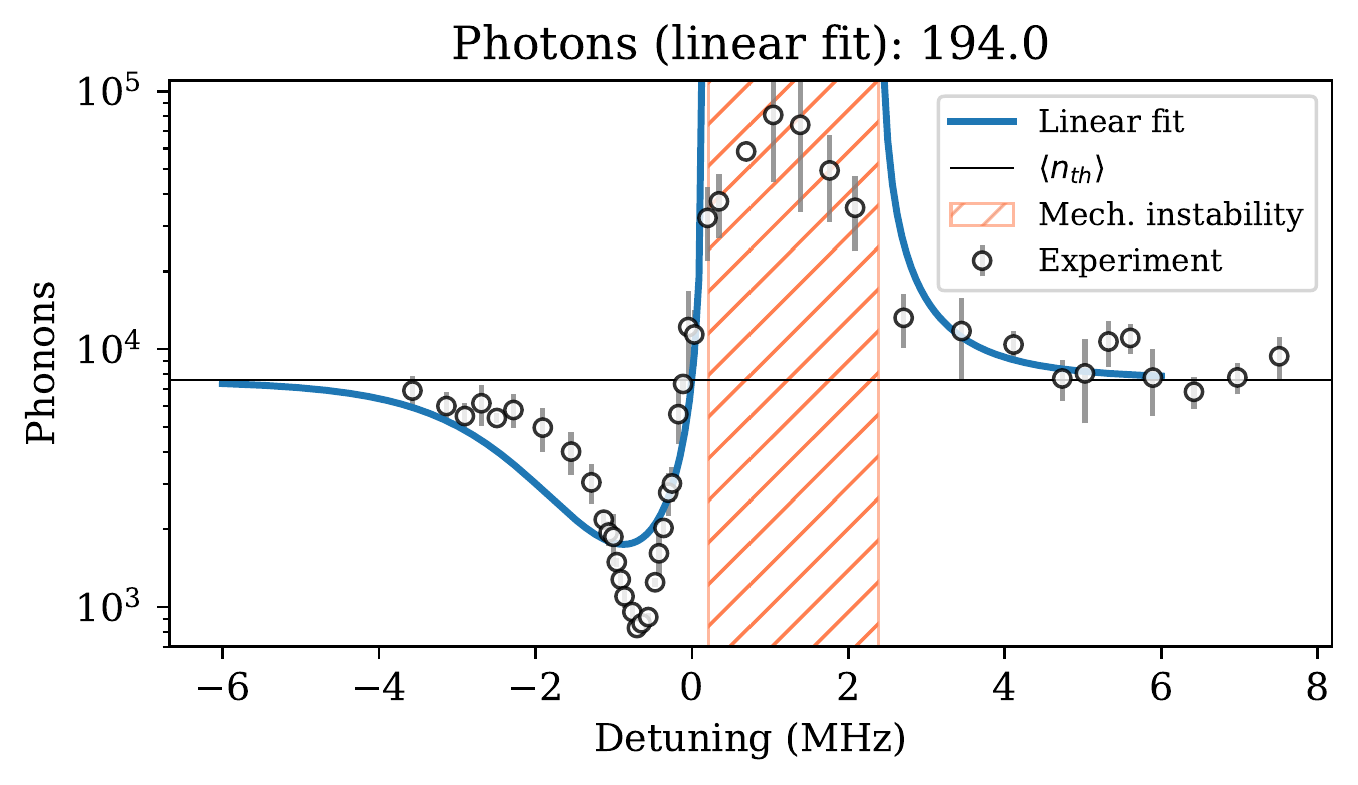}
    \caption{Linear fit on the high power dataset shown in the main article. The linear theory is clearly unable to reproduce the measurement. For the fit, we only consider the cooling parts, so values measured for negative detunings.}
    \label{figS:18_linFit}
\end{figure*}

In contrast to this, we show in Fig.~\ref{figS:18_linFit} a fit to the high power dataset presented in the main article using the linear theory. We see that the theory is clearly unable to describe the measured data well. It overestimates the heating by even entering into the region of mechanical instability, where the predicted mechanical linewidth is below zero. On the other hand it underestimates the cooling, while the photon number predicted by the fit is higher than what the nonlinear theory gives with 126.1. We want to note, that a crucial aspect is how to define the detuning (x-axis) for each measured data point when comparing with the linear theory. The reason is that due to the nonlinearity, the cavity shifts to lower frequencies when probing with a strong pump tone. To obtain the detuning for each data point, we typically use a low power VNA measurement, so we measure the unshifted frequency. This shift is naturally part of the nonlinear theory and can be clearly seen in Fig.~\ref{figS:nonLinCooling_201Hz}. However in the linear case it cannot be included and for Fig.~\ref{figS:18_linFit} we correct this by shifting the data points \SI{1.65}{\mega \hertz}, such that the measurement with no net backaction is exactly at zero detuning, as expected for the linear case. For clarity, we also shift the data shown in the main article by this same amount.

\section{High coupling regime}
\begin{figure*}[h]
    \centering
    \includegraphics[width = \textwidth]{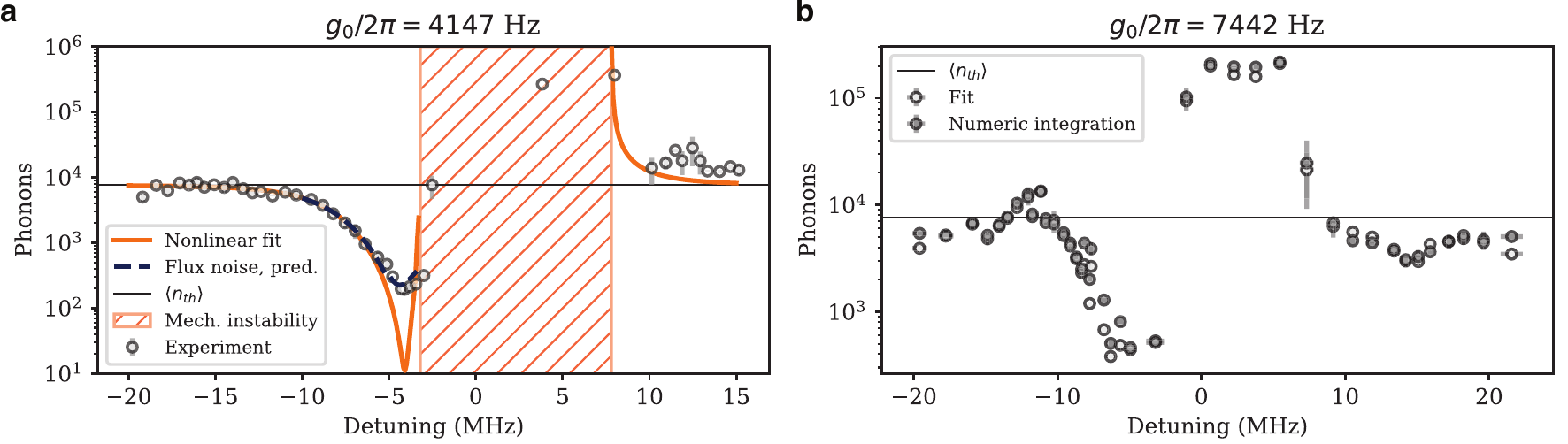}
    \caption{Cooling traces measured at high $g_0$. \textbf{a,} Trace at around $g_0/2\pi = \SI{4}{\kilo \hertz}$. We are able to fit the cooling trace, however it is clearly limited by flux noise. Plotting the prediction of flux noise using the parameters from the fit, together with the flux noise strength discussed in Section~\ref{secS:FN_model} gives good agreement. We also highlight the region, where the mechanical mode is expected to be in the regime of mechanical instability. \textbf{b,} Cooling trace at around $g_0/2 \pi = \SI{7.5}{\kilo \hertz}$. Here, the data quality clearly suffers from flux noise and we are not able to pass the fit through the data using the nonlinear cooling theory. As some spectra are heavily influenced by flux noise, we also show the results from numerical integration, alongside the usual fits to the spectra. There is overall good agreement between both methods, however in the region of most cooling, some differences occur.}
    \label{figS:highg0}
\end{figure*}

In this section we show cooling curves measured at high couplings, to demonstrate our ability of operating our system at those parameters. As flux noise has higher influence at higher couplings, we cannot fit traces at highest couplings and our cooling is limited as well. In Fig.~\ref{figS:highg0} we show two cooling traces, one at a $g_0/2 \pi$ of around \SI{4}{\kilo \hertz}, sub panel (a), and one at a $g_0/2 \pi$ of around \SI{7.5}{\kilo \hertz} (b). For the \SI{4}{\kilo \hertz} trace, which is taken at around half the bistable power, we are still able to perform a fit using the nonlinear theory. However in the region of most cooling, we are clearly limited by flux noise. The flux noise model (Section~\ref{secS:FN_model}), where we again estimate the strength of the flux noise using Eq.~\ref{equS:SigmaFN}, predicts the overall behaviour and also the minimum phonon occupation very well. While we manage to cool to just below 200 phonons starting from around 7600 thermal phonons, according to the nonlinear theory, we would be able to cool to nearly 10 phonons if we had no flux noise. While assuming the same parameters for a linear cavity, we expect to cool to around 90 phonons.

In sub panel (b) we show a cooling trace taken at even higher $g_0/2 \pi$ of nearly \SI{7.5}{\kilo \hertz}. Here, flux noise prevents us from fitting this data. Still we are able to measure a cooling trace, even though the data quality clearly suffers. As the mechanical spectra are also influenced by flux noise, we show the results from numerical integration along the results of fitting the spectra. Here, we see good agreement, only in the region of most cooling - where we are also most sensitive to flux noise - we see slight differences, which shows that we can trust the fit for such high coupling strength.

With the data shown in this section, we demonstrate that we can operate our system at very large couplings of up to \SI{7.5}{\kilo \hertz}. While at those couplings strength, there is too much flux noise to fit the cooling traces to theory, we can still do so at couplings of \SI{4}{\kilo \hertz}. However already at those couplings, our cooling abilities are clearly limited by flux noise.

\section{Kerr anomaly}
\label{secS:KerrAnom}
\begin{figure*}[h]
    \centering
    \includegraphics[width = 9cm]{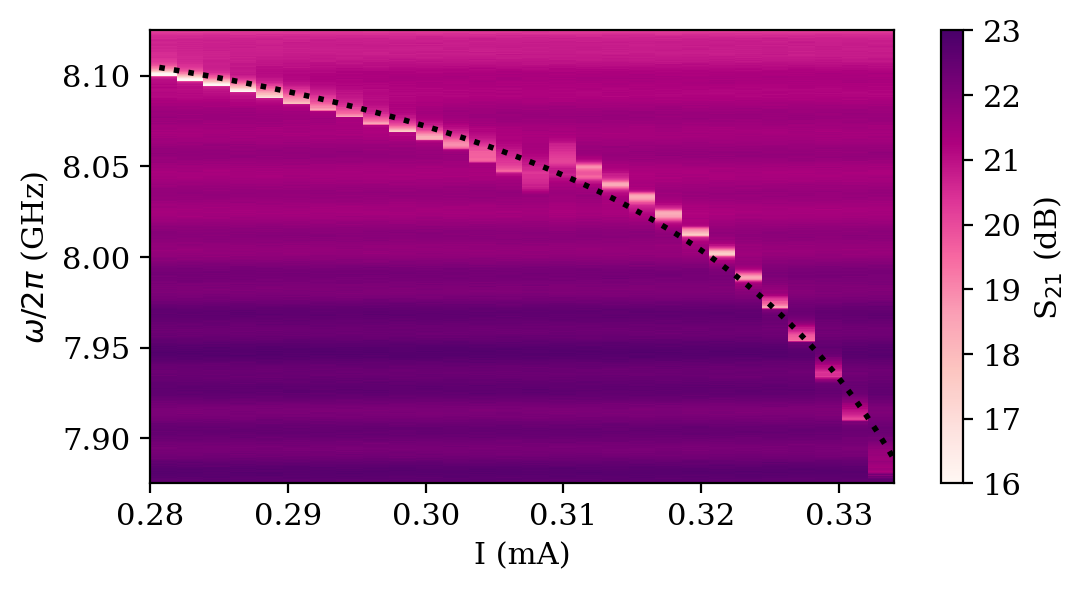}
    \caption{Flux map taken with a strong drive (-30~dBm fridge input power) in the region of the anomalous Kerr. The dotted black line shows the expected cavity frequency when doing a low power measurement. At a bias frequency of around \SI{8.05}{\giga \hertz} we see we see the onset of the anomaly. There the cavity shifts higher compared to the low power case, indicative of a positive Kerr. Afterwards we recover a negative Kerr again.}
    \label{figS:fm_highPow_kerrAno}
\end{figure*}

Here we discuss and present a model for the occurrence of the Kerr anomaly in our system. First, we want to show in which frequency region this anomaly occurs. For this, we show a flux map taken with high power in Fig.~\ref{figS:fm_highPow_kerrAno}. The anomaly begins at around \SI{8.05}{\giga \hertz}, where the response of the cavity gets very shallow in a narrow range of frequencies. Afterwards the cavity frequency is increased compared to the low power case (black dashed line). For even further increasing the flux bias point, it returns to the usual shift to lower frequencies compared to the low power case.

\begin{figure*}[h]
    \centering
    \includegraphics[width = \textwidth]{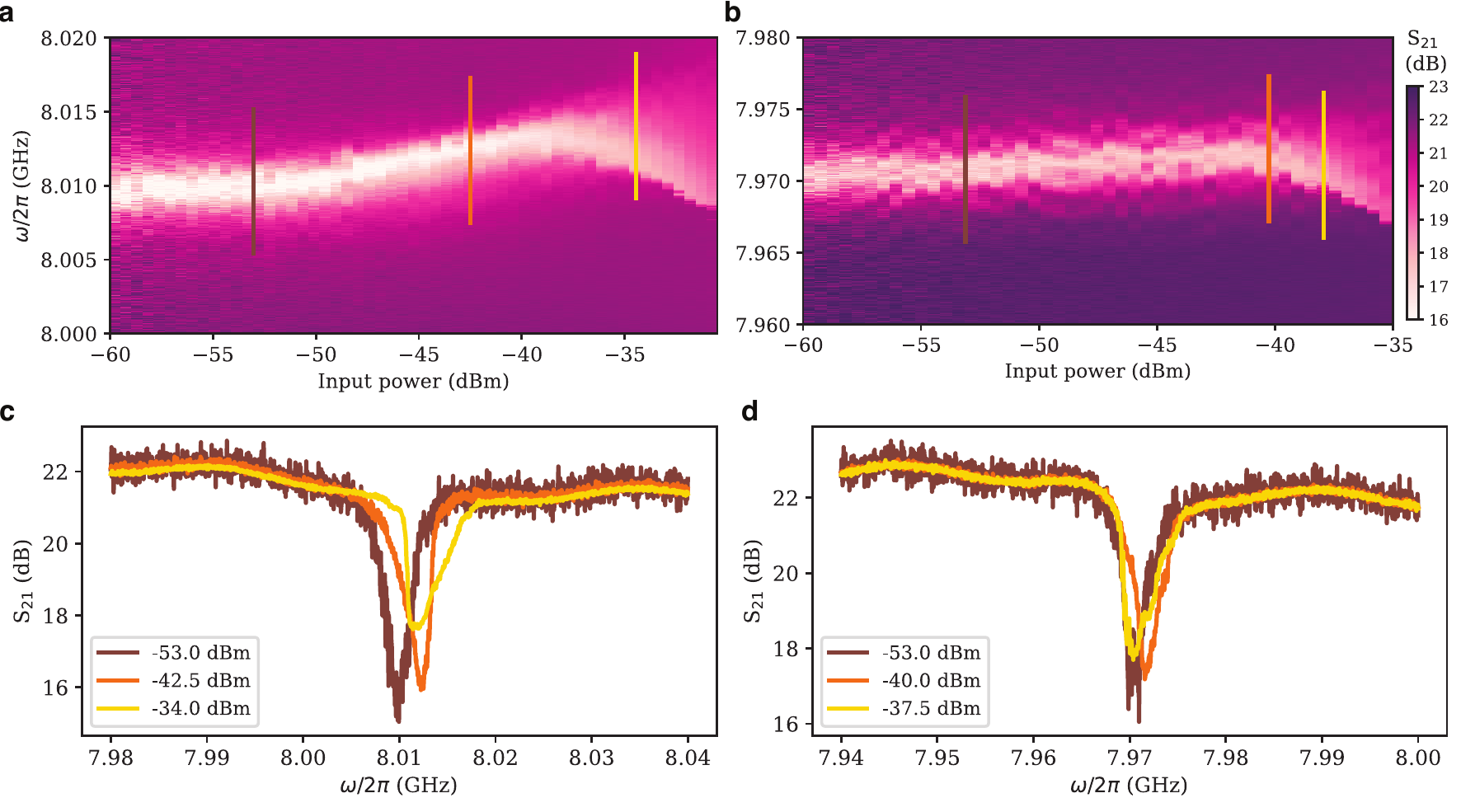}
    \caption{Power sweeps of the cavity for two different flux bias points in the anomalous Kerr region. \textbf{a,} Power sweep at a bias frequency of \SI{8.01}{\giga \hertz} and \textbf{c,} line cuts at different powers. We see that the cavity shifts to higher frequencies for intermediate powers before shifting to lower frequencies for higher powers. At -42.5~dBm we see a steep slope on the right side of the cavity, a clear sign of a cavity with positive Kerr. Going to even higher powers, this changes to the opposite, having the steep part of the cavity at the typical lower frequency side. \textbf{b,d,} Power sweep at the frequency we took the data for the highest cooling we achieve (e.g. Section~\ref{secS:bestCooling}). Here, the effect of the positive Kerr is reduced but it still has an influence, allowing us to work at higher powers when reaching bistability. Remarkably, as it can be seen in (d), at bistability the net frequency shift compared to low power is close to zero, as the shifts from positive and negative Kerr just cancel.}
    \label{figS:KerrAno_overview}
\end{figure*}

Now, we give more details about the behaviour of the cavity in this anomalous region, by showing two power sweeps at different cavity frequencies, Fig.~\ref{figS:KerrAno_overview}. As we can see, especially in the sweep taken at \SI{8.01}{\giga \hertz} (sub panels (a), (c)), the frequency of the cavity first shifts up, due to a positive Kerr and shifts down for even higher powers. This can be also seen in the line cuts plotted below, where in the medium power measurement the cavity has a steep slope on the higher frequency side, indicative of a positive Kerr. For higher powers we obtain, as usual, the steep part on the low frequency side. Thus for high enough powers, the cavity behaves like a usual cavity with negative Kerr and this anomaly just pushes this regime to higher powers, effectively reducing the Kerr. As stated in the main article, this Kerr anomaly is actually beneficial for cooling the mechanical resonator. Typically, the nonlinearity of the cavity increases when going to higher coupling strengths, which allows for fewer photons at bistability. While the backaction strength increases with increasing coupling, operating at higher photon number and still having a nonlinear cavity is even more beneficial. In sub panels (b) and (d) we show such a measurement at a bias point of \SI{7.97}{\giga \hertz}, which we use for measuring the best cooling we currently achieve. Here, the effect of the Kerr anomaly is already less dominant, but we still benefit from the overall reduced Kerr and thus higher photon numbers when reaching bistability. Remarkably, we can operate at similar powers as in the backaction measurements at low $g_0$ (e.g. Fig.~\ref{figS:kerrShift_8p125}), while having a more than ten times bigger coupling strength. However, it should be noted, that we are also more sensitive to flux noise, which limits the increase of backaction we can achieve.

\begin{figure*}[h]
    \centering
    \includegraphics[width = \textwidth]{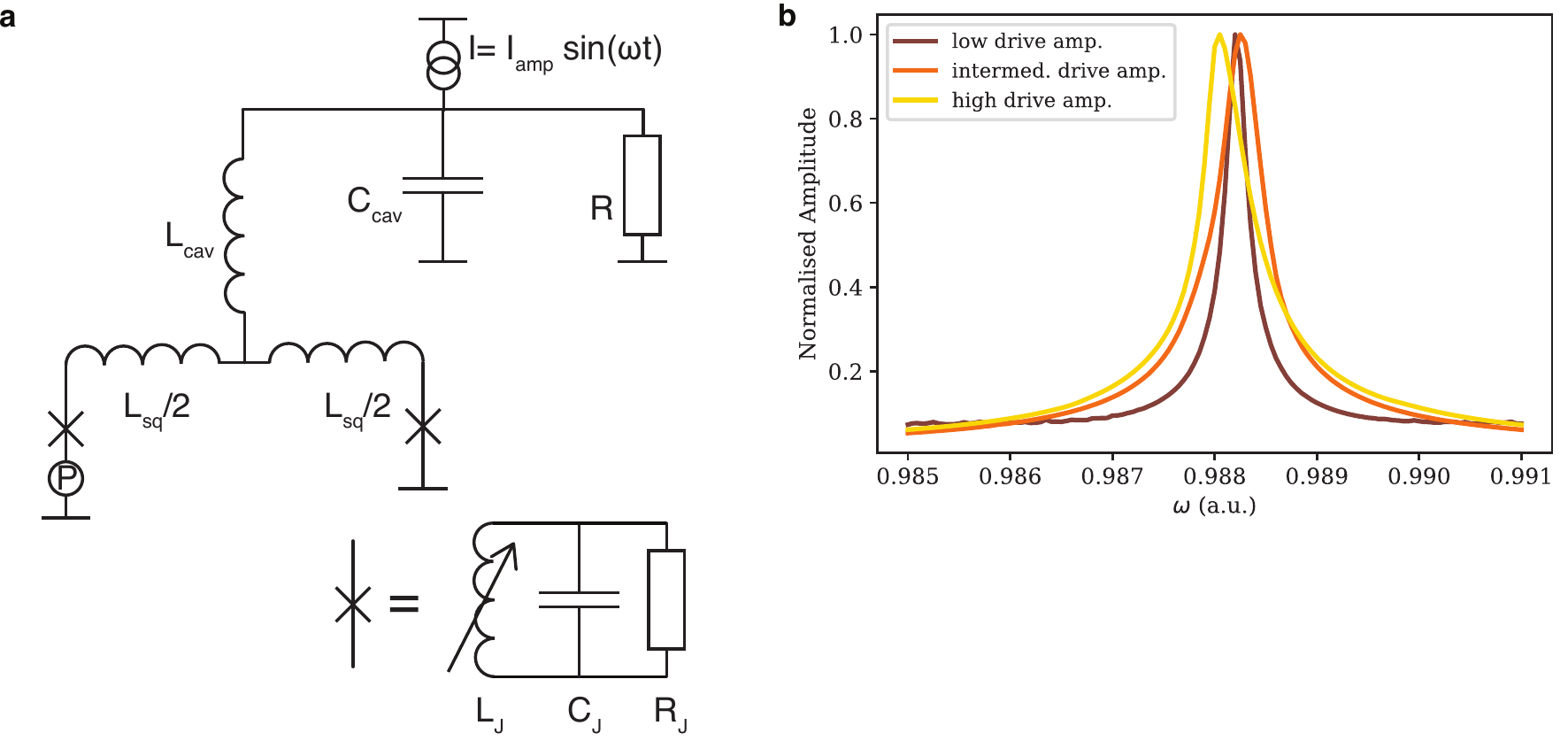}
    \caption{Simulation of our circuit to get better insight into the origin of the Kerr anomaly. \textbf{a,} Model used for simulating the circuit. \textbf{b,} In a certain region of frequencies, we see exactly the same qualitative behaviour as we see in the measurements (Fig.~\ref{figS:KerrAno_overview}). There, the resonance frequency increases first for increasing power, before it decreases, as it would usually do, when increasing the power further.}
    \label{figS:KerrAno_sims}
\end{figure*}

To get a qualitative understanding of the origin of this phenomenon, we model our circuit using lumped elements (Fig.~\ref{figS:KerrAno_sims}a), and recover a similar behaviour as in the experiment. We use a software called PScan~\cite{SkornevNumericalSimulationJosephsonjunction1997,SpolonskyPSCANPersonalSuperconductor1991} and simulate our circuit with the parameters shown in Tab.~\ref{tabS:KerrAno_sims}. Those parameters are only estimates of the setup parameters and while they allow us to get a qualitative understanding of the effect, we cannot draw quantitative conclusions. We use the RCSJ model to describe the Josephson junctions together with the phase balanced method to simulate the circuit. To obtain the results, we use the voltage response of the system to an applied driving current with different frequency and amplitude. In Fig.~\ref{figS:KerrAno_sims}b we show the voltage response for three different current strengths when sweeping the frequency across, where we always normalise to the highest voltage response for a given drive strength, similar to a VNA type of measurement. In a certain range of frequencies, we find very similar behaviour as in the experiment, Fig.~\ref{figS:KerrAno_overview}, with the Kerr being positive in a certain range of powers, before becoming negative again for increasing powers. This behaviour is happening in a region, where the third harmonic of the cavity frequency is very similar to the plasma frequency of the junction. A junction is a nonlinear element with the ability to transfer energy between different modes. As the current-phase relation of the junction is described by a sinusoidal, we expect that, next to the fundamental mode, the third harmonic has most influence, in this case the third harmonic of the cavity mode. Therefore, we believe that the plasma frequency of the junction is the reason for this Kerr anomaly.

\begin{table}[h]
\begin{tabular}{c|c} 

 Parameter & Value \\  \hline
 $L_{cav}$ & \SI{4.96}{\nano \henry} \\  \hline
 $C_{cav}$ &   \SI{144}{\femto \farad} \\ \hline
 $R$ &  \SI{648.2}{\ohm} \\ \hline
 $L_{sq}$ &  \SI{0.0195}{\nano \henry} \\ \hline
 $I_c$ & \SI{1.67}{\micro \ampere} \\  \hline
 $\beta$ & 333420 \\  \hline
 $R_{N}$ & \SI{13100}{\ohm} \\  \hline
 $P$ & \SI{0.13}{\phi0} \\
 
\end{tabular}

\caption{Parameter used to simulate our circuit. The inductance and capacitance of the cavity, $L_{cav}$ and $C_{cav}$, are obtained from finite element simulations. The resistance $R$ is chosen such that it matches the linewidth of the cavity. The geometric inductance of the SQUID, $L_{sq}$, is estimated from the flux map. The critical current of the junction $I_c$ is obtained from the design parameters, where we estimate that it is lowered by a factor of 5 due to the presence of the magnet in its vicinity, obtained from a change in the flux map, when placing the magnet. To fully characterise the junction, we use the McCumber parameter $\beta$, which also depends on the junction capacitance (known from design parameters) and the gap resistance $R_N$. For the shown simulation, we use a phase drop $P$, to set the flux bias point, such that the third harmonic of the cavity is similar to the plasma frequency of the junction. Relating those parameters to the model given in Fig.~2a of the main article, the inductance $L$ given there, is approximately $L_{cav}/2$. The capacitance $C_{cav}$ relates to $C$ and $C_G$, further the coupling given in the main manuscript by $C_C$, is here accounted for by dissipation through $R$, owing to the current driven model used for the simulation.}
\label{tabS:KerrAno_sims}
\end{table}

This complex behaviour prevents us from comparing full cooling traces to the theory, as this would require a quantitative modelling of the power dependent Kerr. Developing such a model is highly challenging, as already for a single power, the photon number in the cavity changes with the probe-cavity detuning, which also changes the power dependent Kerr in this region. Additionally, flux noise is clearly present in this range of coupling strengths of around \SI{2}{\kilo \hertz}, and not only does it influences the cooling traces, but also the frequency sweeps of the cavity.


\section{Supplementary data on best cooling}
\label{secS:bestCooling}
Here we give supplementary information about the measurements where we achieve highest cooling in our system (see Fig.~4(a,b) of the main article), which is at a coupling of around \SI{2.1}{\kilo \hertz}. In contrast to all other measurements, taken at a cryostat temperature of \SI{100}{\milli \kelvin}, we reduce the cryostat temperature to \SI{40}{\milli \kelvin} for these measurements. Next to lowering the bath temperature surrounding our cantilever, we also have further decoupling from the environment, as its linewidth decreases with temperature (Fig.~\ref{figS:tempRamp}c). Due to the large coupling, we are also increasingly sensitive to flux noise. In order to slightly reduce the effect of slow drifts, we increase the bandwidth of the spectrum analyser from the usually used \SI{0.1}{\hertz} to \SI{0.2}{\hertz}, which approximately halves our measurement time per spectrum. To keep the measurement time per data point the same, we double the number of traces taken when measuring with \SI{0.2}{\hertz}. To minimise the sensitivity to flux drifts further, we also re-tune the cavity more frequent for those measurements (i.e. every four traces with \SI{0.2}{\hertz} instead of every five traces with \SI{0.1}{\hertz}). As we still take 8001 data points within the spectrum, we increase the span to \SI{1600}{\kilo \hertz}.
For the spectra shown in the main article, we average all traces within a detuning bin on top of each other, which are typically 40 traces for a \SI{0.2}{\hertz} bandwidth measurement, Sec.~\ref{secS:dataAnalysis}. For the two traces shown in Fig.~4 of the main article, where we measure strong cooling (123 and 14 phonons), we use a power close to the bistable one, but different probe-cavity detunings. For the shown spectra with less cooling, we operate at half the bistable power and are further away from the optimal probe-cavity detuning for cooling.

\begin{figure*}[h]
    \centering
    \includegraphics[width = 9cm]{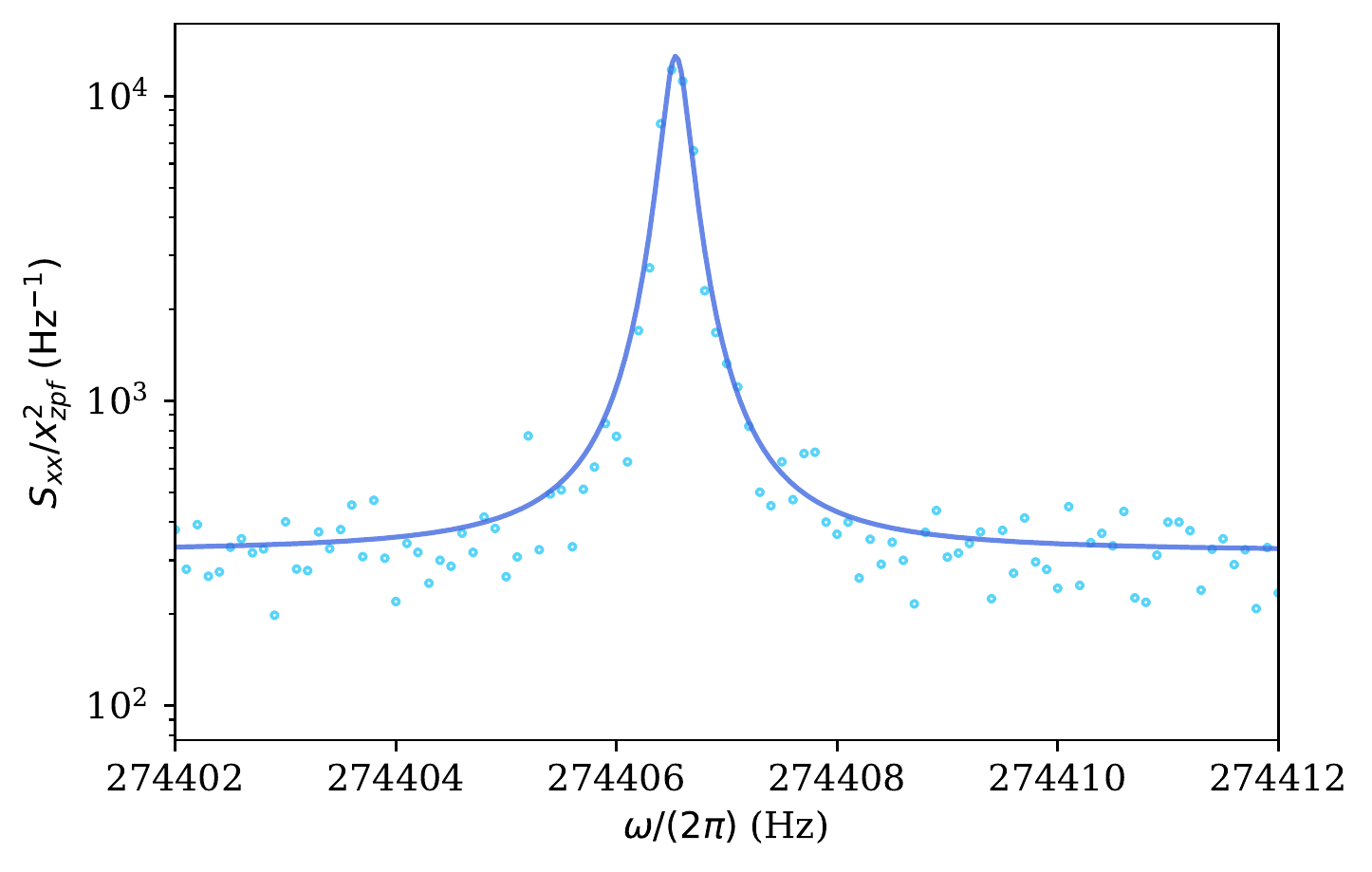}
    \caption{Zoom in to the trace where we do not observe backaction, from the spectrum shown in Fig.~4a of the main article. Due to the large coupling, we still have a decent signal to noise, also at an estimated circulating photon number below 2.}
    \label{figS:bestCool_noBA}
\end{figure*}

In Fig.~\ref{figS:bestCool_noBA} we show a zoom in to the spectrum shown in Fig.~4a of the main article, where we do not have backaction, due to a sufficiently large probe-cavity detuning. Despite having a low number of circulating photons in the cavity, we still have a decent signal due to the large coupling strength.

To determine the photon number in the cavity we would usually use the (nonlinear) fit, Sec.~\ref{secS:nonLinTheo}. As the measurements discussed here suffer from flux noise and on top of that we operate in the Kerr anomalous region (Sec.~\ref{secS:KerrAnom}), we are not able to fit the traces. In order to still get a faithful estimate of the photon number, we use the input photon number known form measurements at lower $g_0$, together with the measured cavity parameters at this coupling. The Kerr has most influence around the resonance, where most backaction occurs. For the estimate of the photon number, we use a fixed value for the Kerr, which resembles the overall frequency shift well, but underestimates the effective Kerr for most cooling.

For the trace of highest cooling we show, Fig.~4b of the main article, flux noise heavily influences the mechanical spectrum, thus we cannot trust the model of a damped harmonic oscillator we usually use. To still obtain an estimate of the phonon number we use a two fold approach, using the numeric integration and the model including flux noise, discussed in Section~\ref{secS:FN_model}. There we also discuss the asymmetry of the spectrum, which ultimately leads to an asymmetry in the tail and constrains the accuracy of the numerical integration. In contrast to typically just fitting the mechanical peak to extract linewidth and frequency, here we simulate the mechanical spectrum using the full model and use the offset, the Kerr, the probe-cavity detuning as well as flux noise strength number as fit parameters. We fix the circulating photon number in the cavity given by the known input photon number, calibrated via a measurement at lower coupling. As we just fit a single spectrum and the fit parameters are strongly correlated, we have to be careful with interpreting the values, however they are within an expected range. The shape of the curve is fitted very well, and key features, like the tail towards lower frequencies being much shallower are reproduced.

\begin{figure*}[h]
    \centering
    \includegraphics[width = \textwidth]{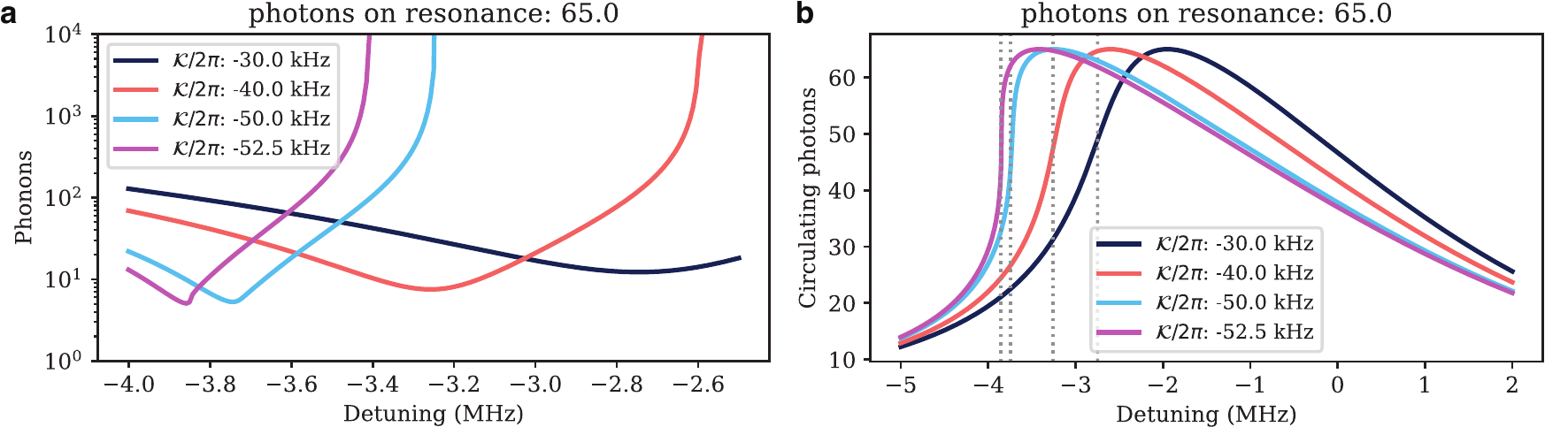}
    \caption{Estimation of the cavity photon number for different Kerr values at the optimal detuning for cooling. \textbf{a,} Cooling traces for a fixed resonance (input) photon number, but different Kerr values. At the highest Kerr (\SI{52.5}{\kilo \hertz}), we are at 99\% of the bistable photon number. There we would expect to cool (without flux noise) to around 5 phonons. \textbf{b,} Cavity photon number against detuning for different Kerr values. We further indicate (dashed) the detuning for which we achieve best cooling. The photon number at the optimal detuning is similar for all Kerr values, despite the change of the cavity lineshape.}
    \label{figS:19p8_PhotEsti}
\end{figure*}

It is also challenging to determine the cavity photon number, due to the Kerr anomaly. Performing the approach we discussed above, we get an estimate, which we give in the main manuscript. Here, we want to investigate the influence of a different Kerr on the cavity photon number, as due to the Kerr anomalous region, the Kerr increases with drive strength. We know that we are operating close to bistability. Furthermore, we can use the input known from lower $g_0$ measurements and use this to estimate the Kerr. Additionally, using the parameters from fitting the spectrum from the model including flux noise, we expect to cool the mechanical mode to below 10 phonons assuming no flux noise. In Fig~\ref{figS:19p8_PhotEsti}a we plot theoretical cooling traces, using the cavity parameters at this coupling together with different values for the Kerr. In Fig~\ref{figS:19p8_PhotEsti}b we show the cavity response and indicate the detunings for best cooling. Depending on the Kerr, we get a photon number of 42 for the highest value of Kerr, where we would operate at 99 \% of the bistable Kerr, and up to 49 photons for the lowest Kerr. Those numbers are close to the 52(5) estimated with the method discussed previously, however slightly lower. This difference can be discussed by the Kerr anomaly, where the Kerr is positive for low drive strengths (e.g. see Fig.~\ref{figS:KerrAno_overview}b), requiring more photons to reach bistability or simply flux noise leading to an uncertainty in the detuning. Further, we see that the Kerr only has a slight influence on the photon number for optimal detuning for the cooling.

\begin{figure*}[h]
    \centering
    \includegraphics[width = 9cm]{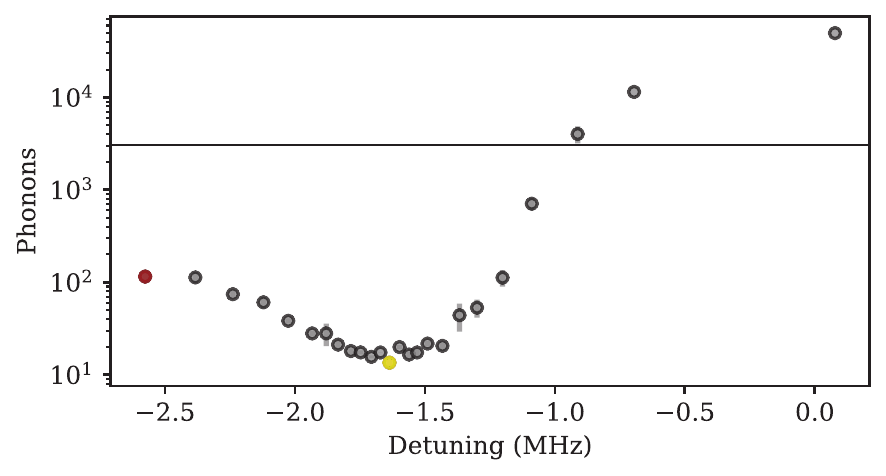}
    \caption{Full cooling trace measured at a $g_0/2 \pi$ of \SI{2.1}{\kilo \hertz}, where we observe strong cooling. Two spectra of this trace (coloured points) are show in Fig.~4 of the main article, the line symbolises the thermal occupation. As we cannot trust the usual fit anymore, the phonon numbers shown here are obtained from numerical integration.}
    \label{figS:19p8_coolTrace}
\end{figure*}

In Fig.~\ref{figS:19p8_coolTrace} we show the full cooling trace for the measurement closest to bistability and indicate which points are used for the plots in the main article (Fig. 4(a,b)). As we cannot trust the usual damped harmonic oscillator model, we show data obtained via numerical integration here. We also note, that the bottoming out at largest cooling, clearly indicates limiting effects from flux noise.

\section{Supplementary data re-scaling the power spectrum for the best cooling plots}
\label{secS:bestCoolingRescale}
For Fig.~4a,b in the main article we rescale the power spectrum to units of mechanical zero point motion. For this we use the calibration tone~\cite{SgorodetksyDeterminationVacuumOptomechanical2010a} (or also in the supplementary of~\cite{SzoepflSinglePhotonCoolingMicrowave2020a}). The calibration tone is used in all measurements and required for extracting the coupling strength/phonon number. Essentially it arises from frequency modulating our probe tone, which scans the cavity response and tells us about the transduction of our system. As in the measurement we only have access to the intensity fluctuations, $S_{II}$, from the probe tone, arising from the mechanical oscillation when probing the cavity. The calibration allows us to convert to frequency fluctuations, $S_{\omega \omega}$, of the cavity. Those are then related via the coupling to the position fluctuations, $S_{xx}$, of the mechanical mode.

The following relation allows to convert from intensity fluctuations to position fluctuations in terms of $x_{\text{ZPM}}$:
\begin{equation}
    \frac{S_{xx}}{x_{\text{ZPM}}^2} = \left( \frac{\phi_0 \omega_{mod}}{2} \frac{1}{S_{II} (\omega_{mod}) \text{ENBW}} \frac{1}{g_0^2} \right) S_{II} (\omega).
    \label{equS:scaleFac}
\end{equation}
Here, $\phi_0$ concerns the modulation index of the frequency modulation and $\omega_{mod}$ its frequency. $S_{II} (\omega_{mod})$ is the signal height of the calibration and $\text{ENBW}$ the normalisation with the bandwidth of the spectrum analyser (which is in contrast to all other quantities given in real frequency~\cite{SgorodetksyDeterminationVacuumOptomechanical2010a}). The normalisation with the coupling $g_0$ arises due to the conversion from $S_{\omega \omega}$ to $S_{xx}$ and $S_{II} (\omega)$ is the measured intensity spectrum.
\\As seen in the main manuscript, the noise floor decreases with increasing power as well as measuring closer to resonance. This might seem counter-intuitive, however arises due to the probably unusual normalisation to the spectra in terms of $x_{\text{ZPM}}$. It can be seen in the above equation,~\ref{equS:scaleFac}, that as the calibration tone increases, the noise floor will decrease. Thus the better transduction allows to resolve increasingly small fluctuations of the mechanical mode, resulting in a decreasing noise floor.

\section{Checking the calibration routine and estimation of attenuation between sample and HEMT}
\label{secS:caliCheck}

To extract the optomechanical coupling strength, we are using a frequency modulated calibration tone~\cite{SgorodetksyDeterminationVacuumOptomechanical2010a}, the Gorodetsky method, already briefly explained in Sec.~\ref{secS:bestCoolingRescale}. The frequency modulated tone scans the transduction of the cavity, which leads to a sideband and is then used as a calibration for the mechanical mode. This method can be verified by calculating the transduction coefficient from the response of the cavity. 

The coupling of the mechanical mode to the cavity leads to a frequency shift of the cavity. To measure this, we are using a fixed frequency probe tone, which undergoes an amplitude and/or phase modulation (depending on the probe cavity detuning) due to the optomechanical coupling.
\begin{figure*}[h]
    \centering
    \includegraphics[width = 17cm]{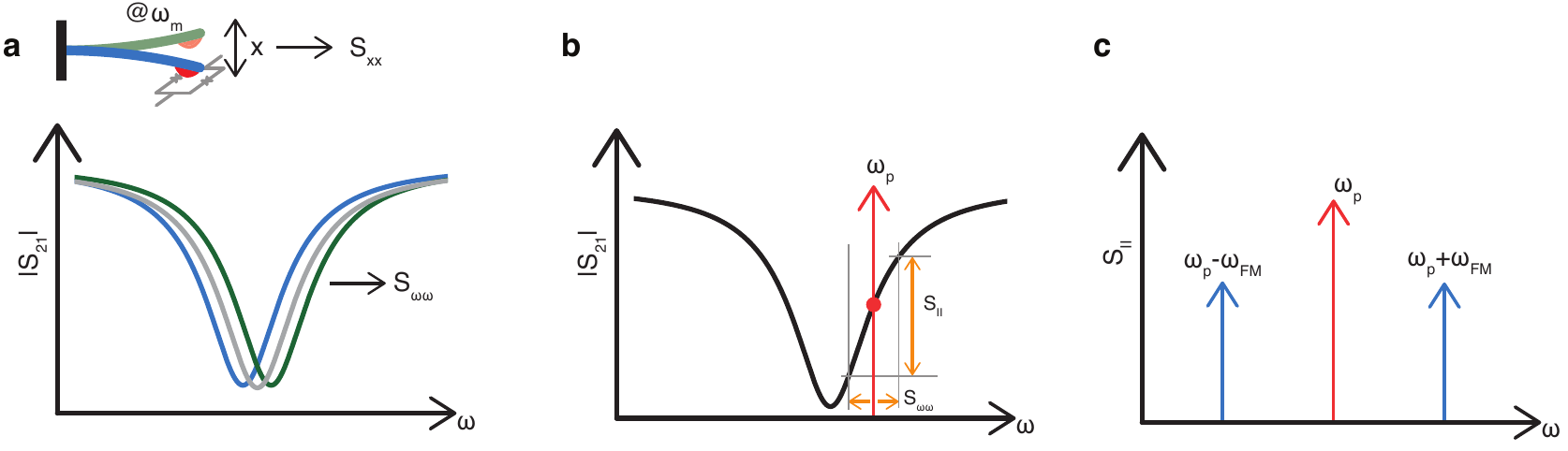}
    \caption{Illustration on how the mechanical position signal, $S_{xx}$, is converted to the measured intensity fluctuations $S_{II}$. \textbf{a.} Position change of the cantilever leads to a frequency shift of the cavity. \textbf{b.} This leads to amplitude (or phase) modulation of a fixed frequency probe tone with the mechanical frequency. The amplitude of the modulation depends on the slope (transduction of the cavity), as well as the interaction strength and population of the mechanical mode. \textbf{c.} Taking the spectrum, two symmetrical sidebands with a spacing of the mechanical frequency appear around the pump tone. The relative height compared to the pump tone allows to determine $g_0/2 \pi \times \sqrt{n_m}$, using the transduction of the cavity (see text for details).}
    \label{figS:GamChain_sketch_part1}
\end{figure*}
Fig.~\ref{figS:GamChain_sketch_part1} shows a sketch of how the position fluctuations of the mechanical mode lead to the modulation of this probe tone. This modulation leads to sidebands which are measured with the spectrum analyser. Crucially, the relative height between the carrier and the sidebands also depends on $g_0/2 \pi \times \sqrt{n_m}$, which allows to extract the coupling strength/phonon occupation. Conveniently both peaks are equivalently impacted by the amplification chain, thus exact knowledge of the amplification chain is not required. However, the height of the sidebands also depend on the amplitude and phase slope of the cavity at the frequency of the pump, which can be extracted by fitting the response of the cavity.

As a first step, it is useful to consider the general response of a fixed frequency probe tone, which scans a frequency modulated cavity. Such a frequency modulation leads to an amplitude and a phase modulation of the probe tone
\begin{equation}
    u_p = A_p (1+ \eta \sin{ (\omega_m t)}) \cos{ (\omega_p t + \nu \sin { (\omega_m t) } )}.
    \label{equS:AmpPhaseMod}
\end{equation}
Here $A_p$ is the amplitude of the probe tone at $\omega_p$. The modulation indices, $\eta$ and $\nu$, depend on the slope of the cavity amplitude and phase as well as the amplitude of the modulation
\begin{align}
\begin{split}
    \eta &= \frac{\partial \vert S_{21} \vert}{\partial \omega} \bigg|_{\omega = \omega_p} \cdot \Delta \omega^{\text{amp}} =: \alpha \cdot \Delta \omega^{\text{amp}}
    \\
    \nu &= \frac{\partial \text{arg} (S_{21})}{\partial \omega} \bigg|_{\omega = \omega_p} \cdot \Delta \omega^{\text{amp}} =: \beta \cdot \Delta \omega^{\text{amp}},
\end{split}
\label{eqs:AmplitudeModu}
\end{align}
where $\alpha$ and $\beta$ are the slope of the magnitude and the phase at the probe frequency. Here we are using the normalised (background subtracted) transmission, such that we have a transmission of unity far off resonance. The modulation amplitude depends on the interaction strength and the mechanical population
\begin{equation}
    \Delta \omega^{\text{amp}}_{\text{mech}} = 2 \sqrt{n_{m}} g_0.
    \label{equS:dOm_mech}
\end{equation}
In case of the calibration tone, the modulation amplitude is conversely given by the amplitude of the development of the frequency modulation ($\text{Dev}$)
\begin{equation}
    \Delta \omega^{\text{amp}}_{\text{cali}} = \text{Dev}.
    \label{equS:dOm_cali}
\end{equation}

To obtain the components at the individual frequencies we again considering Eq.~\ref{equS:AmpPhaseMod}. First, it is useful to introduce the Bessel functions and re-write Eq.~\ref{equS:AmpPhaseMod} as
\begin{equation}
    u_p = A_p (1+ \eta \sin{ (\omega_m t)}) \Sigma_n J_n(\nu) \cos{ (\omega_p t + n \omega_m t )},
    \label{equS:AmpPhaseMod_Bessel}
\end{equation}
where $J_n$ is the Bessel function of $n$-th order. Assuming that the modulation is small and using trigonometric identities we finally arrive at
\begin{equation}
    u_p = A_p \left( \cos{ (\omega_p t)} + \frac{\nu}{2} \cos{((\omega_p + \omega_m)t)} - \frac{\nu}{2} \cos{((\omega_p - \omega_m)t)} \right) + \frac{\eta}{2} A_p \left( \sin{((\omega_p + \omega_m)t)} + \sin{((\omega_p - \omega_m)t)} \right),
    \label{equS:AmpPhaseMod_Simpli}
\end{equation}
which is essentially combining the components arising due to the amplitude and the phase modulation. Now we can do a Fourier transform, and compare the spectral components. For this we took into account, that the spectrum is usually folded by the spectrum analyser as only positive frequencies are measured, which leads to a factor of two
\begin{align}
\begin{split}
    S_{II}^p &= \vert 2 u_p (\omega = \omega_p) \vert^2 = A_p^2
    \\
    S_{II}^m &= \vert 2 u_p (\omega = \omega_p + \omega_m) \vert^2 = \frac{\eta^2 + \nu^2}{4} A_p^2.
\end{split}
\label{eqs:AmpPhaseMod_FT}
\end{align}
Before comparing the signal strength, we have to take into account, that the mechanical peak has a finite linewidth, $\Gamma_m$, and the area below the peak has to be integrated, which yields~\cite{SNote1}
\begin{equation}
    S_{II}^m \rightarrow \int_{-\infty} ^{+\infty} \frac{S_{II}}{\text{ENBW}} \frac{d \omega}{2 \pi} = \frac{1}{2 \pi} \frac{S_{II} (\omega_m)}{\text{ENBW}} \frac{\pi \Gamma_m}{2} = \frac{S_{II} (\omega_m)}{\text{ENBW}} \frac{ \Gamma_m}{4}.
    \label{equS:IntOm_M}
\end{equation}
Several key observations are required to get this equation. We are interested in the power in the mechanical peak. This power is given by the area below the peak, which is therefore integrated. However for the integration we have to consider the power spectral density (PSD), which we obtain by dividing the measured power spectrum by the measurement bandwidth (ENBW). The ENBW is naturally given in Hz, which leads to the additional $2 \pi$ factor. This is in contrast to $\Gamma_m$, which is given in angular frequency, while the peak value is unchanged whether using a spectrum displayed in natural or angular frequencies. 

In contrast to this, for the pump tone it is sufficient to consider the peak height as it is a $\delta$-peak, and its height is invariant under different bandwidth
\begin{equation}
    S_{II}^p \rightarrow S_{II} (\omega_p).
    \label{equS:IntOm_P}
\end{equation}
Thus, we obtain the power below the pump tone simply from its value in the power spectrum.

Now, we can also verify the Gorodetsky calibration method, described in~\cite{SgorodetksyDeterminationVacuumOptomechanical2010a}. For this we use the second line of Eq.~\ref{eqs:AmpPhaseMod_FT} and replace $\nu$ and $\eta$ according to \ref{eqs:AmplitudeModu}. One time we use the modulation of the calibration (a $\delta$-peak, Eq.~\ref{equS:dOm_cali}) and one time of the mechanical mode (Eq.~\ref{equS:dOm_mech}). With this, we obtain after re-arranging
\begin{align}
\begin{split}
    \text{Dev}^2 &= \frac{4}{(\alpha^2 + \beta^2) A_p^2} S_{II} (\omega_{mod})
    \\
    g_0^2 n_{m} &= \frac{1}{(\alpha^2 + \beta^2) A_p^2} \frac{S_{II} (\omega_m)\Gamma_m/4}{\text{ENBW}},
\end{split}
\label{eqS:CaliCheck1}
\end{align}
where $\omega_{mod}$ is the frequency of the modulation. From the ratio of these two equations and some re-arranging we find
\begin{equation}
    g_0^2 = \frac{1}{2 n_{m}} \frac{\text{Dev}^2}{2} \frac{S_{II} (\omega_m) \Gamma_m/4}{S_{II} (\omega_{mod}) \text{ENBW}}
    \label{eqS:CaliCheck2}
\end{equation}
which is identical to the expression in~\cite{SgorodetksyDeterminationVacuumOptomechanical2010a}. For completeness it should be mentioned that only the $\text{ENBW}$ is in Hz, while all other quantities ($\text{Dev}, \omega_m, \Gamma_m$ and $\omega_{mod}$) are given in angular frequency.

However in this part one of the goals is to verify this calibration method by using the slope of the cavity response. This can be done by using the measured ratio between power in the carrier peak and in the mechanical sideband. It should be noted, that both signals undergo the same amplification chain between the sample and the spectrum analyser, such that there is no need for considering the amplification chain for this calculation. For taking the ratio, we use Eq.~\ref{eqs:AmpPhaseMod_FT} and take the areas below the peaks (Eqs.~\ref{equS:IntOm_M} and~\ref{equS:IntOm_P}) into account
\begin{equation}
    \frac{S_{II} (\omega_p)}{\frac{S_{II} (\omega_m)}{\text{ENBW}} \frac{\Gamma_m}{4}} = \frac{A_p^2}{\frac{A_p^2 (\eta^2 + \nu^2)}{4}}.
    \label{eqS:CaliCheck_NoMix1}
\end{equation}
Here we see that the pump amplitude at the sample, $A_p$, drops out. Now, we can include the amplitude of the modulation for $\eta$ and $\nu$ by using Eqs.~\ref{eqs:AmplitudeModu} and~\ref{equS:dOm_mech}. With this we find after re-arranging
\begin{equation}
    n_{m} g_0^2 = \frac{S_{II} (\omega_m) \Gamma_m/4}{(\alpha^2 + \beta^2) S_{II} (\omega_p) \text{ENBW}}.
    \label{eqS:CaliCheck_NoMixFinal}
\end{equation}

\begin{figure*}[h]
    \centering
    \includegraphics[width = 9cm]{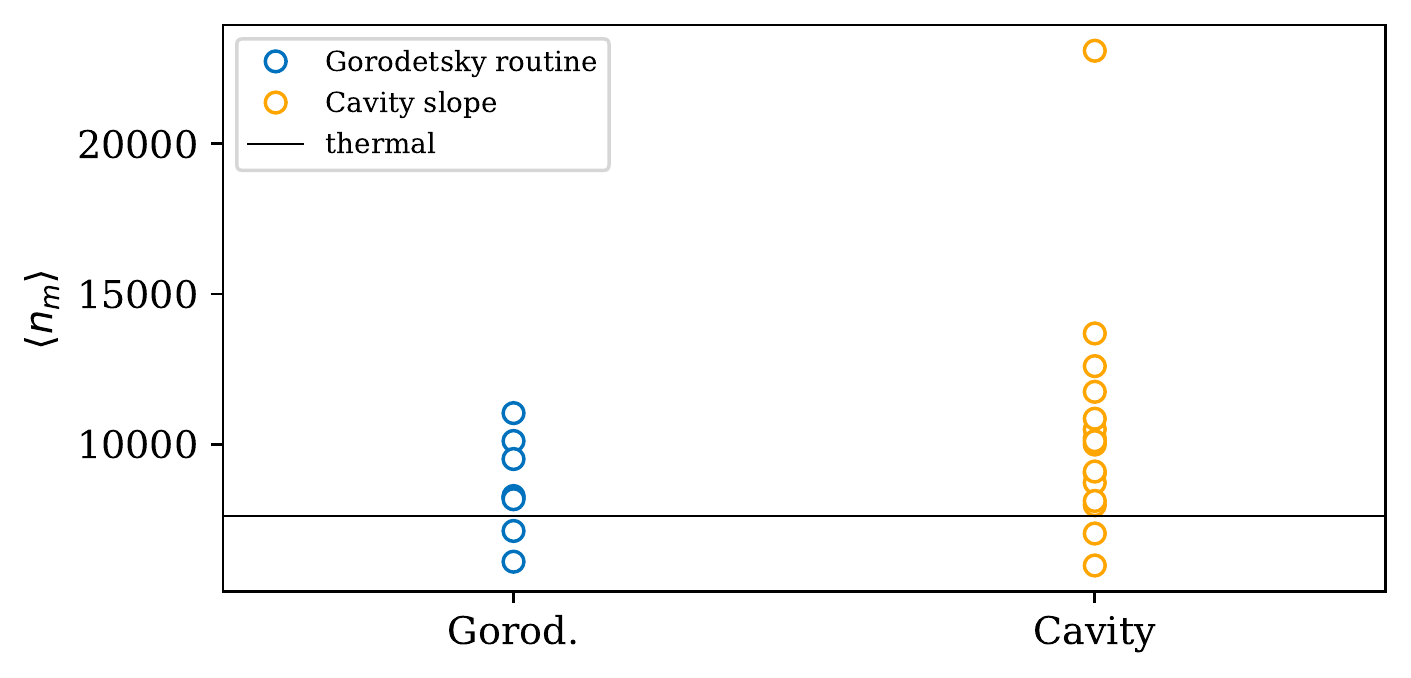}
    \caption{Comparison of the Gorodetsky calibration method~\cite{gorodetksyDeterminationVacuumOptomechanical2010a} and using the slope of the cavity response for the calibration (Eq.~\ref{eqS:CaliCheck_NoMixFinal}). We always measured on resonance and used the same cavity frequency (and thus coupling) for both measurements. We see very good agreement between both methods and have thus verified the calibration routine.}
    \label{figS:Goro_vs_Slope}
\end{figure*}

Generally, this is already sufficient to verify the calibration routine. However we have to additionally take into account, that we are measuring in a notch configuration~\cite{SprobstEfficientRobustAnalysis2015,SzoepflCharacterizationLowLoss2017}, where the signal from the cavity can scatter forwards and backwards. For the mechanical sideband, we can assume that the signal scatters equally forwards and backwards and thus we only measure half the signal in our $S_{21}$ measurement. For the probe tone, the ratio between forward and backward scattered traces is given by the measured $S_{21}$ value of the cavity response. Including this, we can verify the Gorodetsky calibration routine. This is done by taking multiple measurements of the (thermalised) mechanical mode. One time with the mixer and following the Gorodetsky calibration routine using a frequency modulation probe tone and one time, as discussed here, by using the cavity slope and the relative height between the probe tone and the sidebands to extract the coupling. The results are plotted in Fig.~\ref{figS:Goro_vs_Slope}, where we see good agreement and have therefore verified the calibration routine. However, this shows once more the beauty of the calibration method, which already takes the cavity line shape into account and includes everything in a single measured spectrum. We also observe a larger spread in the data taken with the cavity slope method, which already shows its higher sensitivity to changes in the experimental setup. For instance flux noise could occur during the measurement of the mechanical mode, which changes the cavity frequency and thus the measured transmission. For the Gorodetsky method this is not an issue as the calibration is measured simultaneously to the mechanical peak, while for the cavity slope it degrades the quality of the data measured.

Using the measurements of the mechanical mode it is also possible to estimate the attenuation between the sample and the HEMT. For this we only have to assume that our amplification chain is limited by the added noise from the HEMT, which we did in a separate measurement described in the following. In a first step we characterised the room temperature output amplification chain after the HEMT. For doing so, we used a matched load one time at room temperature and one time at \SI{77}{\kelvin}. From this we could extract an equivalent noise temperature of \SI{599.8}{\kelvin} at the input of the room temperature amplification chain and a gain of around \SI{61}{\decibel}. These are reasonable numbers, and the gain is in the expected range, given that we are using two room temperature amplifiers and have several attenuating elements (Section~\ref{secS:setup}). Now we can use those parameters in combination with the HEMT parameters (according to the datasheet the HEMT has a gain of around \SI{35}{\decibel}) and compare the estimated to the measured noise floor. While we measure a noise floor of -111.4~dBm, we estimated one of -106.2~dBm, where the difference might arise from additional unaccounted attenuation or a slightly lower gain of the HEMT. According to the datasheet, the HEMT has a noise temperature of \SI{4.2}{\kelvin}. We can also assign a theoretical noise temperature to the room temperature amplification chain at the input of the HEMT, by dividing this noise temperature (\SI{599.8}{\kelvin}) through the HEMT gain. Then we find value of \SI{0.75}{\kelvin} at the HEMT input, which shows that our noise floor is limited by HEMT noise.

\begin{figure*}[h]
    \centering
    \includegraphics[width = \textwidth]{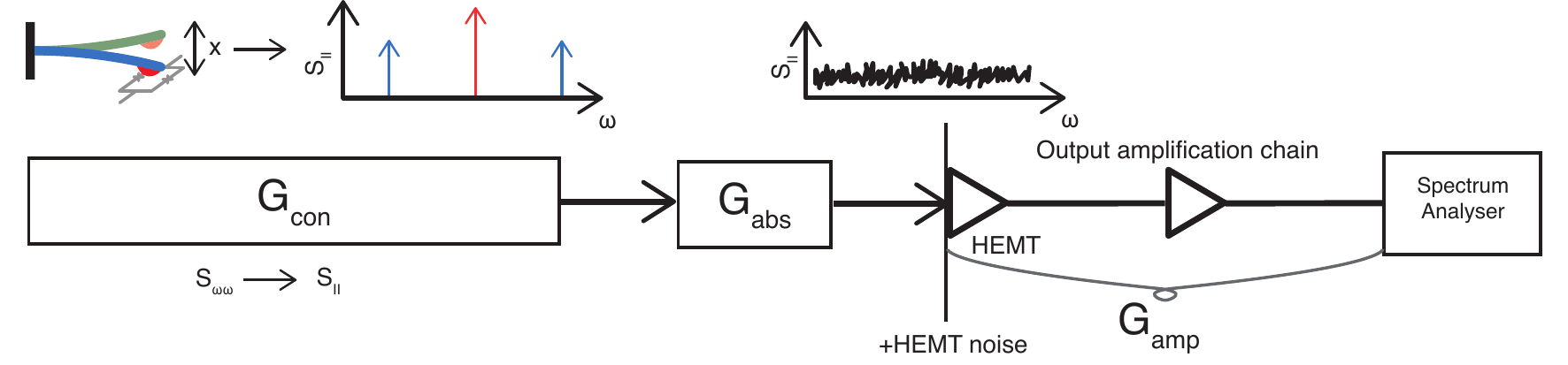}
    \caption{Illustration of the output amplification chain. The position noise of the mechanical mode is converted to the measured intensity fluctuations (described in detail in Fig.~\ref{figS:GamChain_sketch_part1}). The position noise is first converted to frequency fluctuations of the cavity, which is proportional to the coupling strength. Then these frequency fluctuations are converted to intensity fluctuations, which is described by a conversion factor, $G_{con}$. Afterwards the signal undergoes some attenuation, $G_{abs}$, before it goes through the amplification chain. First it is amplified by the HEMT, which also dominates the noise floor. Afterwards it goes through the setup at room temperature, including additional amplifiers. All amplification/attenuation from the HEMT input onwards is combined by $G_{amp}$.}
    \label{figS:GamChain_sketch_part2}
\end{figure*}

The output path the signal takes is illustrated in Fig.~\ref{figS:GamChain_sketch_part2}. In a first step, position noise from the mechanical mode is converted to frequency and then intensity noise, shown in more detail in Fig.~\ref{figS:GamChain_sketch_part1}. The conversion between position and frequency noise is basically given by the coupling: $S_{\omega \omega}^m = g_0^2 / x_{zpm}^2 S_{xx}^m = 4 g_0^2 n_m$. To convert then to intensity, the conversion factor is required
\begin{equation}
    S_{II}^m = G_{con} S_{\omega \omega}^m.
    \label{eqS:Gcon1}
\end{equation}
To obtain $G_{con}$ it is convenient to re-express the second line of Eq.~\ref{eqs:AmpPhaseMod_FT}
\begin{equation}
    S_{II}^m = \frac{\eta^2 + \nu^2}{4} A_p^2 = \frac{\alpha^2 + \beta^2}{4} A_p^2 \left( \Delta \omega^{\text{amp}} \right) ^2 = G_{con} \left( \Delta \omega^{\text{amp}} \right) ^2 = G_{con} S_{\omega \omega}^m
    \label{eqS:Gcon2}
\end{equation}
For this we used that $\eta$ and $\nu$ depend on the slope of the cavity response (given by $\alpha$ and $\beta$, Eq.~\ref{eqs:AmplitudeModu}) and the amplitude of the frequency modulation, given by $S_{\omega \omega}$.

Now, we need to consider the rest of the output chain. This is however straight forward, as the complete signal undergoes the full amplification/attenuation of the output chain. Thus the measured signal at the spectrum analyser and the intensity fluctuations measured directly at the cavity relate as
\begin{equation}
    S_{SA} = G_{abs} G_{amp} S_{II}
    \label{eqS:Gamp_Gabs}
\end{equation}
It is useful to make a distinction between $G_{abs}$ and $G_{amp}$, as usually the noise floor is dominated by the HEMT noise. For $G_{abs}$ we expect several dB of loss, while for $G_{amp}$ we expect high amplification due to our HEMT and two room temperature amplifiers (Section~\ref{secS:setup}). As we are indeed limited by the noise floor of the HEMT, we can use the noise temperature of the HEMT to estimate $G_{amp}$. For this we find values of around \SI{91}{\decibel}, which implies a HEMT gain of around \SI{30}{\decibel}, in agreement with the separated measurement discussed previously.

\begin{figure*}[h]
    \centering
    \includegraphics[width = 9cm]{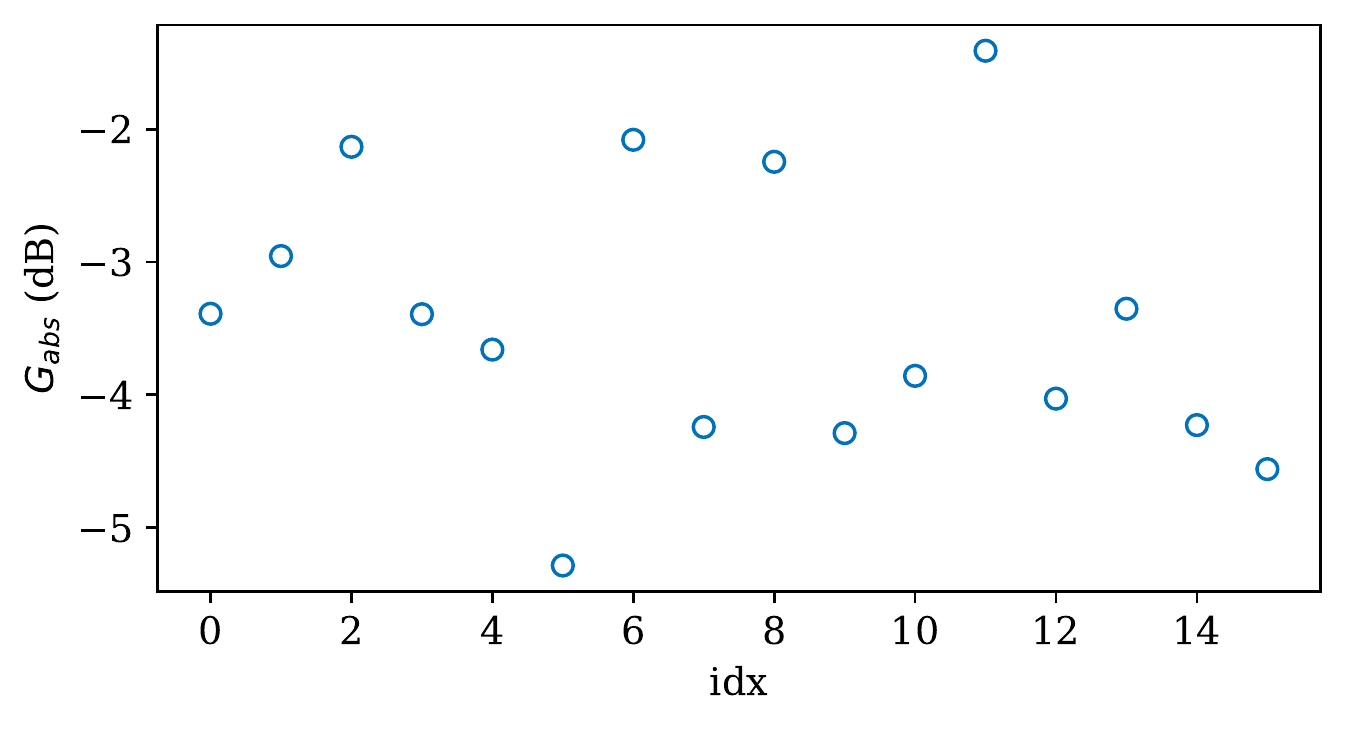}
    \caption{Estimated sample-HEMT attenuation for several measurements. The attenuation is around \SI{3.5}{\decibel}, which is a reasonable value, compatible with the expected attenuation form the isolatros and cables.}
    \label{figS:Gabs}
\end{figure*}

The only unknown value is then $G_{abs}$, which we are able to calculate using Eq.~\ref{eqS:Gamp_Gabs} and~\ref{eqS:Gcon2}. Additionally we will again use, that the mechanical peak is a Lorentzian peak with finite area (Eq.~\ref{equS:IntOm_M}). Thus we find
\begin{equation}
    \frac{1}{G_{abs} G_{amp}} S^m_{SA} = \frac{1}{G_{abs} G_{amp}} \frac{S_{SA} (\omega_m) \Gamma_m/4}{\text{ENBW}} = A_p^2 \frac{\alpha^2 + \beta^2}{4} 4 n_{m} g_0^2,
    \label{eqS:Gabs_final1}
\end{equation}
which can be re-arranged to arrive at
\begin{equation}
    G_{abs} = \frac{1}{G_{amp}} \frac{1}{(\alpha^2 + \beta^2) A_p^2} \frac{S_{SA} (\omega_m) \Gamma_m/4}{\text{ENBW}} \frac{1}{n_{m} g_0^2}.
    \label{eqS:Gabs_final2}
\end{equation}
Here $n_{m}$ can be further replaced by $k_b T/ \hbar \omega_m$. We see, that it is crucial to know the power at the sample, $A_p^2$, for obtaining a faithful estimation for the attenuation factor. For this we use a calibration obtained from the cooling traces which is also verified by the known line attenuation. Still we want to emphasise that this still has some uncertainty. Furthermore we have to be careful again, as we are measuring in the notch configuration, where the signal is scattered in both directions. Also taking this into account, the results for the estimated sample-HEMT attenuation are plotted in Fig.~\ref{figS:Gabs}. While the estimated attenuation varies between different measurements, it seems to be around \SI{3.5}{\decibel}. This is a reasonable value, as we expect an attenuation of \SI{0.6}{\decibel} to \SI{1.1}{\decibel} per isolator (according to the data sheet) and the additional attenuation likely comes from a number of cables before the HEMT. We also observe a spread of around \SI{3}{\decibel}, which again shows the sensitivity of the data to the slope of the cavity, already observed in Fig.~\ref{figS:Goro_vs_Slope}.

\end{CJK*}

\end{document}